\documentclass[fleqn,usenatbib]{mnras}

\usepackage{newtxtext,newtxmath}
\usepackage[T1]{fontenc}
\usepackage{xcolor}

\DeclareRobustCommand{\VAN}[3]{#2}
\let\VANthebibliography\thebibliography
\def\thebibliography{\DeclareRobustCommand{\VAN}[3]{##3}\VANthebibliography}

\usepackage{graphicx}	
\usepackage{amsmath}	
\usepackage{overpic}

\newcommand{\teff}{$T_{\text{eff}}$}
\newcommand{\logg}{$\log g$} 
\newcommand{\logtau}{$\log (\tau_{500})$}
\newcommand{\taufive}{$\tau_{500}$}
\newcommand{\meanBz}{$\langle B_z\rangle$}
\newcommand{\equals}{\,=\,}

\title[Photospheric spectral variability due to faculae]{Spectral variability of photospheric radiation due to faculae II: Facular contrasts for cool main-sequence stars.}
\author[C. M. Norris et al.]
{Charlotte M. Norris$^{1}$, Yvonne C. Unruh$^{1}$\thanks{E-mail: y.unruh@imperial.ac.uk (YCU)}, Veronika Witzke$^{2}$, Sami K. Solanki$^{2}$, Natalie A. Krivova$^{2}$, \newauthor
Alexander I. Shapiro$^{2}$, 
Kok Leng Yeo$^{2}$, Robert  Cameron$^{2}$ and Benjamin Beeck$^{2}$
\\
$^{1}$Department of Physics, Imperial College London, London SW7 2AZ, UK\\
$^{2}$Max Planck Institute for Solar System Research, Justus-von-Liebig-Weg 3, 37077 G\"ottingen, Germany\\
}

\date{Accepted 18th April 2023. Received XXX; in original form 24th March 2023}

\pubyear{2023}

\begin{document}
\label{firstpage}
\pagerange{\pageref{firstpage}--\pageref{lastpage}}
\maketitle

\begin{abstract}
Magnetic features on the surface of stars, such as spots and faculae, cause stellar spectral variability on time-scales of days and longer. For stars other than the Sun, the spectral signatures of faculae are poorly understood, limiting our ability to account for stellar pollution in exoplanet transit observations. 
Here we present the first facular contrasts derived from magnetoconvection simulations for K0, M0 and M2 main-sequence stars and compare them to previous calculations for G2 main-sequence stars. We simulate photospheres and immediate subsurface layers of main-sequence spectral types between K0 and M2, with different injected vertical magnetic fields (0\,G, 100\,G, 300\,G and 500\,G) using MURaM, a 3D radiation-magnetohydrodynamics code. We show synthetic spectra and contrasts from the UV (300\,nm) to the IR (10\,000\,nm) calculated using the ATLAS9 radiative transfer code. The calculations are performed for nine viewing angles to characterise the facular radiation across the disc. The brightness contrasts of magnetic regions are found to change significantly across spectral type, wavelength and magnetic field strength, leading to the conclusion that accurate contrasts cannot be found by scaling solar values. This is due to features of different size, apparent structure and spectral brightness emerging in the presence of a given magnetic field for different spectral types.
\end{abstract}

\section{Introduction}
Magnetic activity is thought to be the main source of variability of late-type stars on time-scales of a day and longer. Over 96\% of the total solar irradiance (TSI) variability on these time-scales is reproduced by models attributing  the variability to magnetic features on the solar surface (including dark spots and bright faculae \citep{Krivova2006, Yeo2017}. The radiation observed from each magnetic feature on the surface of the Sun changes as the feature evolves. Additionally, the radiation is modulated due to the movement of the feature across the observed disc caused by rotation. Therefore, both the appearance and disappearance of features on the surface, as well as their location on the disc, affect the radiative output of a star.

Although faculae are small compared to spots, and often low contrast at disc centre, they strongly influence the TSI and solar spectral irradiance (SSI) and, due to their large number and longer lifetimes, lead to the Sun being brighter in times of high magnetic activity \citep[see, e.g.,][]{Solanki2013b}. 
Therefore, being able to characterise the radiation from faculae is important for fully understanding the variability of the Sun and likely also of other stars. Stellar variability in turn is important to understand as a noise source for planetary characterisation, particularly when using the transit method \citep[for a recent overview, see, e.g.,][]{Rackham2022SAG}. The presence of stellar activity has two main effects on the transit lightcurves. On the one hand, the occultation of active regions leads to ``stellar noise'' with bumps and troughs due to occulted dark and bright regions  \citep[see, e.g.,][]{Oshagh+2014faculae,Kirk2016,Espinoza+2019}. On the other hand, the change in the radiation emitted from the unocculted stellar disc affects the transit depth through what is often termed the ``transit light-source effect'' (TLSE) \citep[see][]{rackham+2018TLSE1,rackham2019TLSE2}. 

Small-scale magnetic features on stars other than the Sun are currently unresolvable, leaving the Sun as our main source of information regarding faculae.  However, even on the Sun, the contrasts of faculae are difficult to measure, resulting in measurements having only been taken in a few wavelength bands \citep{Chapman1977,Ermolli2013,Yeo2013}. To obtain full spectra of these features, model atmospheres and radiative transfer methods must be employed. 

One dimensional atmospheric models have been used to synthesise spectra for quiet-Sun and solar facular regions  \citep[see, e.g.,][]{Fontenla1993, Unruh1999}. More recently, one dimensional photospheric models have been used by \citet{Witzke2018} to explore the effects of metallicity on facular contrasts for stars with solar effective temperature. To calculate intensity spectra at different viewing angles, the optical depth is artificially adjusted to account for the longer path length traversed when the stellar atmosphere is viewed towards the limb. This method does not account for the corrugated nature of the granular structure in the atmosphere, or the 3D nature of the faculae. Therefore, while these models capture the overall disc-integrated properties of faculae reasonably well, they do not reproduce the observed limb-dependent facular contrasts. To obtain accurate centre-to-limb variations of magnetic regions on stars, 3D models must be employed. 

A range of such models now exist, including Stagger \citep[][]{Galsgaard1996,Magic2013,CubasArmas+Fabbian2021}, CO5BOLD \citep[][]{Freytag+2012,Salhab+2018}, and MURaM \citep[see][and Sec.~\ref{sect:muram} for more detail]{Vogler2005,Beeck2013}. These have been employed to model line profile shapes and brightness variations on late-type stars. While some differences remain between the different magnetoconvection models, there is generally good agreement of the general granulation properties and observables \citep[see, e.g.,][for a comparsion of Stagger, CO5BOLD and MURaM]{Beeck2012}. \par
In \citet{Norris2017}, henceforth paper \textsc{i}, 3D model atmospheres were used to obtain spectra from the UV to the IR for magnetised regions on G2V spectral type stars at different viewing angles on the stellar disc. Here, we employ the same methods as in paper \textsc{i}, described in section~\ref{sect:methods}, for spectral types of K0V, M0 and M2V. This paper explores the differences in radiation emitted from simulated magnetic regions on various main-sequence spectral types from M2 to G2.
The spectral types compared in this paper are characterised by different effective temperatures, \teff, and strengths of gravitational acceleration, $\log g$, defined in Table~\ref{table:SpectralTypes}. \par
Images of the emergent intensities at a range of wavelengths and viewing angles are presented in Sec.~\ref{sec:images}. Mean contrast spectra with respect to field-free simulations are explored in Sec.~\ref{sect:contrasts}. The mean contrast spectra are used to compare different spectral types and activities for regions of otherwise similar properties (e.g.,~the number of granules).
Our findings are discussed and summarised in Sec.~\ref{sec:discussion}
\par

\begin{table}
\caption{Spectral types and simulation parameters. Effective temperatures, \teff, and their standard deviations are for the field-free simulations and are derived from disc-integrated synthetic intensity spectra calculated with ATLAS9 in this study (see text); the sample standard deviations include a correction factor, $N-1$. 
The gravitational field strength is given in cgs units in column 3. The total depth of the cubes is listed in column 4; the horizontal extent is given in column 5. Columns 6 and 7 list the resolution in the vertical ($z$) and horizontal directions; column 8 gives the number of snapshots, $N$. All cubes encompass 512 by 512 pixels in the horizontal directions. }
\centering
\centering
\begin{tabular}{c c c c c c c c}
\hline\hline
\!Spectral\! & field-free & \logg & depth & width & \multicolumn{2}{c}{Resolution} & snap- \\ 
type & \teff[K] & [cgs] & [km] & [km] & $z$ & $x/y$ & shots\\
[0.5ex]
 \hline
G2V & 5847 $\pm$ 6 & 4.44 & 3000 & 9000 & 10.0 & 17.6 &  10 \\ 
K0V & 4903 $\pm$ 6 & 4.61 & 1800 & 6000 & 6.0 & 11.7 & 6 \\ 
M0V & 3906 $\pm$ 1 & 4.83 & 1000 & 2500 & 4.0 & 4.9 & 6 \\ 
M2V & 3676 $\pm$ 1 & 4.83 & \ 800 & 1500 & 3.2 & 3.0 & 6 \\ 
\hline
\end{tabular}
\label{table:SpectralTypes}
\end{table}

\section{Synthesising Spectra} \label{sect:methods}
\subsection{MURaM simulations}
\label{sect:muram}
The mean spectra presented in this paper were derived from 3D snapshots of the surface of late-type stars. The snapshots are taken from different simulation runs that were either magnetic field-free or had different spatially averaged magnetic flux densities. The grids of temperature, pressure, density, magnetic field and velocity for "box-in-a-star" simulations, containing the photosphere and upper layers of the convection zone, were produced using the MURaM magneto-hydrodynamics code \citep{Vogler2005}. MURaM allows for tuning of the effective temperature of the atmosphere by adjusting the constant entropy density inflow through the bottom boundary \citep{Beeckthesis}; {\logg} can also be set, allowing for the simulation of different spectral types. The boundaries for the box in the horizontal direction are periodic. The simulations used here are as described in \cite{Beeck2013}, with the upper boundary closed for flows. They use the OPAL equation-of-state \citep{Rogers1994,Rogers1996}. Heating and cooling rates are calculated by solving the radiative transfer equation in four bins following the approach by \cite{Nordlund1982}. Opacities in the four bins are derived from the ATLAS9, solar metallicity, opacity distribution functions \citep[ODFs --][]{Kurucz1993,CastelliKurucz2001}.\par

Simulations were performed for four combinations of effective temperature and gravitational acceleration to represent stellar spectral types between G2V and M2V (see Table~\ref{table:SpectralTypes}). The grid of simulations used for the G2, K0, M0 and M2 spectral types in this study have been presented in \citet{Beeck2013,Beeck2013b,Beeck2015,Beeck2015b} and \citet{Beeckthesis}. Results of the spectral synthesis from G2 simulations have been described in \cite{Yeo2017} and \cite{Norris2017}. Here we present equivalent calculations for K0, M0 and M2 main-sequence stars. Each spectral type has a different horizontal and vertical resolution chosen such that the simulation box contains a similar number of granules (approximately 25 for field-free simulations). All simulations are 512 by 512 pixels in the horizontal direction while the number of depth pixels changes from one simulation run to another. Table~\ref{table:SpectralTypes} shows the parameters used for each spectral type. The table includes the magnetic field-free effective temperature (see below), logarithmic gravitational acceleration (\logg) used in the MURaM simulations, the depth and extent of the simulation cubes, as well as the vertical ($z$) and horizontal ($x/y$) resolution. 

\begin{table}
\centering
\caption[Contrasts of magnetic snapshots]{List of magnetic simulations. Columns one and two list the main-sequence spectral types and mean magnetic field; columns 3 and 4 list the disc-integrated bolometric contrast and its standard deviation with respect to the non-magnetic snapshots. Column 5 gives the difference between the effective temperature (as measure of the bolometric flux) of the listed magnetic and the non-magnetic run. The last column lists the number of snapshots used for the facular intensities; see Tab.~\ref{table:SpectralTypes} for the non-magnetic snapshots.}

\begin{tabular}{c c c c c c}
\hline\hline
Spectral & {\meanBz} & $c_{\rm bol}$ & $\sigma_c$ & $\Delta T$ 
 & snap- \\
 type   &   [G]     &               &    & [K] & shots \\[0.5ex]
 \hline
 G2V & 100 & 0.016 & 0.006 & +24 & 10 \\ 
G2V & 300 & 0.039 & 0.009 & +56 & 10 \\ 
G2V & 500 & 0.050 & 0.007 & +72 & 10 \\ 
K0V & 100 & 0.019 & 0.006 & +23 & 6 \\ 
K0V & 300 & 0.033 & 0.007 & +40 & 12 \\ 
K0V & 500 & 0.027 & 0.008 & +32 & 6\\ 
M0V & 100 & 0.004 & 0.001 & +4 & 6 \\ 
M0V & 300 & 0.005 & 0.001 & +5 & 14 \\ 
M0V & 500 & 0.000 & 0.001 & 0 & 6\\
M2V & 100 & 0.002 & 0.002 & +2 & 6 \\ 
M2V & 300 & -0.002 & 0.002  & $-$2 & 12 \\ 
M2V & 500 & -0.014 & 0.001 & $-$13 & 6\\
 \hline
\end{tabular}
\label{tab:contrast}
\end{table}
Magnetic field-free (henceforth hydrodynamic or field-free) simulations are run for several hours, a time comparable to the Kelvin-Helmholtz time scale of the simulation boxes. After this time, the simulations are independent of the initial conditions. The simulations are then run further and at least six snapshots are selected at roughly constant time intervals for a given spectral type. On average this time interval is 6 minutes for field-free simulations. This time interval is chosen to allow the granulation patterns to evolve sufficiently between snapshots to make them independent. However, on the time-scales of these snapshots, granules tend to re-appear in similar locations, thus, the granule positions will be correlated across the snapshots \citep{Cegla2018}. While the average total radiative flux of the snapshots is close to the average over the whole relaxed portion of the simulation, longer time series \citep[see, e.g.,][]{Salhab+2018,Thaler+Spruit2014} are necessary to accurately determine the intrinsic variability of the bolometric flux along with the bolometric flux deficits or enhancements due to the magnetic flux. 

Translated into effective temperature, the standard deviation for the field-free snapshots is given in Table~\ref{table:SpectralTypes} along with the mean effective temperature, \teff. The \teff\ listed here has been derived from disc-integrated spectral intensities calculated with ATLAS9 (see Sec.~\ref{sect:spectralsynth} for details). In a first step, the spectral intensities were integrated between 149.5\,nm and 160\,000\,nm to yield bolometric intensities.
As our calculations include limb distances between $\mu=0.2$ and $\mu=1.0$, a  3-parameter non-linear limb darkening law \citep{Sing2009} was then used to extrapolate intensities to the limb and integrated over the stellar disc to yield bolometric fluxes. The effective temperatures listed for the K0, M0 and M2 simulations are slightly different from the temperatures given in \cite{Beeck2013} mainly due to the different number of frequency bins and angles in ATLAS9 and MURaM. The G2V simulations used here are those presented in \cite{Norris2017} and \cite{Yeo2017} and differ from the ones in \cite{Beeck2013}. \par
Homogeneous vertical magnetic fields of 100~G, 300~G\footnote{The K0, M0 and M2 simulations with {\meanBz\equals}300~G were not presented in \citet{Beeck2013,Beeck2013b,Beeck2015,Beeck2015b} and are new to this publication.} and 500~G are injected into one of the relaxed hydrodynamic snapshots to represent a range of activity levels. 
For each initial average vertical magnetic field, \meanBz, the simulations are allowed to relax. To determine whether simulations are relaxed, the effective temperature, \teff, and the horizontally averaged magnetic energy density of the simulation are confirmed to be quasi-stationary; see \citet{Beeckthesis} for more information on this relaxation criterion. As in the case of the field-free simulations, snapshots are taken at intervals that allow for sufficient evolution of the granulation pattern. As the number of surface elements is large and synthesising spectra is time expensive, we use between six and fourteen snapshots for each of our chosen values of the average magnetic field. The average bolometric flux for these snapshots is close to the average over the whole interval of time when the simulation is in a relaxed state. The bolometric contrasts, the standard deviations and effective temperature differences of the magnetic simulations with respect to the non-magnetic simulations are listed in Tab.~\ref{tab:contrast}. 

\subsection{Spectral Synthesis}
\label{sect:spectralsynth}
Emergent intensities from the ultraviolet (UV) to the far infrared (FIR) are synthesised using the radiative transfer code, ATLAS9 \citep{Kurucz1992, Castelli1994}. Rays are laid through each pixel of the simulation box, producing a grid of 1D atmospheres. For viewing angles away from disc centre, the rays are inclined at an angle, $\theta$, to the local normal. These sight lines are pivoted at the mean geometrical depth ($z$) where the disc-centre optical depth at 500~nm ({\taufive}$_{,\, \text{DC}}$) is unity. This causes surface features to appear at a similar $x$/$y$ position in the snapshot at each limb distance, $\mu$\equals$\cos\theta$. We assume that intensities are a function of viewing angle only. As the extent of the simulation boxes is small compared to the stellar radius, we can neglect changes in the viewing angle within a box.  Furthermore, at the low spectral resolution considered here, the position of the box relative to the rotation axis of the star does not need to be taken into account.

Inclined rays are calculated between $\mu$~=~1.0 and 0.2 in increments of 0.1 by 2D linear interpolation along the line of sight at depth intervals that match that of the simulation. For each ray the column mass, electron number density and continuum absorption coefficient at 500~nm are derived. For use in ATLAS9, these parameters are interpolated on to an evenly spaced {\logtau} grid with $256$ points between {\logtau}~=~$-5.0$ and $2.5$, where most of the radiation is produced.\par
Synthetic spectra are produced for each sight line of the MURaM box using solar metallicity ODF tables to account for the opacity of the millions of atomic and molecular spectral lines in a stellar atmosphere, particularly in the UV \citep{CastelliKurucz2001,Castelli2004}. In paper \textsc{i}, outdated ODF tables were used. These have been updated here to the ODFs presented in \citet{Castelli2004}, with additionally updated H$_2$O contributions produced in 2012. These ODFs use abundances from \citet{Grevesse1998}, as well as TiO lines from \citet{Schwenke1998}. Other improvements for the new ODFs are described in \citet{Castelli2004}. For each pixel, we calculate spectra ranging from 149.5~nm to 160\,000~nm. The spectra are calculated for 1040 wavelength bins that correspond to the resolution provided by the ATLAS9 ODFs. The resolution of the spectra thus ranges from 1~nm in the UV, to 2~nm in the visible, and up to 20\,000~nm in the far infrared. 

The spectral synthesis calculations are performed under the assumption of local thermodynamic equilibrium (LTE). This assumption becomes less valid high in the atmosphere, leading to intensities calculated for wavelengths below 300~nm to deviate significantly from more accurate, but time intensive, non-LTE calculations. We thus focus on wavelengths longward of 300~nm here (except for the computation of the bolometric intensities and \teff{} values). Further calculations are under way to explore non-LTE effects for wavelengths below 300\,nm \citep{tagirov2023}. 

\section[Stellar spectral type and disc centre snapshots]{Intensity images for different spectral types and magnetic flux densities}
\label{sec:images}
\subsection{Disc-centre images}
\label{sect:disccentre}
\begin{figure*}
\centering
\vspace*{3.5ex}

\begin{overpic}[width=\textwidth]{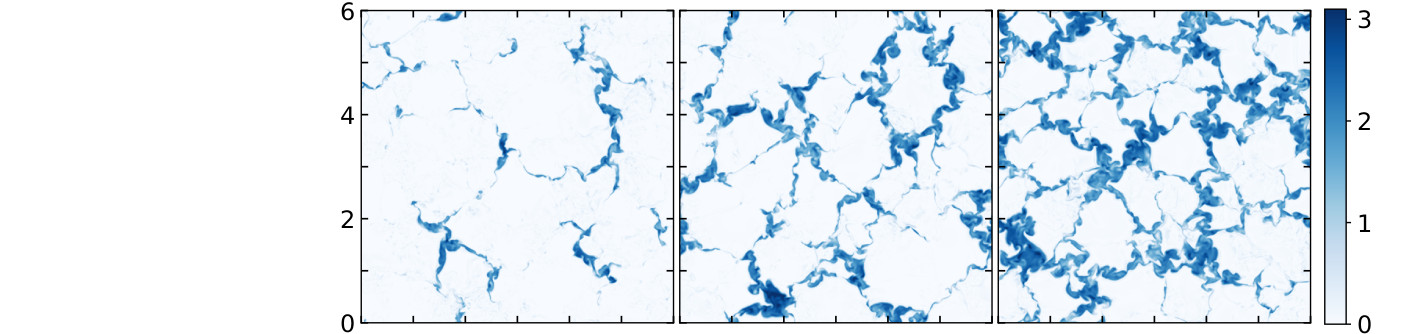}
    \put(11,23.7){\textsf{\Large hydro  \hspace{5.6em} $\langle B_z \rangle =$\,100\,G \hspace{4.2em} $\langle B_z \rangle =$\,300\,G \hspace{4.2em} $\langle B_z \rangle =$\,500\,G }}
\end{overpic}
\begin{overpic}[width=\textwidth,trim={0 2 0 3},clip=]{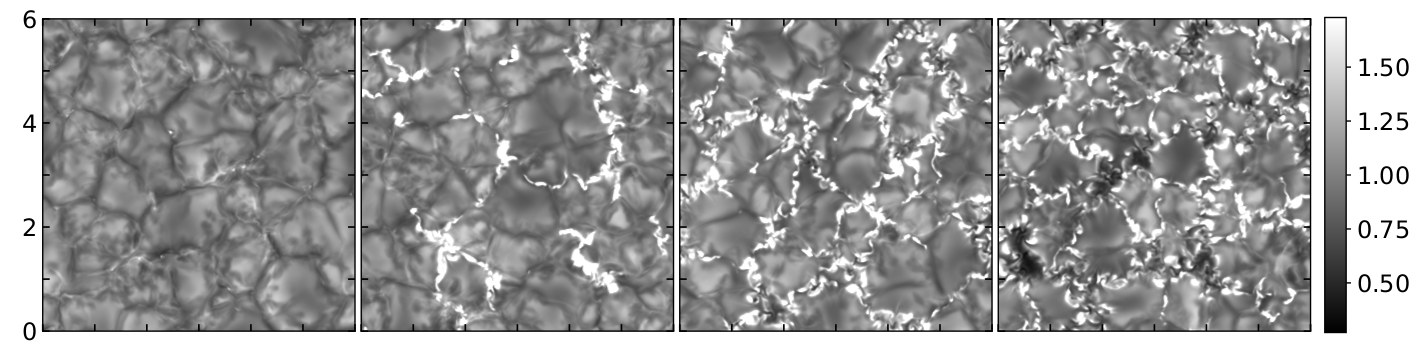}
    \put(4,20.5){\textsf{\textbf{\large \textcolor{white}{388\,nm}}}}
\end{overpic}
\begin{overpic}[width=\textwidth,trim={0 2 0 5},clip]{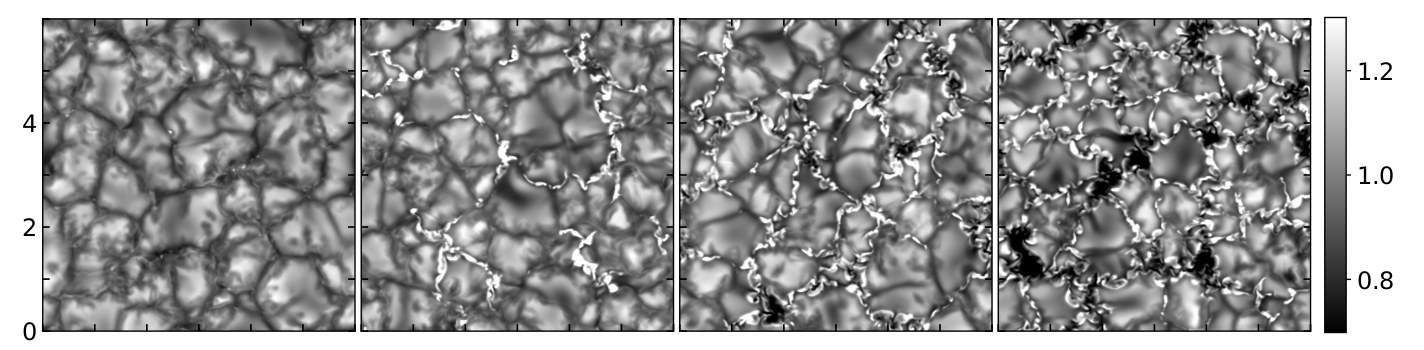}
    \put(4,20.5){\textsf{\textbf{\large \textcolor{white}{602\,nm}}}}
\end{overpic}
\begin{overpic}[width=\textwidth,trim={0 2 0 5},clip]{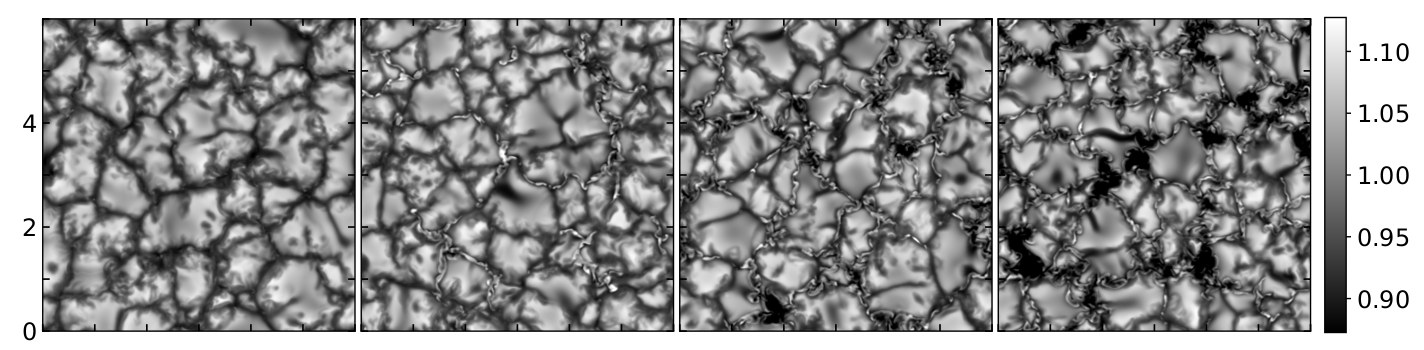}
    \put(3.5,20.5){\colorbox{lightgray}{\textsf{\textbf{\large \textcolor{white}{1610\,nm}}}}}
\end{overpic}
\begin{overpic}[width=\textwidth,trim={0 0 0 5},clip]{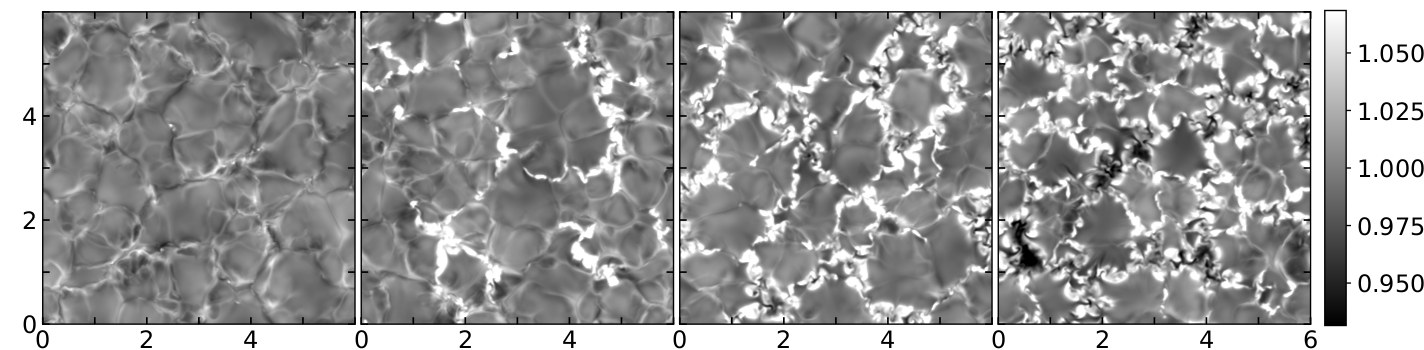}
    \put(4,21){\textsf{\textbf{\large \textcolor{white}{8040\,nm}}}}
\end{overpic}
\caption[]{Emergent intensities for K0 snapshots. {\it Top row:} unsigned vertical magnetic flux density (in units of kG) at a depth where \taufive\ is unity. {\it Rows 2 to 5:} disc-centre emergent intensities at 388~nm, 602~nm, 1610~nm and 8040~nm. From left to right, the images are for field-free, \meanBz\equals 100\,G, 300\,G and 500\,G, respectively. The $x$ and $y$-axes show the size of the snapshots in Mm. Intensities have been normalised by the mean intensity of the field-free simulation (see Tab.~\protect{\ref{tab:SpecInt_SN}}). Greyscales for intensity images saturate at $\pm 2 \sigma_\lambda$, where $\sigma_\lambda$ is the standard deviation of the intensity within a snapshot at wavelength $\lambda$.} 
\label{fig:images_K0}
\end{figure*}

\begin{figure*}
\centering
\vspace*{3.5ex}

\begin{overpic}[width=\textwidth]{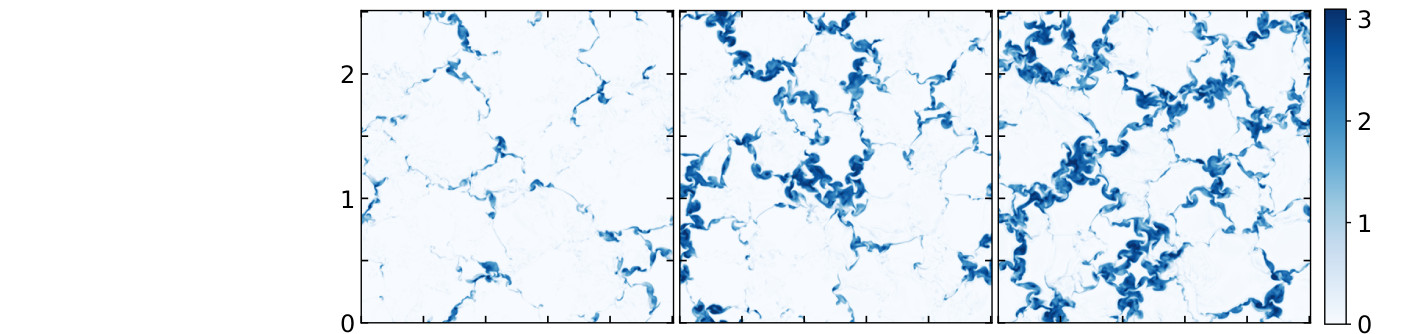}
    \put(11,23.7){\textsf{\Large hydro  \hspace{5.6em} $\langle B_z \rangle =$\,100\,G \hspace{4.2em} $\langle B_z \rangle =$\,300\,G \hspace{4.2em} $\langle B_z \rangle =$\,500\,G }}
\end{overpic}
\begin{overpic}[width=\textwidth,trim={0 1 0 4},clip]{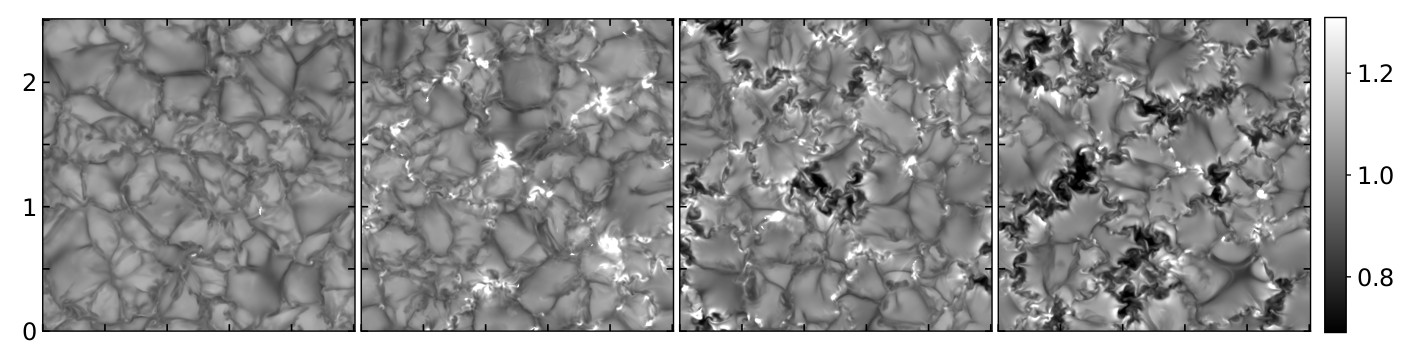}
    \put(4,20){\textsf{\textbf{\large \textcolor{white}{388\,nm}}}}
\end{overpic}
\begin{overpic}[width=\textwidth,trim={0 1 0 4},clip]{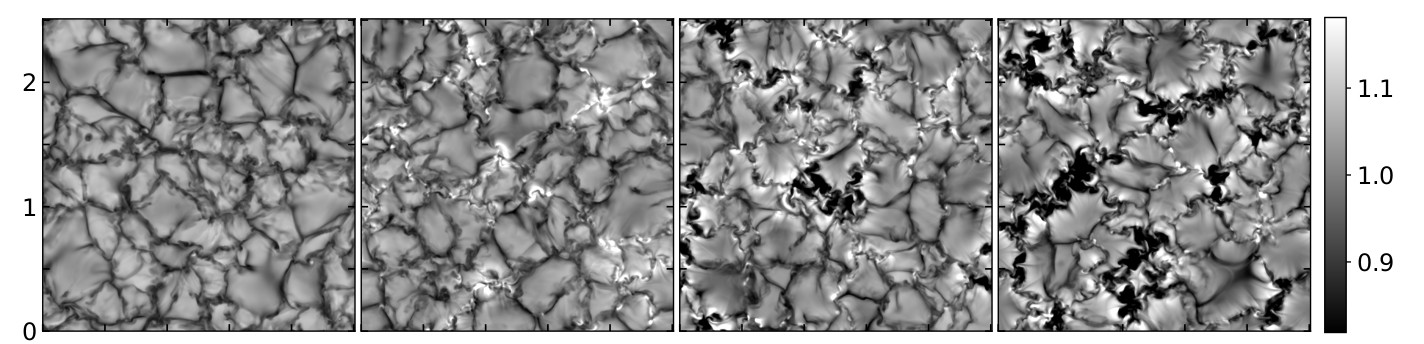}
    \put(4,20){\textsf{\textbf{\large \textcolor{white}{602\,nm}}}}
\end{overpic}
\begin{overpic}[width=\textwidth,trim={0 1 0 4},clip]{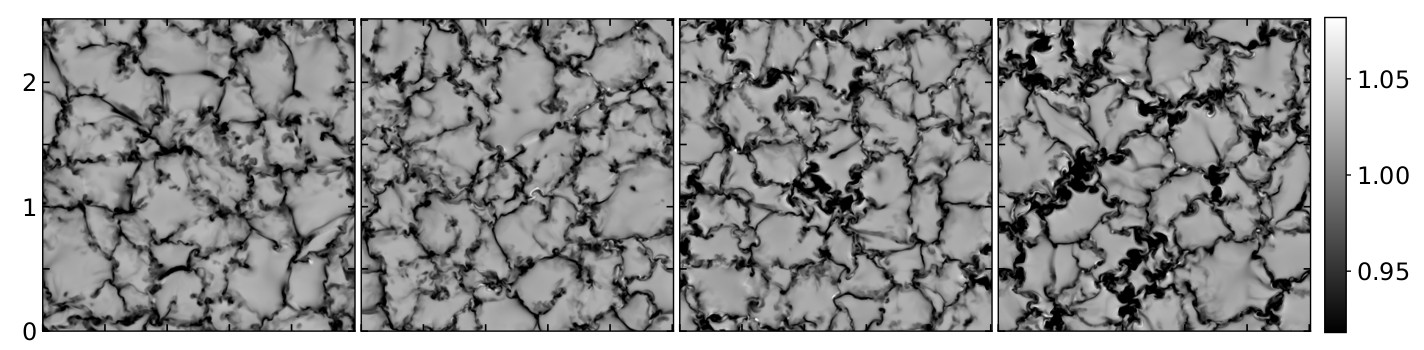}
    \put(4,20){\textsf{\textbf{\large \textcolor{white}{1610\,nm}}}}
\end{overpic}
\begin{overpic}[width=\textwidth,trim={0 0 0 4},clip]{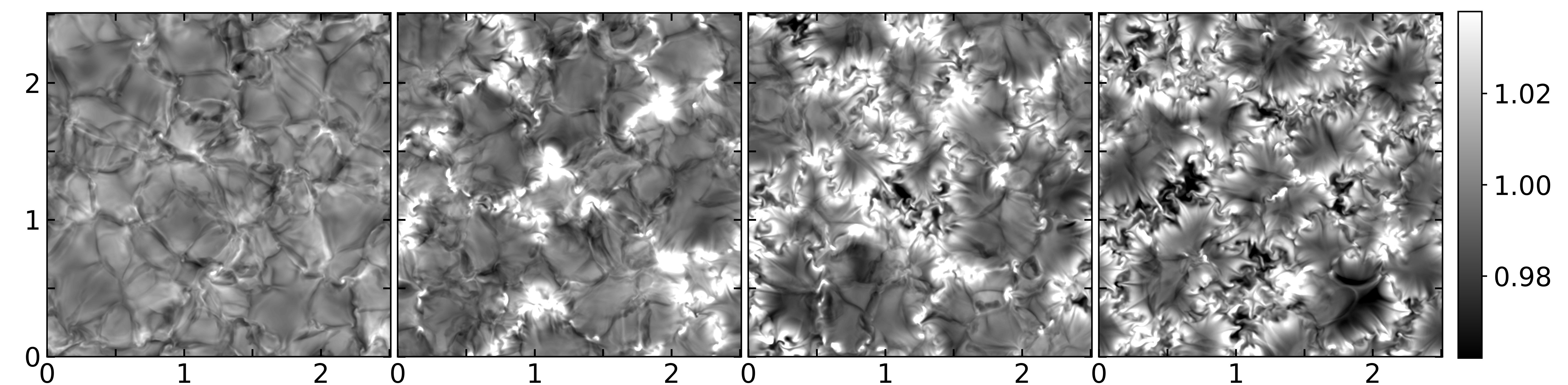}
    \put(4,21){\textsf{\textbf{\large \textcolor{white}{8040\,nm}}}}
\end{overpic}
\caption[]{As Fig.\,\ref{fig:images_K0}, but for M0V simulations.}
\label{fig:images_M0}
\end{figure*}

\begin{figure*}
\centering
\vspace*{3.5ex}

\begin{overpic}[width=\textwidth]{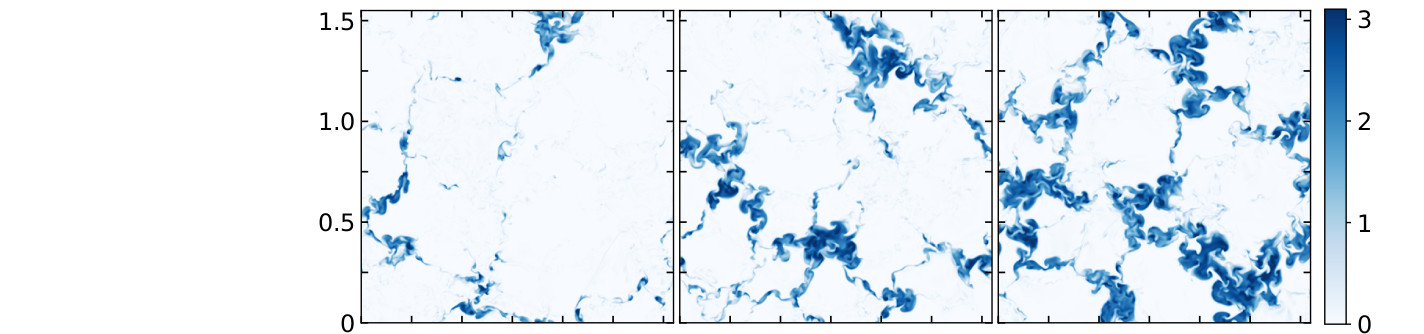}
    \put(11,23.7){\textsf{\Large hydro  \hspace{5.6em} $\langle B_z \rangle =$\,100\,G \hspace{4.2em} $\langle B_z \rangle =$\,300\,G \hspace{4.2em} $\langle B_z \rangle =$\,500\,G }}
\end{overpic}
\begin{overpic}[width=\textwidth]{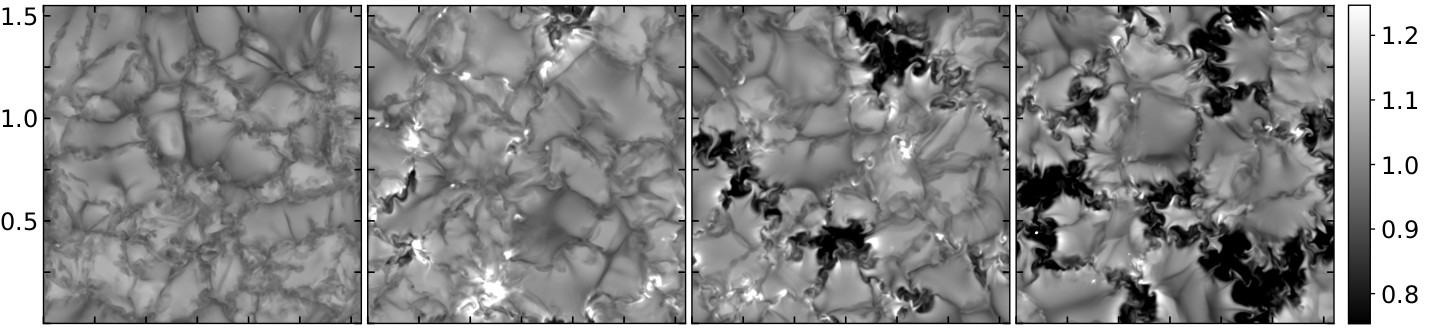}
    \put(4,20){\textsf{\textbf{\large \textcolor{white}{388\,nm}}}}
\end{overpic}
\begin{overpic}[width=\textwidth]{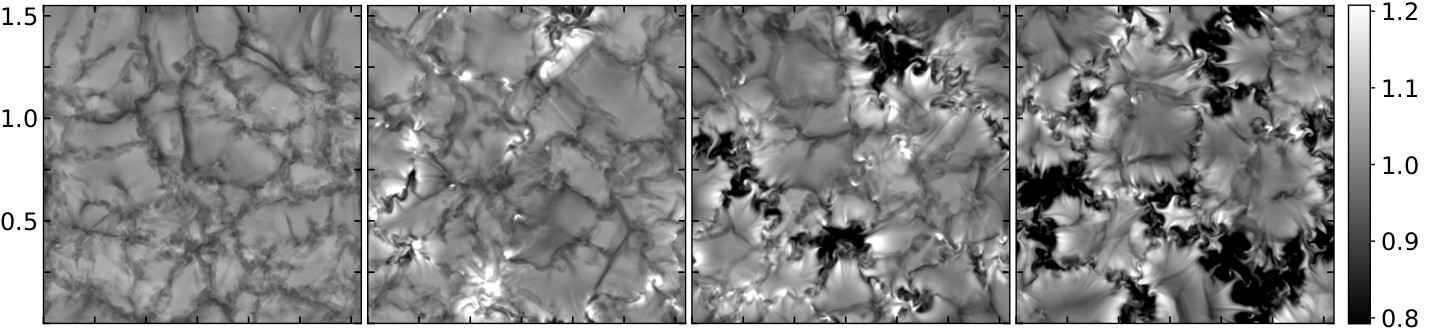}
    \put(4,20){\textsf{\textbf{\large \textcolor{white}{602\,nm}}}}
\end{overpic}
\begin{overpic}[width=\textwidth]{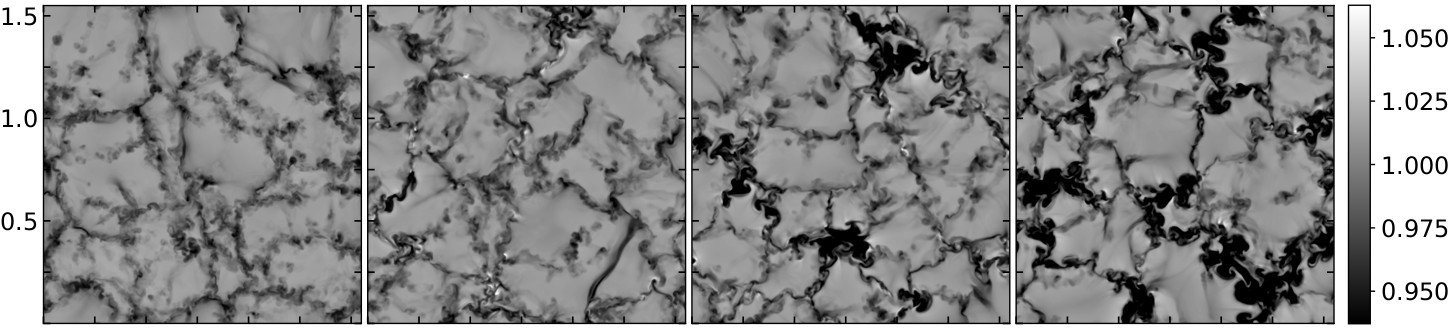}
    \put(4,20){\textsf{\textbf{\large  \textcolor{white}{1610\,nm}}}}
\end{overpic}
\begin{overpic}[width=\textwidth]{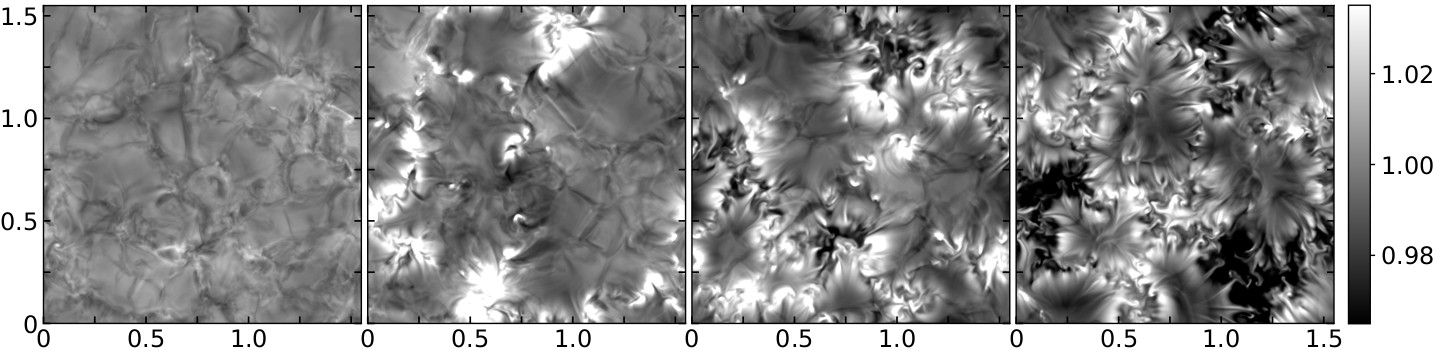}
    \put(4,21){\textsf{\textbf{\large \textcolor{white}{8040\,nm}}}}
\end{overpic}
\caption[]{As Fig.\,\ref{fig:images_K0}, but for M2V simulations.}
\label{fig:images_M2}
\end{figure*}

Figures~\ref{fig:images_K0} to \ref{fig:images_M2} show disc-centre images for K0V, M0V and M2V simulations, respectively. These spectral types were chosen as examples of how the surface structure of a star changes with {\teff} and {\logg}, both in the field-free and magnetic cases (comparable images for the G2 simulations were presented in paper~{\sc i}). The top row in each of the figures shows the unsigned vertical magnetic field at an optical depth at 500~nm, {\taufive}, of one. Images of emergent intensity for four different wavelength bins are shown to represent a variety of behaviours in typically distinct wavelength regions: 388~nm, 602~nm, 1610~nm and 8040~nm from top to bottom\footnote{For the images presented here, the intensity bin widths range from 2~nm in the visible to 40~nm in the infrared; specifically, the intensity bins are 387~nm to 389~nm, 601~nm to 603~nm, 1605~nm to 1615~nm, and 8020~nm to 8040~nm.}. 
Due to the opacity minimum at $\sim$\,1600\,nm, the images at 1610~nm allow us to see deepest into the stellar photospheres. Radiation at 600\,nm typically forms at intermediate depth (though the presence of TiO shifts the formation depth to higher layers for the M2 stars), while radiation at 388\,nm (due to a dense forest of lines) and 8040\,nm is predominantly formed in the upper photosphere.

The columns in Figs.~\ref{fig:images_K0} to \ref{fig:images_M2} show, from left to right, field-free, {\meanBz}{\equals}100~G, \meanBz{\equals}300~G and {\meanBz}{\equals}500~G snapshots. The colour bars to the right of the magnetic field images indicate the magnetic flux density in kG. 
For ease of comparison, the intensities of the snapshots have been normalised to the mean intensity of the field-free snapshot at each of the displayed wavelengths. The greyscales have been adjusted so that they saturate at $2\sigma_\lambda$, where $\sigma_\lambda$ is the standard deviation of intensities $I_\lambda$ at the wavelength for which the snapshot is shown  (this is done purely for display purposes; the monochromatic intensities $I_\lambda$ tend to show non-Gaussian distributions, see Sec.~\ref{sec:histograms}). The brightness temperatures corresponding to the mean intensities of the snapshots at the different wavelengths are given in Tab.\,\ref{tab:SpecInt_SN}. 

For all spectral types, granulation, caused by convection, is seen. The size of these convective cells, or granules, changes with spectral type. In the mid-IR, the K0V and G2V snapshots (the latter of which have been discussed in paper~{\sc i} and are not shown here) granules appear darker than the intergranular lanes. This is known as reversed granulation (see, e.g.,~images at 8040~nm in the bottom row of Fig.~\ref{fig:images_K0}). This occurs due to the reversal of horizontal temperature fluctuations in the middle to upper photosphere due to the balance of adiabatic expansion and radiative heating, as described in \citet{Cheung2007}. This reversed granulation also appears in the UV (not shown here) for K0V to M2V stars. The intergranular lanes appear wider in the NUV and mid IR than in the visible and NIR due to the pressure in the atmosphere being lower at the height were the NUV/MIR radiation is formed, leading to more diffuse features. 

When magnetic field is injected into the simulation, convective motions of the plasma sweep the flux into the intergranular lanes. For all stars, with an injected vertical magnetic field, {\meanBz}, of 100~G, mainly bright features are present, as seen in the second column of Fig.\,\ref{fig:images_K0} to Fig.\,\ref{fig:images_M2}. For the K0-star simulations, these bright features stand out particularly clearly at 388~nm (due to the temperature-sensitive CN lines) where they can be more than 50\% brighter than the mean intensity. For the cooler M2-type stars, contrasts are only slightly larger at 388~nm than at 602~nm, as the CN bands become relatively less important and TiO features in the visible gain in prominence. At 8040~nm, contrasts are typically very low (of the order of 5\%, see bottom rows of Figs~\ref{fig:images_K0} to \ref{fig:images_M2}). As radiation emerges higher in the atmosphere where gas pressure is low, the bright magnetic features take up more of the surface area and stand out very clearly against the subdued granulation patterns. 

In some selected wavelength regions such as near 1.6\,$\mu$m, the otherwise bright features appear dark or show very low contrast compared to the granules. This is due to 1.6~$\mu$m being close to the H$^-$ opacity minimum, allowing radiation from deeper in the star to emerge. At these wavelengths, radiation seen in the granules will be produced in deeper and hotter layers that are closer in geometrical depth to where radiation is seen emerging from magnetic features, leading to only weak intensity contrasts. In the deeper layers, the magnetic features are also less visible, because they are narrower than higher in the atmosphere \citep[e.g.,][]{Solanki+1999}. 

For {\meanBz}{\equals}300\,G and even more so for 500\,G, larger dark features emerge in addition to bright features for all spectral types; these vary in size and number and also according to spectral type, as discussed in the following paragraphs. The resulting (mean) contrast of a magnetic region is produced by the sum of these dark and bright features; mean contrasts will be discussed further in Sec.\,\ref{sect:contrasts}. 

K0 stars (Fig.~\ref{fig:images_K0}) have noticeable bright features for {\meanBz}{\equals}100\,G that increase in number as {\meanBz} increases. For \meanBz{\equals}300\,G and 500~G the bright magnetic features are fragmented into smaller sections and for \meanBz\equals 500\,G most of the intergranular lanes are filled with magnetic features. Although the small magnetic features are bright at most wavelengths presented, they are very low contrast, if not dark, around 1.6~nm. A small number of dark features develop for \meanBz{\equals}300\,G; these increase in size and number as the magnetic flux density increases. These dark features stand out clearly at 602~nm and 1610~nm; at 388\,nm they appear somewhat more diffuse, while at 8040\,nm the emission is dominated by the extended bright features that surround them. Therefore, at different wavelengths the balance between dark and bright features varies, resulting in changes in the overall contrast of the magnetic region. For example, the mean contrast (with respect to the mean hydrodynamic intensity) of the {\meanBz}{\equals}500\,G snapshot is positive at 388~nm and 8040~nm, and negative for the same snapshot at 602~nm and 1610~nm (see also Tab.~\ref{tab:SpecInt_SN}).\par
For M0 and M2 snapshots with {\meanBz}{\equals}100\,G, magnetic features have low contrast or are dark at 1610~nm, but have high contrasts for 388~nm, 602~nm and 8040~nm. For both M-type stars with {\meanBz}{\equals}500\,G large dark features are present, with very few bright features. For M0 snapshots (Fig.\,\ref{fig:images_M0}), the dark features are elongated and more filament-like in shape. They, similarly to the dark features for the K0 snapshot, become less dark at 8040~nm than at 602~nm or 1610~nm. For M2 simulations (Fig.~\ref{fig:images_M2}), the dark features are larger relative to the granule sizes, and are dark throughout the NUV to the NIR. At which magnetic flux densities dark features first appear depends on the spectral type. Considering the intermediate \meanBz\equals 300~G simulations, almost no dark features are present for the K0V simulations. The corresponding M0V simulations show some prominent dark features (that somewhat resemble the K0V 500\,G simulations) while the M2V simulations are predominantly dark. 

\subsection{Images at a limb distance of 0.5}
\begin{figure}
\centering
\vspace*{3ex}

\begin{overpic}[width=.48\textwidth,trim={0 8 0 0},clip=]{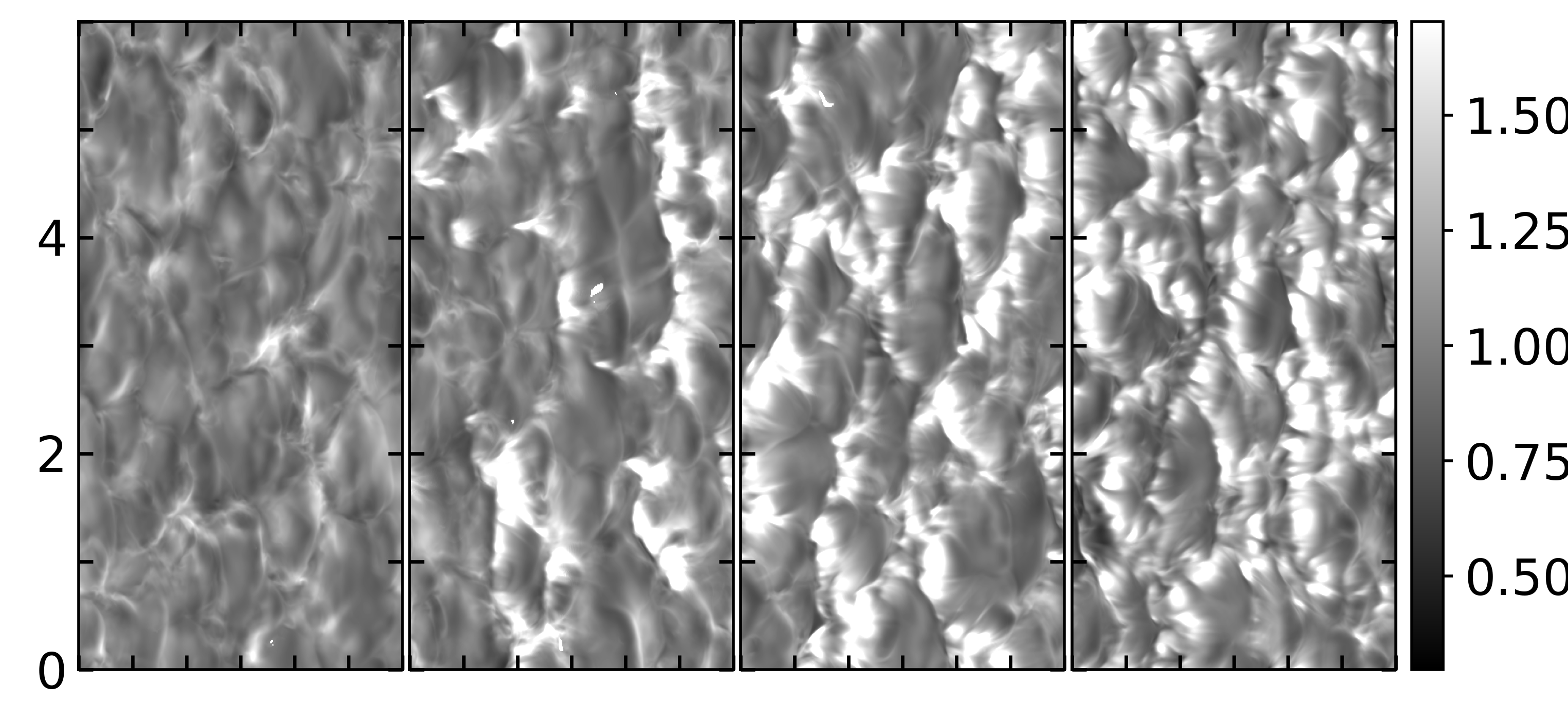}
    \put(12,44.5){\textsf{hydro  \hspace{2.em} $\langle B_z \rangle =$\,100\,G \hspace{.4em} $\langle B_z \rangle =$\,300\,G \hspace{.4em} $\langle B_z \rangle =$\,500\,G }}
    \put(7,38){\textsf{\textbf{\large \textcolor{white}{388\,nm}}}}
\end{overpic}
\begin{overpic}[width=.48\textwidth,trim={0 8 0 4},clip=]{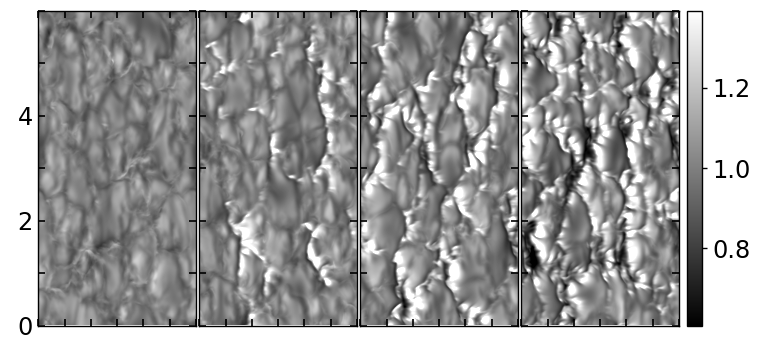}
    \put(7,38){\textsf{\textbf{\large \textcolor{white}{602\,nm}}}}
\end{overpic}
\begin{overpic}[width=.48\textwidth,trim={0 8 0 4},clip=]{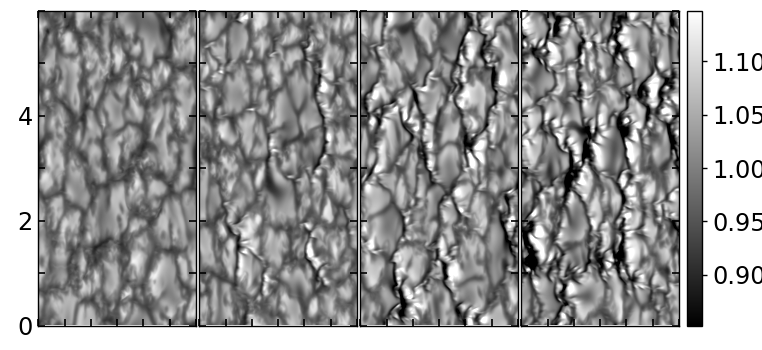}
    \put(7,38){\textsf{\textbf{\large \textcolor{white}{1610\,nm}}}}
\end{overpic}
\begin{overpic}[width=.48\textwidth,trim={0 0 0 4},clip=]{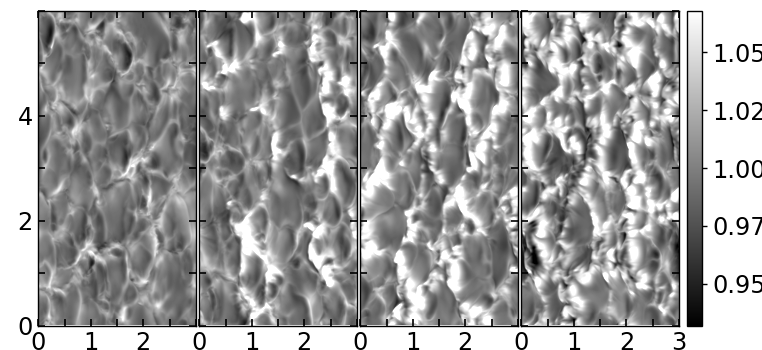}
    \put(7,40){\textsf{\textbf{\large \textcolor{white}{8040\,nm}}}}
\end{overpic}
\caption[]{Emergent intensities at $\mu=0.5$ from K0 simulated atmospheric snapshots. As for the corresponding disc-centre images (Figs~\protect{\ref{fig:images_K0}} to \protect{\ref{fig:images_M2}}), the intensities have been normalised by the mean intensity of the field-free simulation; the greysales saturate at $2\sigma_\lambda$. The equivalent brightness temperatures for the snapshots shown here are listed in Tab.~\protect{\ref{tab:SpecInt_muSN}}.}
\label{fig:images_muK0}
\end{figure}

\begin{figure}
\centering
\vspace*{3ex}

\begin{overpic}[width=.48\textwidth,trim={0 8 0 0},clip=]{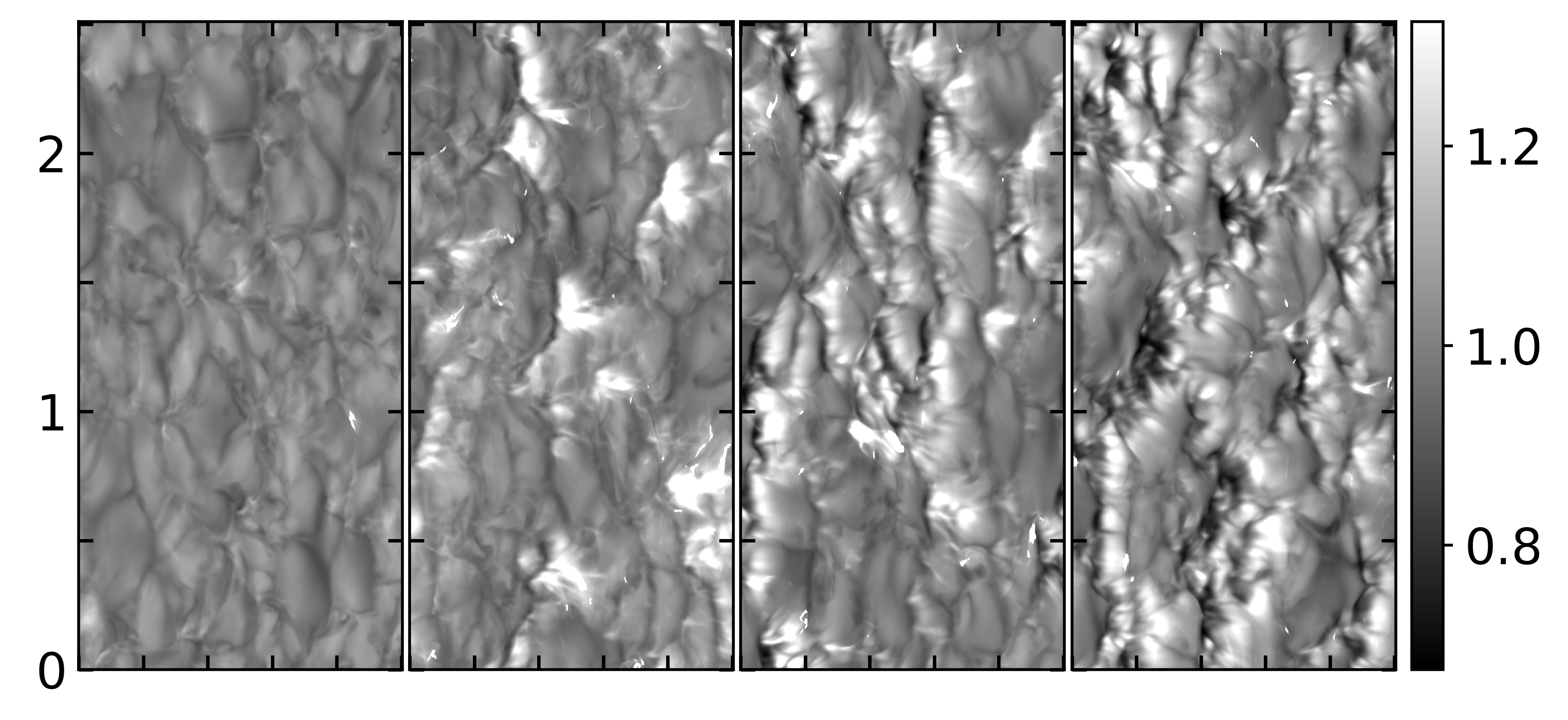}
    \put(12,44.5){\textsf{hydro  \hspace{2.em} $\langle B_z \rangle =$\,100\,G \hspace{.4em} $\langle B_z \rangle =$\,300\,G \hspace{.4em} $\langle B_z \rangle =$\,500\,G }}
    \put(7,38){\textsf{\textbf{\large \textcolor{white}{388\,nm}}}}
\end{overpic}
\begin{overpic}[width=.48\textwidth,trim={0 8 0 4},clip=]{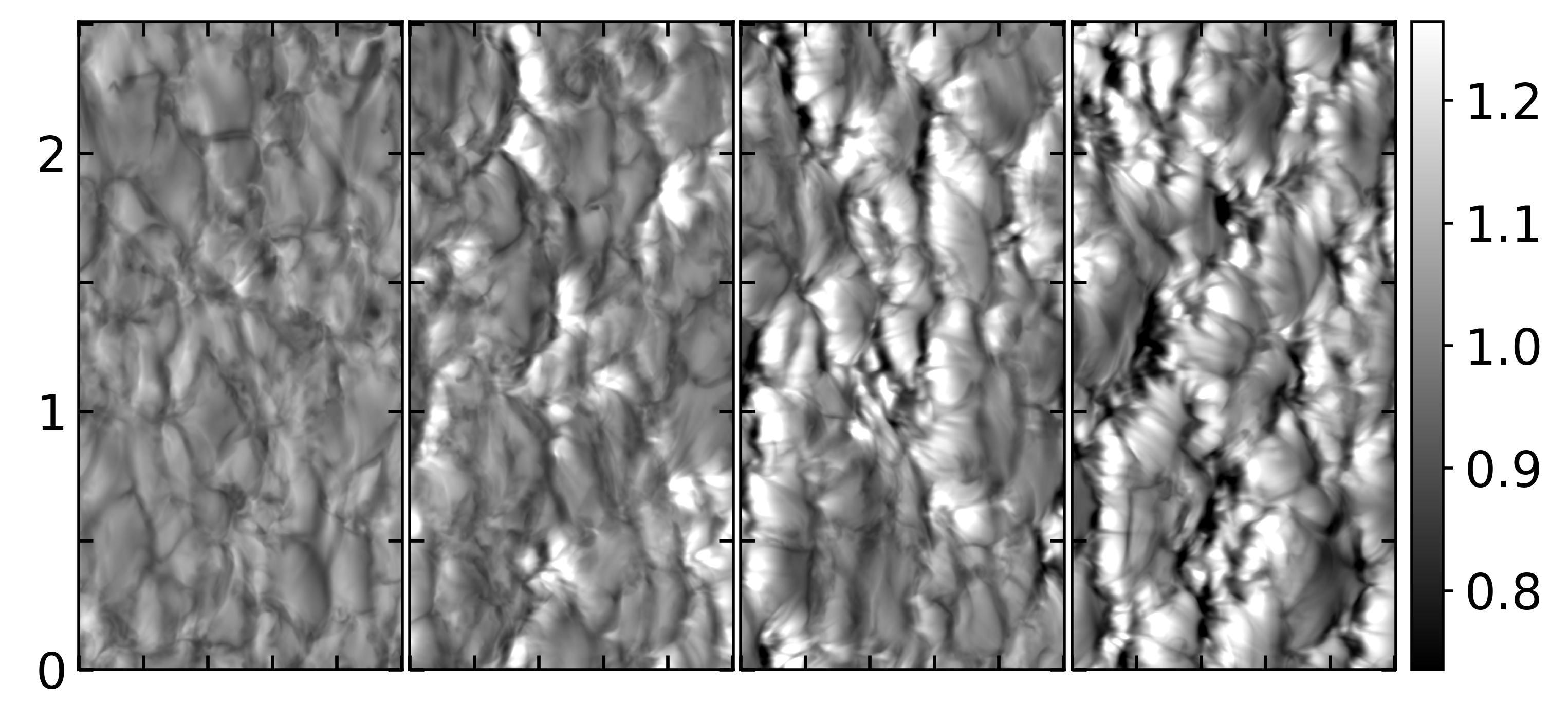}
    \put(7,38){\textsf{\textbf{\large \textcolor{white}{602\,nm}}}}
\end{overpic}
\begin{overpic}[width=.48\textwidth,trim={0 8 0 4},clip=]{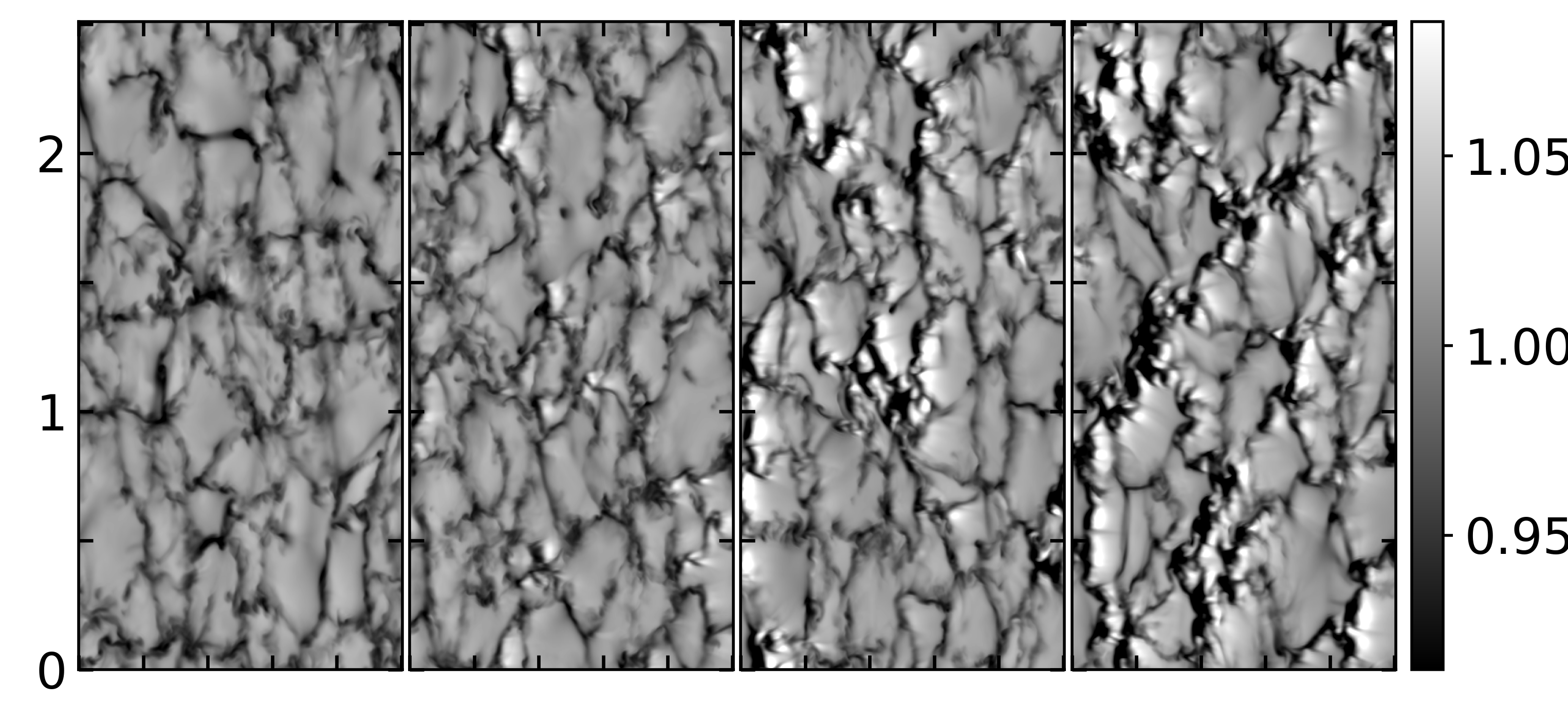}
    \put(7,38){\textsf{\textbf{\large \textcolor{white}{1610\,nm}}}}
\end{overpic}
\begin{overpic}[width=.48\textwidth,trim={0 0 0 4},clip=]{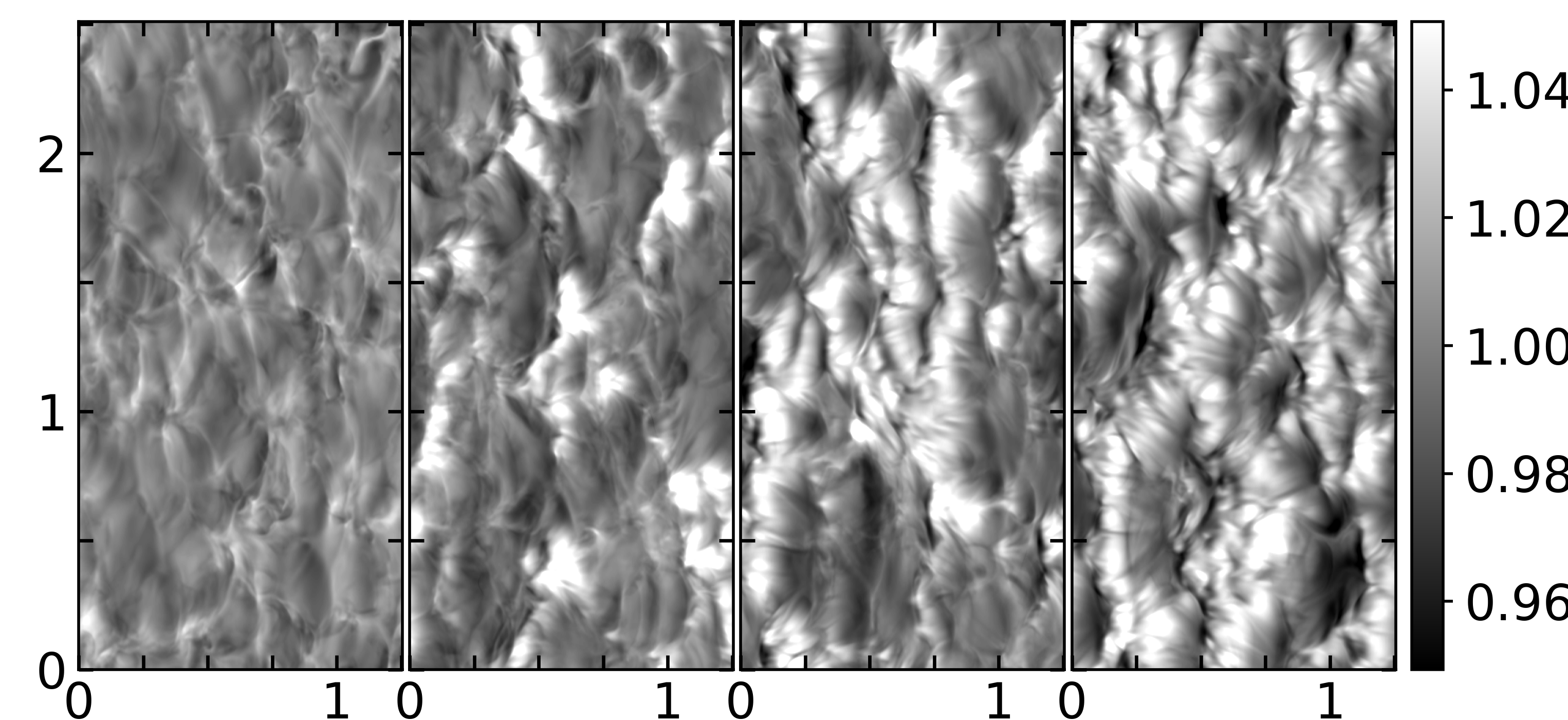}
    \put(7,40){\textsf{\textbf{\large \textcolor{white}{8040\,nm}}}}
\end{overpic}
\caption[]{Emergent intensities at $\mu=0.5$ from M0 simulated atmospheric snapshots; see Fig.~\protect{\ref{fig:images_muK0}}.}
\label{fig:images_muM0}
\end{figure}

\begin{figure}
\centering
\vspace*{3ex}

\begin{overpic}[width=.48\textwidth,trim={0 8 0 0},clip=]{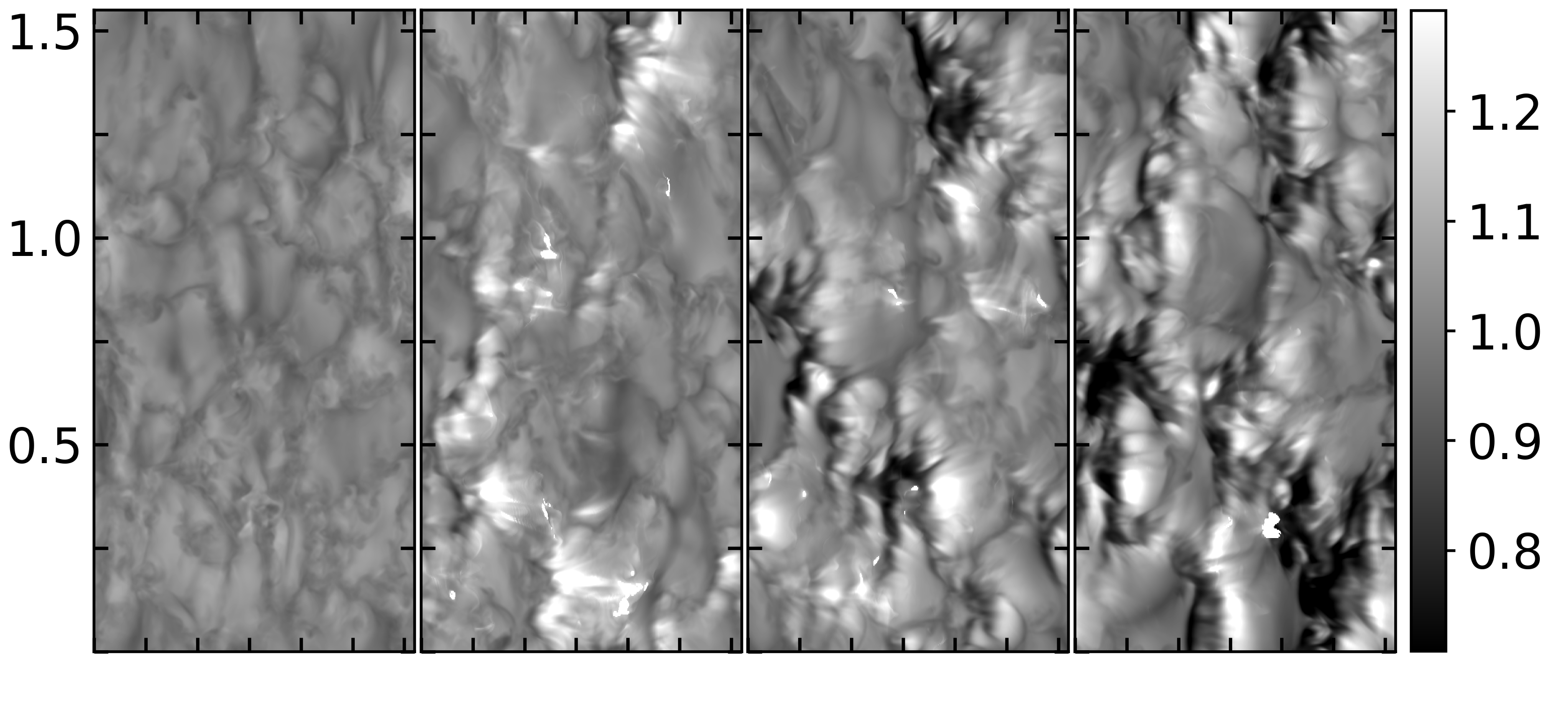}
        \put(12,44.5){\textsf{hydro  \hspace{2.em} $\langle B_z \rangle =$\,100\,G \hspace{.4em} $\langle B_z \rangle =$\,300\,G \hspace{.4em} $\langle B_z \rangle =$\,500\,G }}
    \put(7,38){\textsf{\textbf{\large\textcolor{white}{388\,nm}}}}
\end{overpic}
\begin{overpic}[width=.48\textwidth,trim={0 8 0 0},clip=]{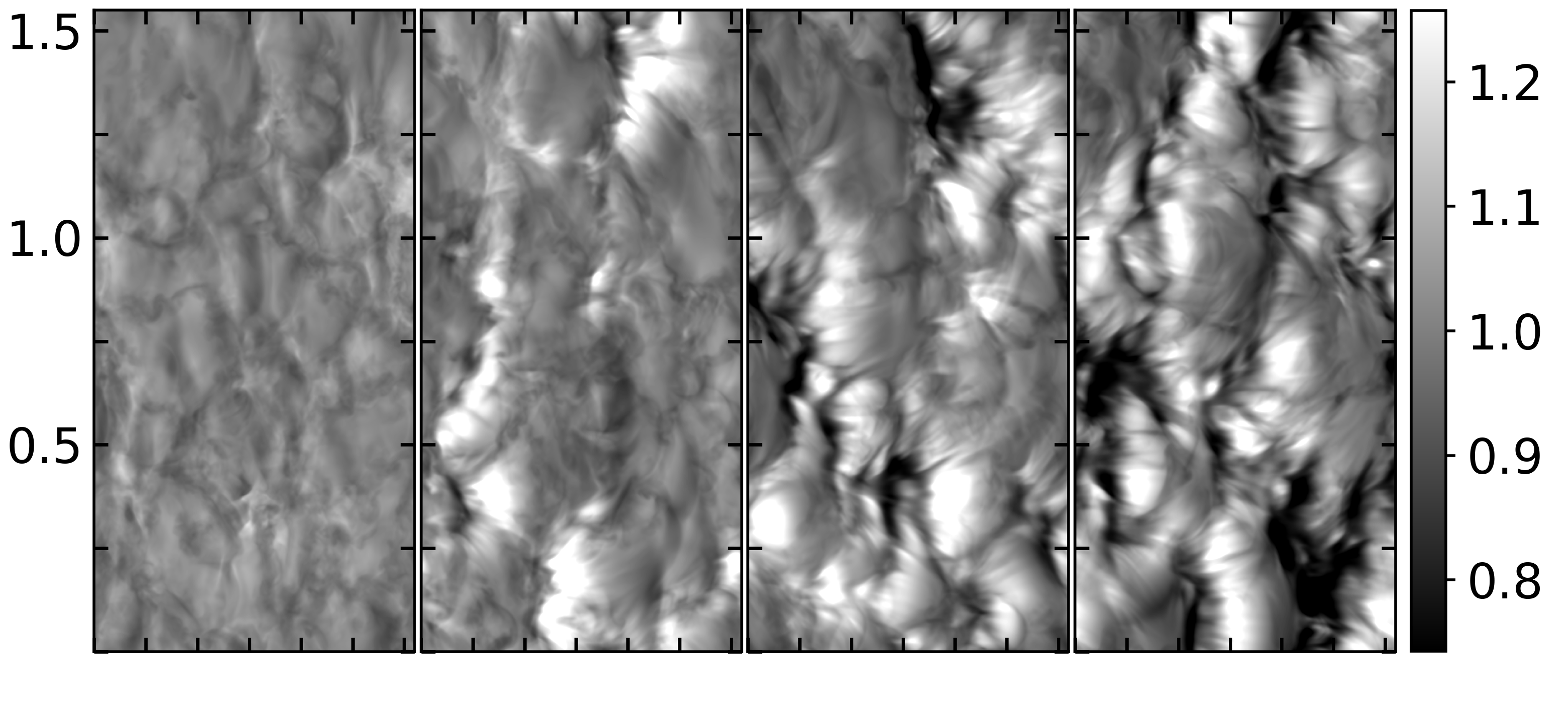}
    \put(8,38){\textsf{\textbf{\large \textcolor{white}{602\,nm}}}}
\end{overpic}
\begin{overpic}[width=.48\textwidth,trim={0 8 0 0},clip=]{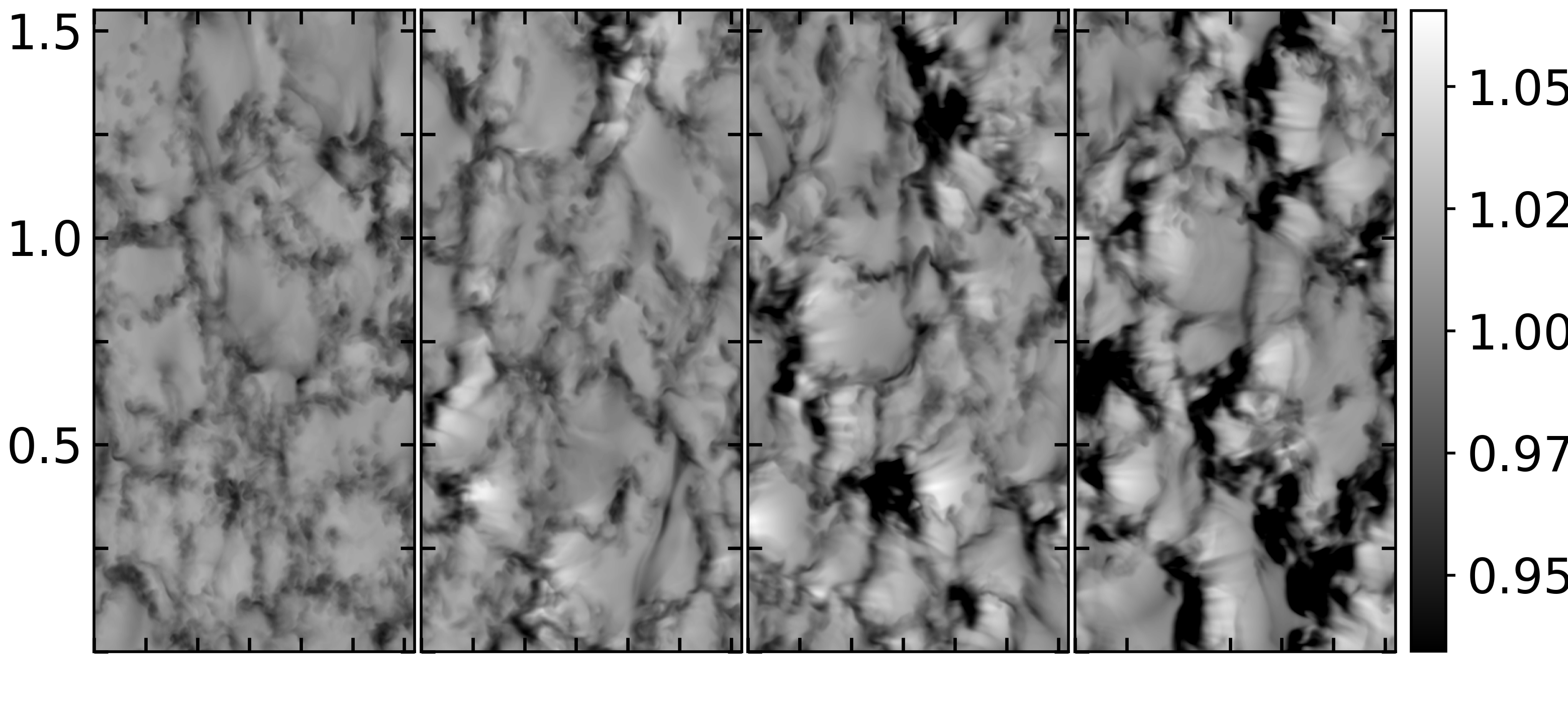}
    \put(8,38){\textsf{\textbf{\large \textcolor{white}{1610\,nm}}}}
\end{overpic}
\begin{overpic}[width=.48\textwidth,trim={0 0 0 0},clip=]{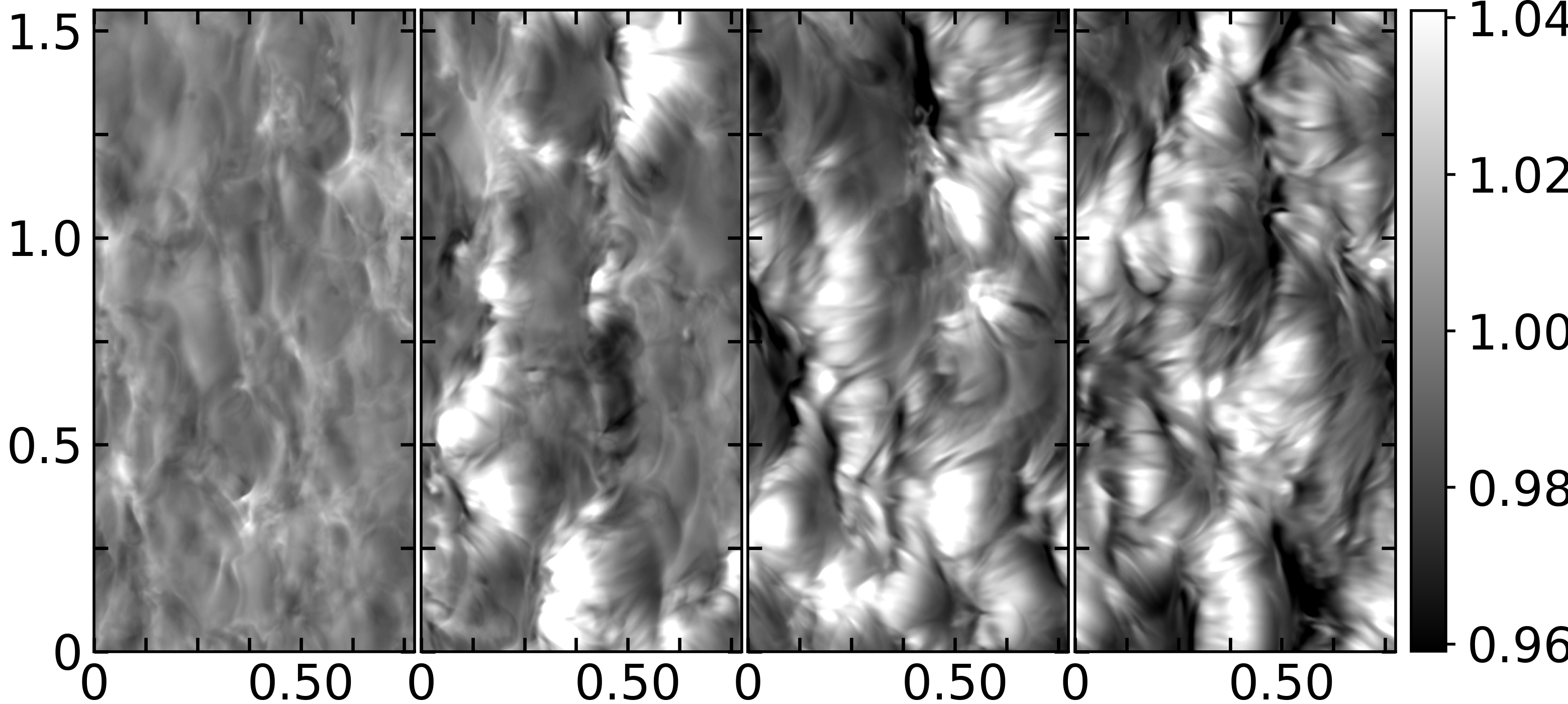}
    \put(8,40){\textsf{\textbf{\large \textcolor{white}{8040\,nm}}}}
\end{overpic}
\caption[]{Emergent intensities at $\mu=0.5$ from M2 simulated atmospheric snapshots; see Fig.~\protect{\ref{fig:images_muK0}}.}
\label{fig:images_muM2}
\end{figure}

Figures~\ref{fig:images_muK0} to \ref{fig:images_muM2} show the same snapshots as Figs.~\ref{fig:images_K0} to \ref{fig:images_M2}, but now seen at a limb distance of $\mu=\cos\theta=0.5$, corresponding to a limb angle of $60^\circ$. The images have been normalised by the mean intensity of the field-free simulations at $\mu=0.5$ (see Tab.\,\ref{tab:SpecInt_muSN} for the corresponding brightness temperatures). The inclined viewing angle highlights the corrugated aspect of the granulation. Relative to disc centre, a larger fraction of the hot walls of the magnetic features can be seen, leading to enhanced brightening. As for the disc-centre images, features seen at deeper atmospheric layers (i.e., in the NIR) are less extended and show little brightening. 

For snapshots with \meanBz\,$=$\,100\,G, images are typically dominated by bright features; an exception here are images in the NIR at the opacity minimum. For the 300\,G and 500\,G shapshots we see a mix of dark and bright features with some bright pixels exceeding the mean disc centre intensity. For the K0V and M0V simulations the contribution of the bright features dominates even at \meanBz\ of 500\,G, though at most wavelengths the mean contrasts are highest for the \meanBz\,$=$\,300\,G. For the M2V simulations, dark features dominate at \meanBz\,$=$\,500\,G. To illustrate the range of intensities and their distribution, we consider histograms of the images in more detail in the next section. 


\subsection{Intensity histograms}
\label{sec:histograms}
\begin{figure*}
\vspace*{.2ex}
\vbox{\begin{overpic}[height=.149\textheight,trim={15 15 5 5},clip]{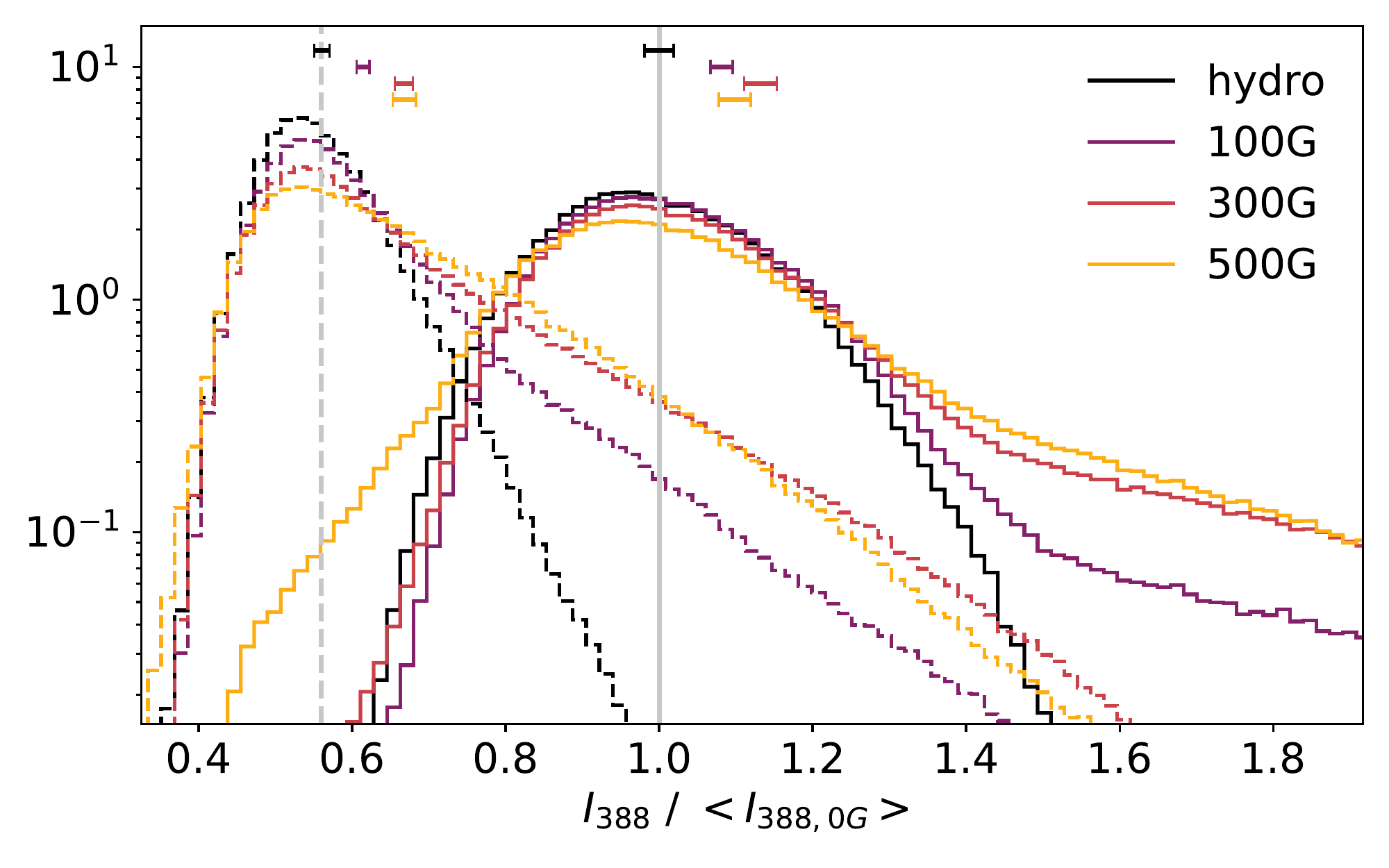}
    \put(52,61){\textsf{\Large K0}}
\end{overpic}
\begin{overpic}[height=.149\textheight,trim={5 15 5 5},clip=true]{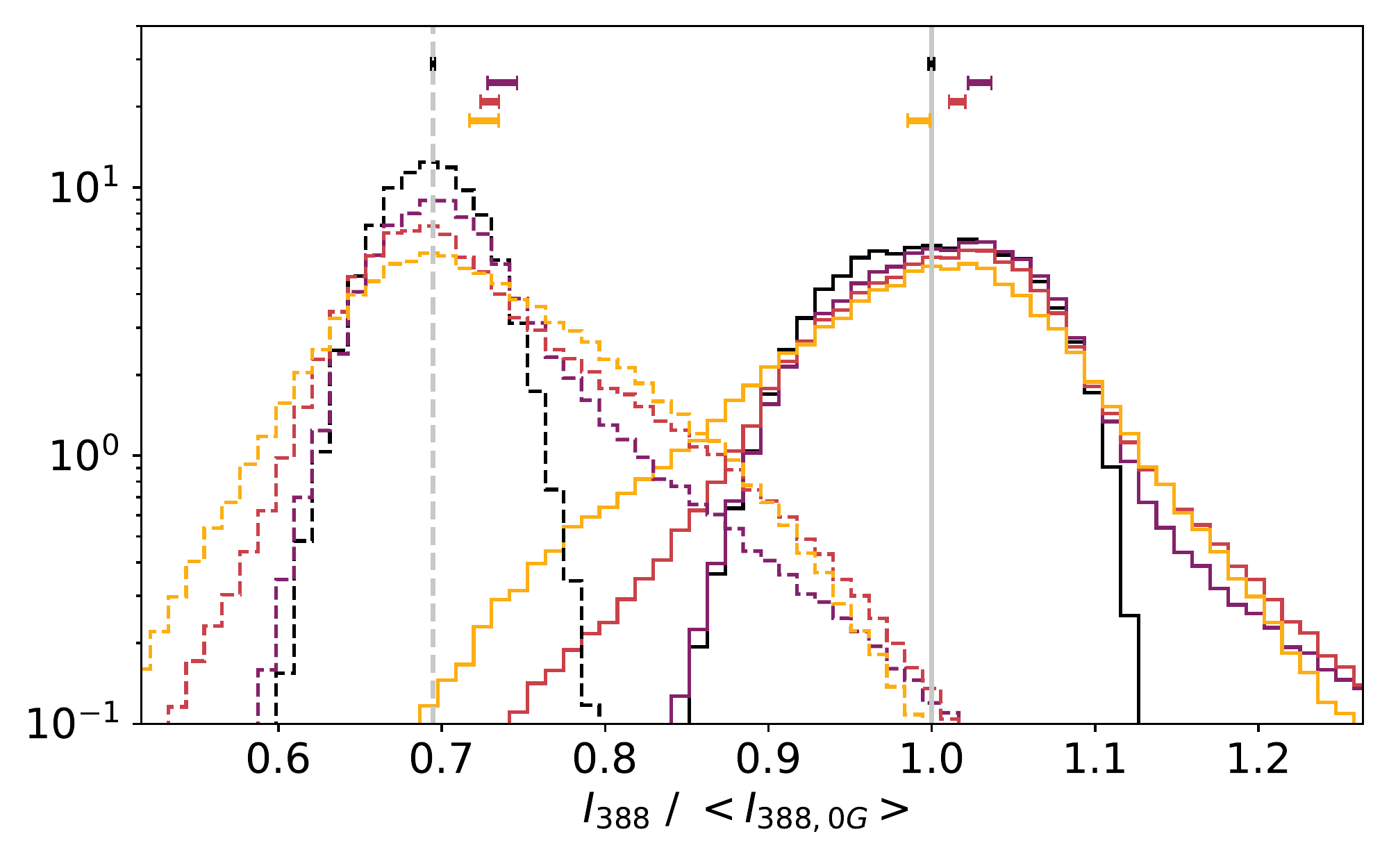}
    \put(52,60){\textsf{\Large M0}}
\end{overpic}
\begin{overpic}[height=.149\textheight,trim={5 15 5 5},clip]{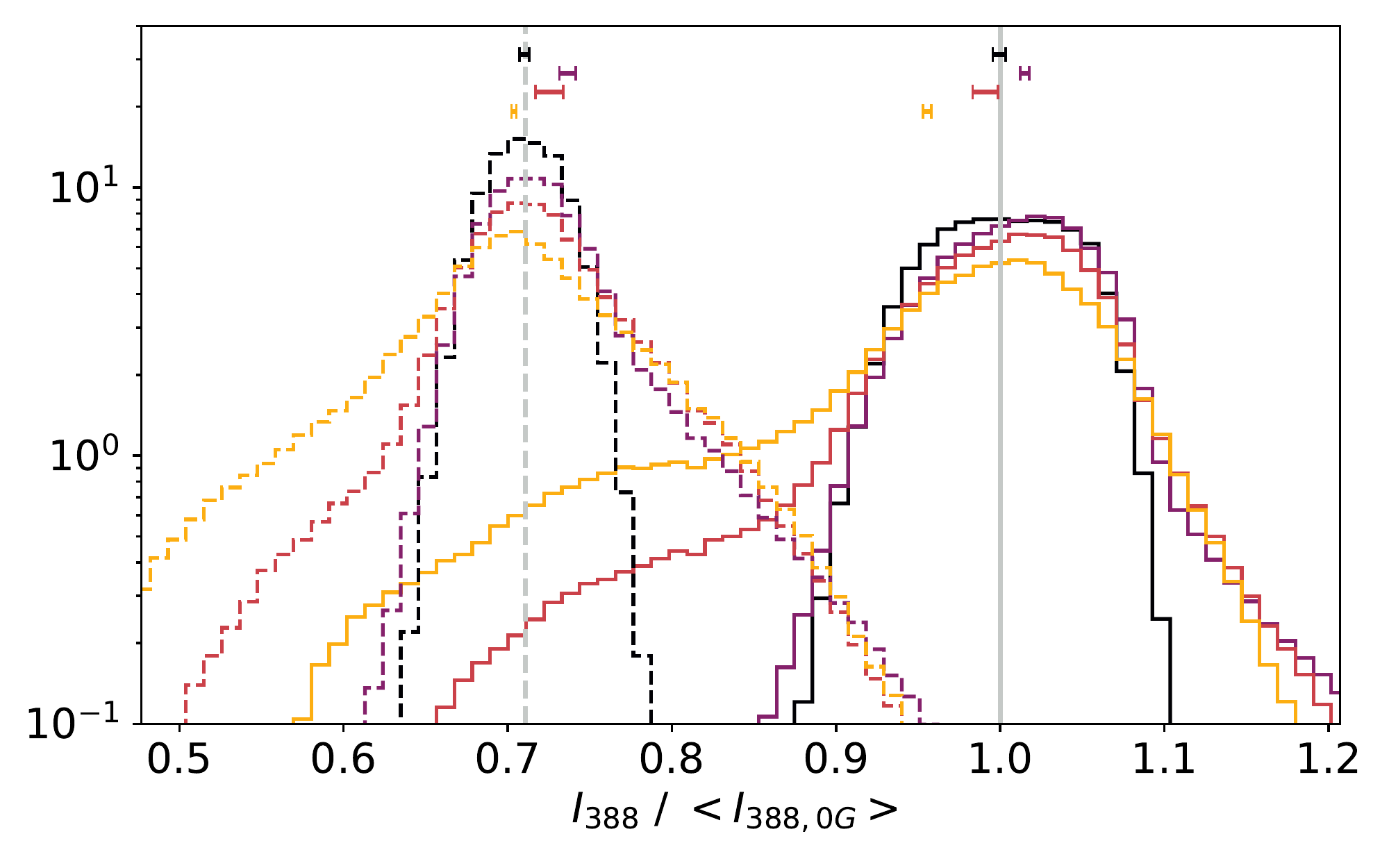}
    \put(52,60){\textsf{\Large M2}}
\end{overpic}
}

\vbox{\includegraphics[height=.149\textheight,trim={15 15 5 5},clip]{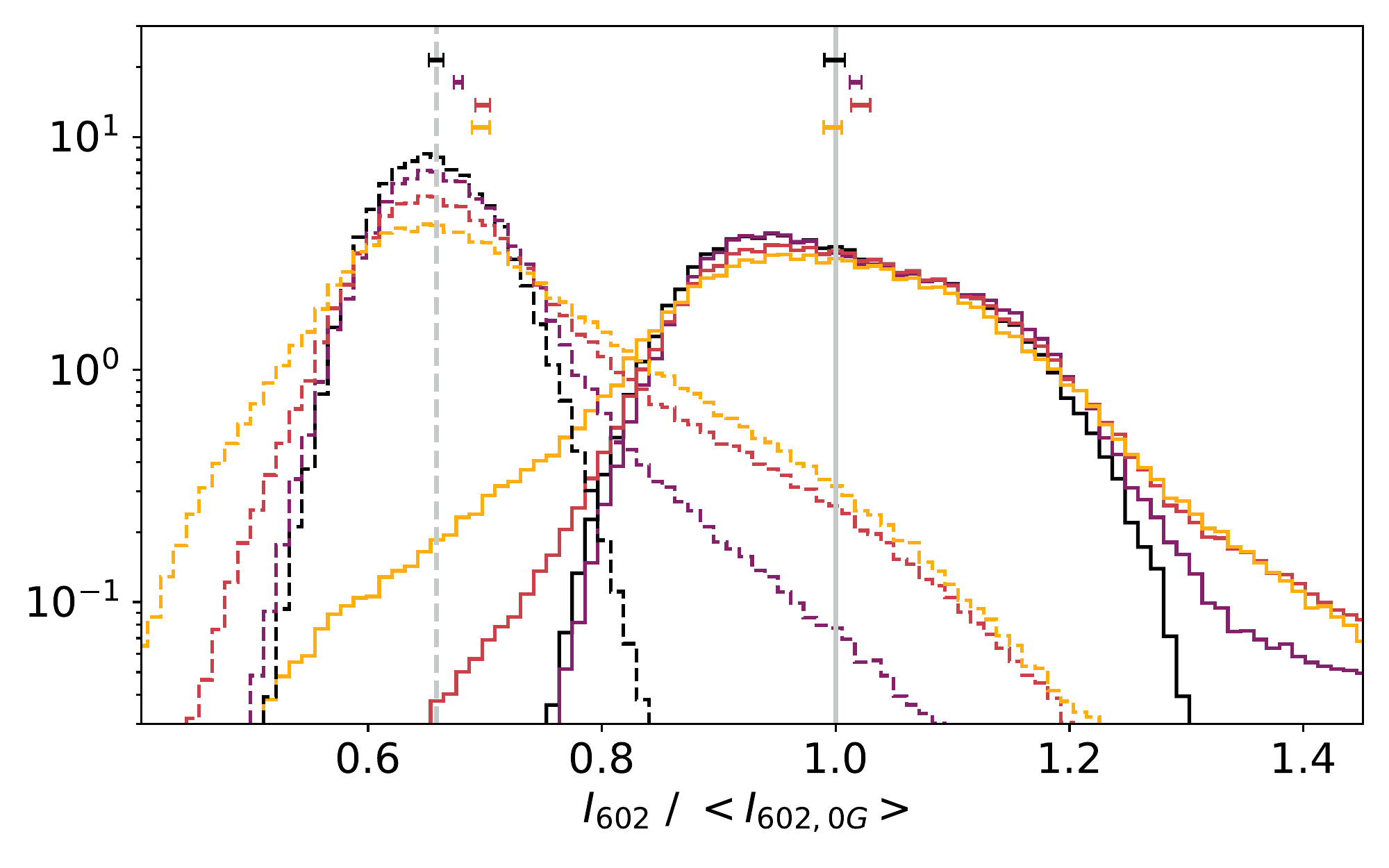}
\includegraphics[height=.149\textheight,trim={5 15 5 5},clip]{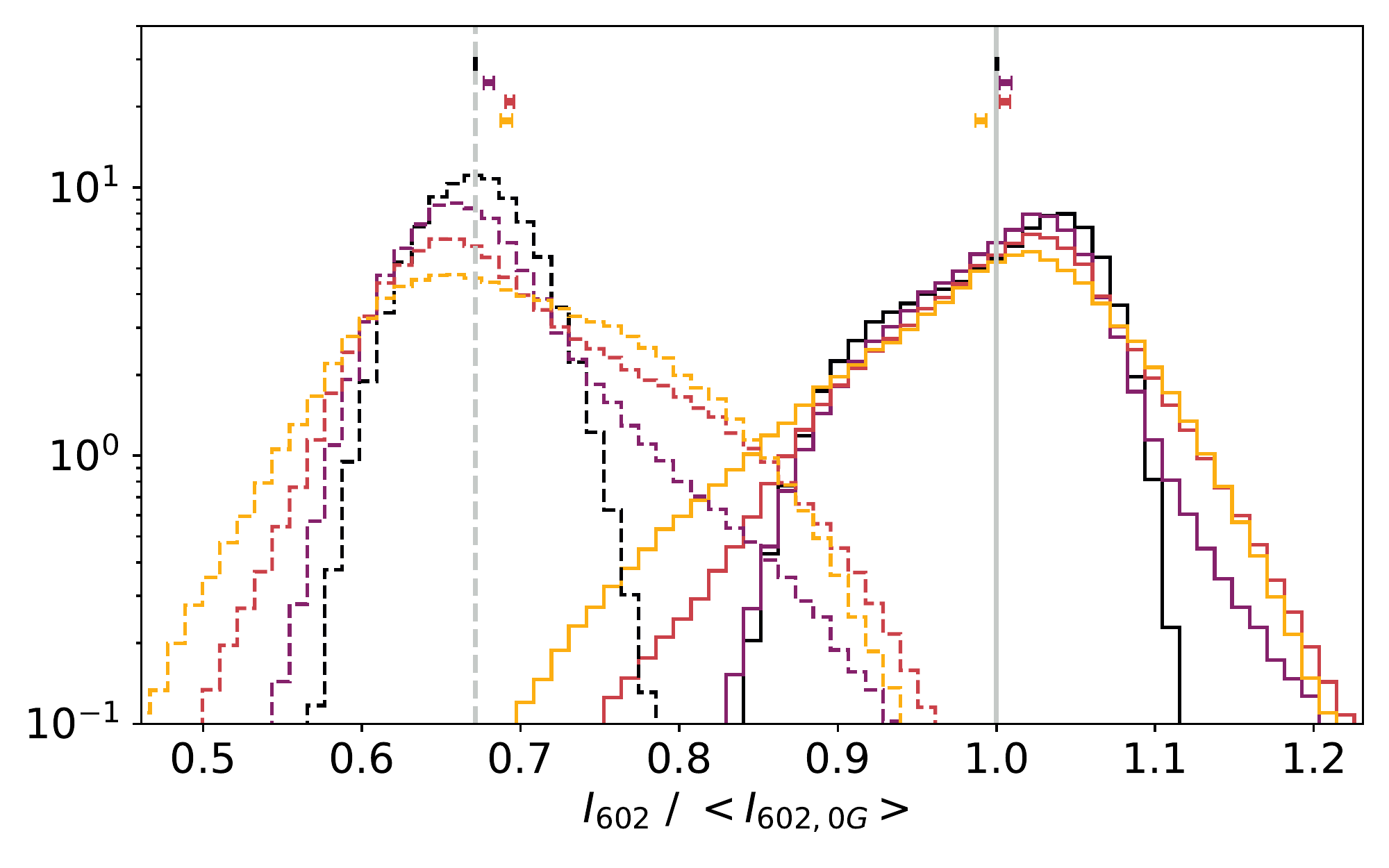}
\includegraphics[height=.149\textheight,trim={5 15 5 5},clip]{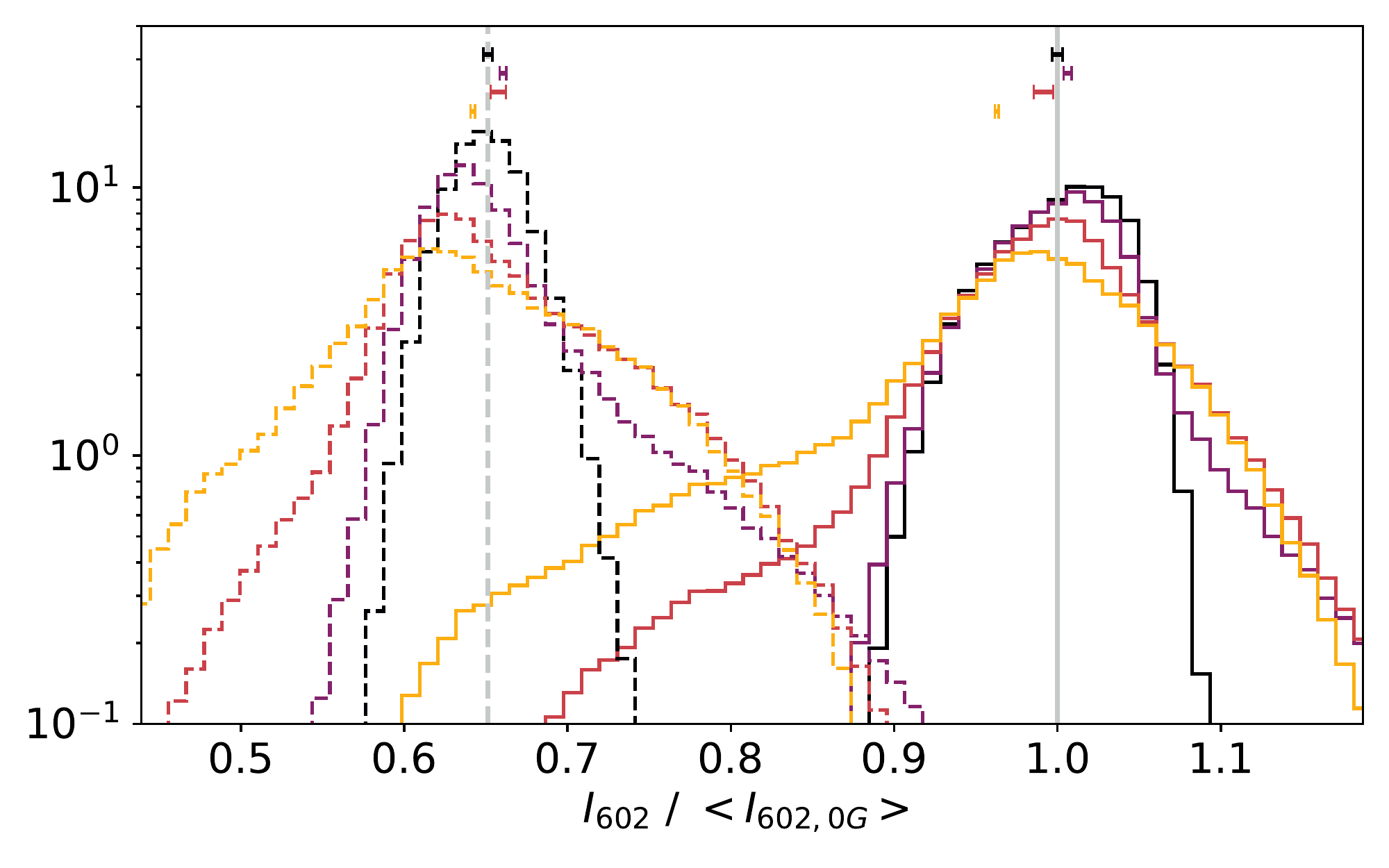}}
\vbox{\includegraphics[height=.149\textheight,trim={15 15 5 5},clip]{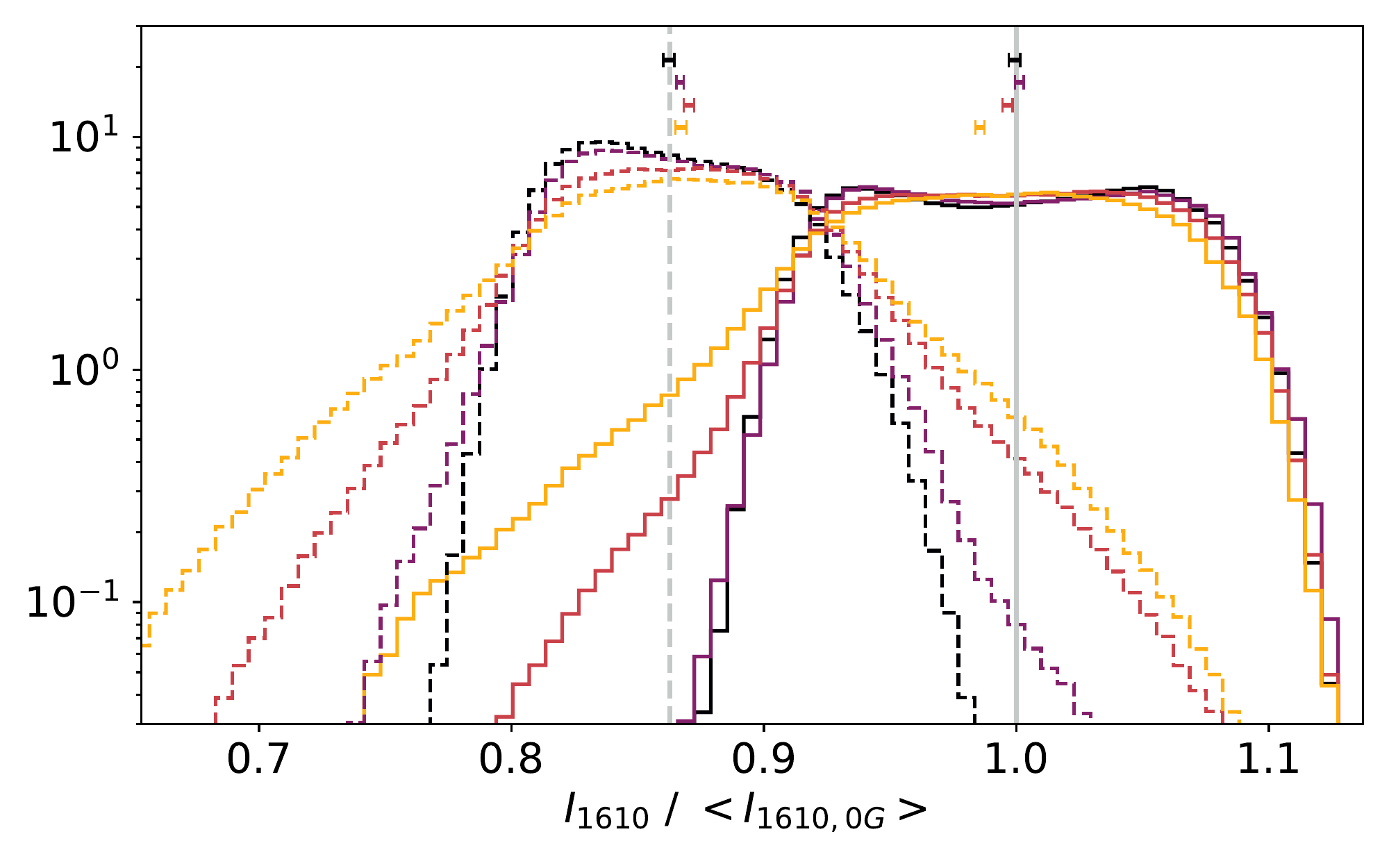}
\includegraphics[height=.149\textheight,trim={5 15 5 5},clip]{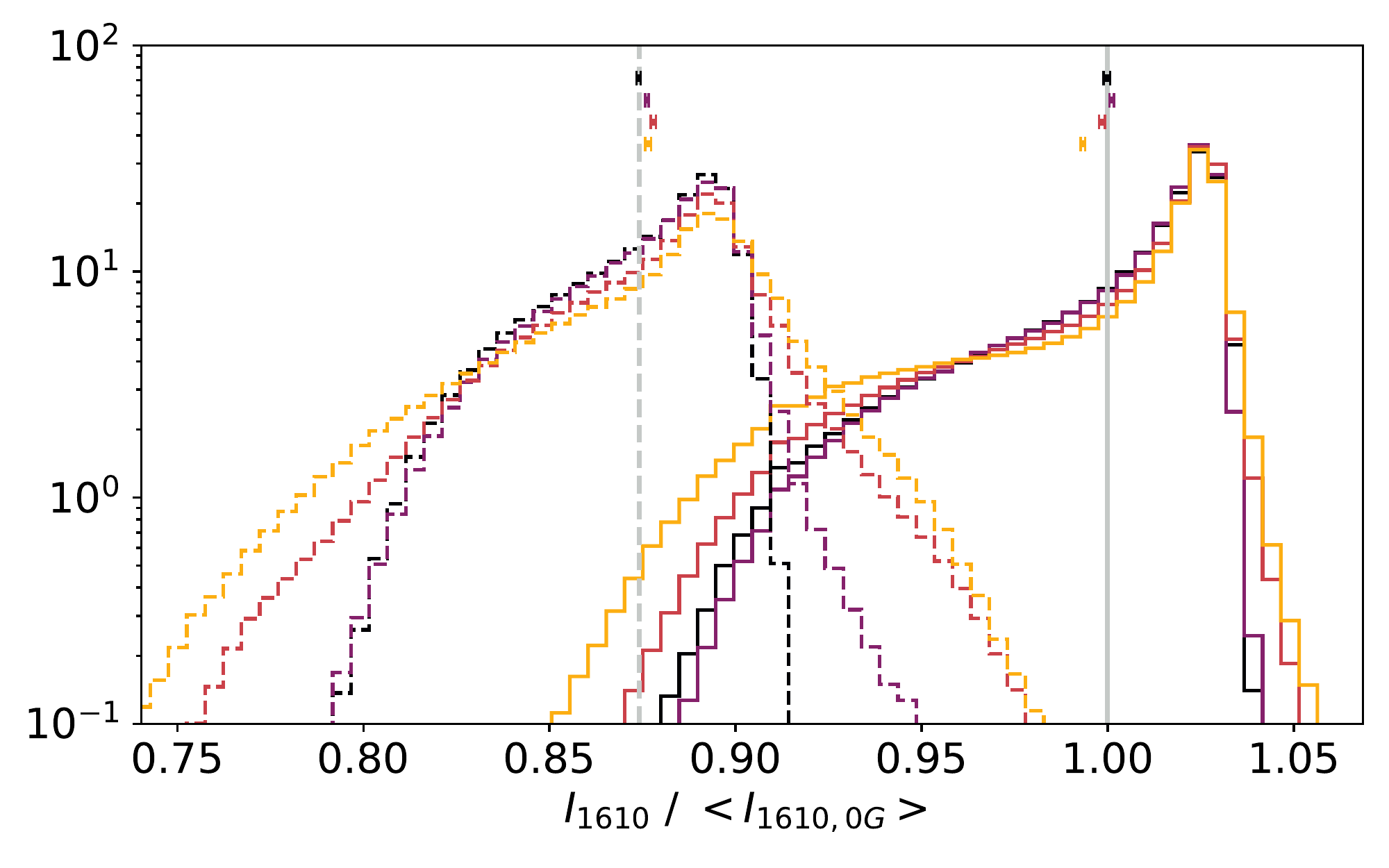}
\includegraphics[height=.149\textheight,trim={5 15 5 5},clip]{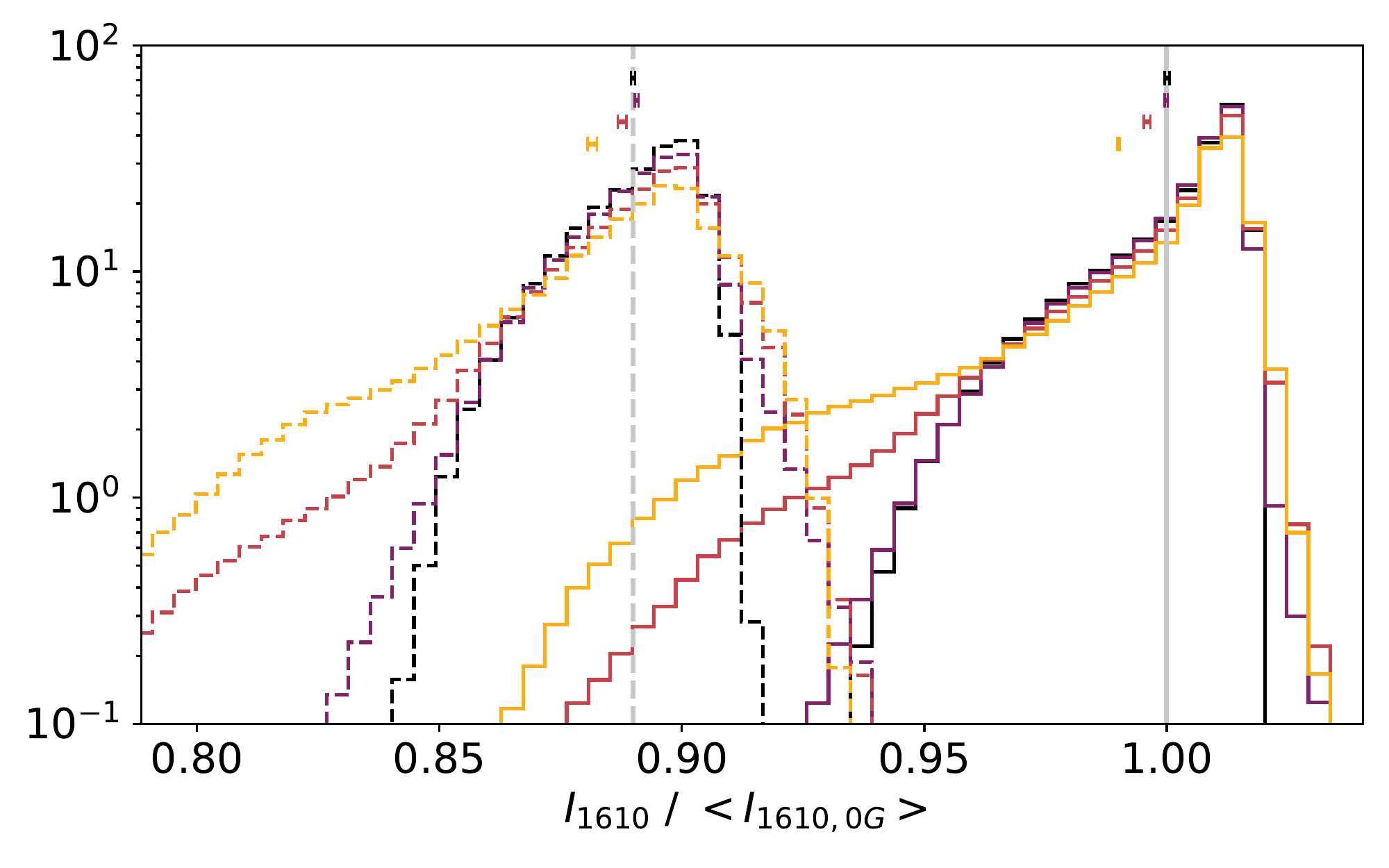}}
\vbox{
\includegraphics[height=.149\textheight,trim={15 15 5 5}]
{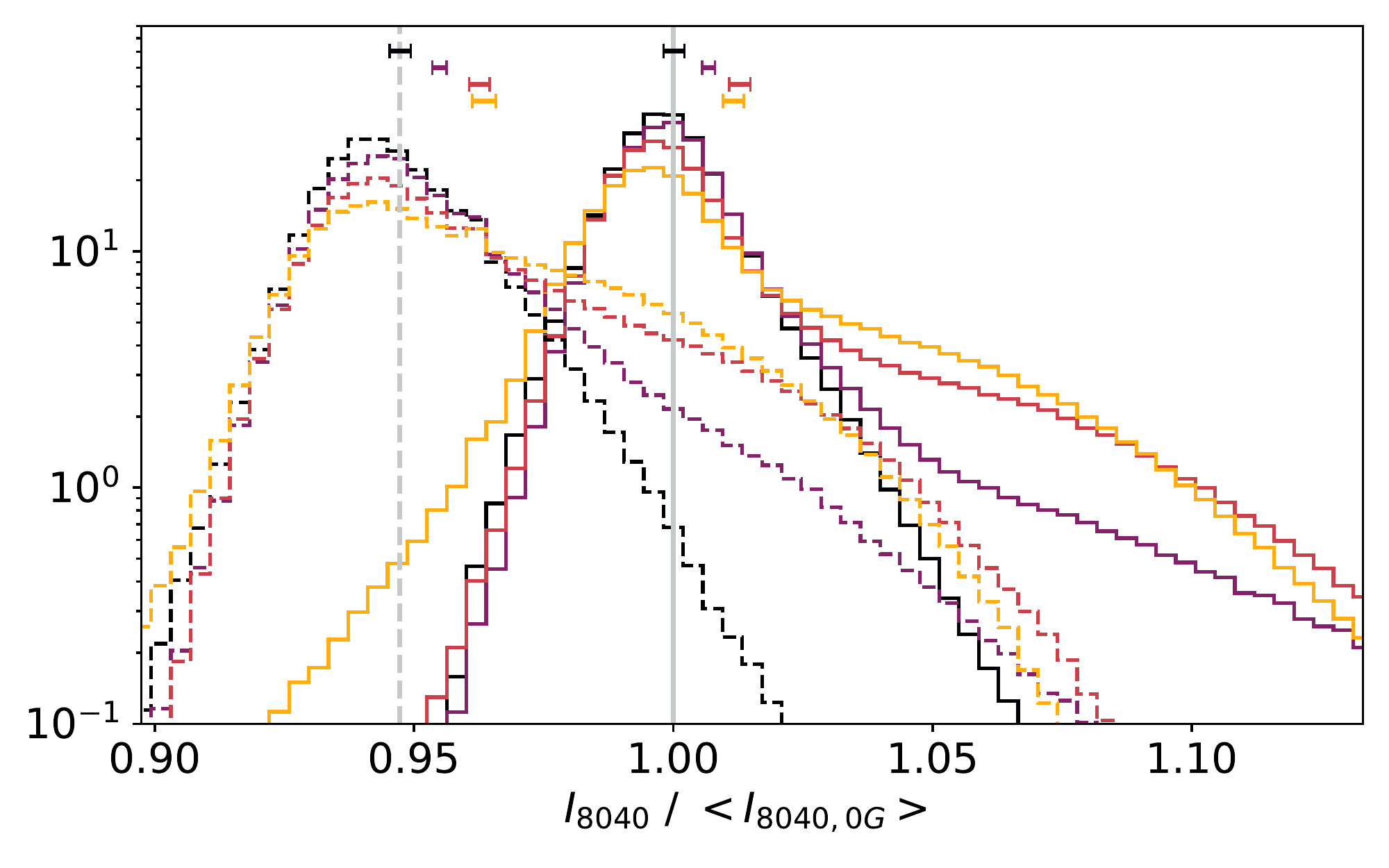}
\includegraphics[height=.149\textheight,trim={5 15 5 5},clip]{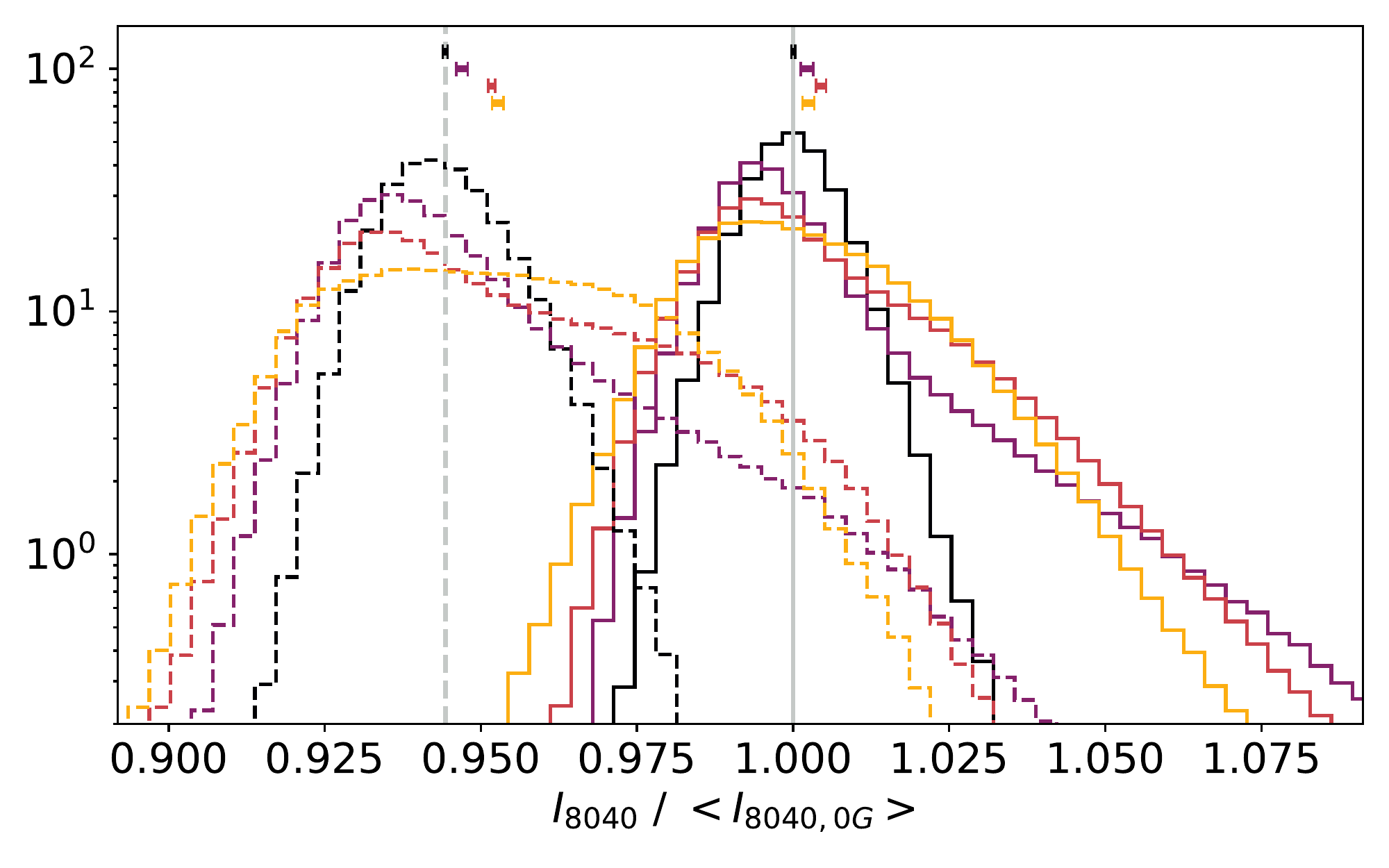}
\includegraphics[height=.149\textheight,trim={5 15 5 5},clip]{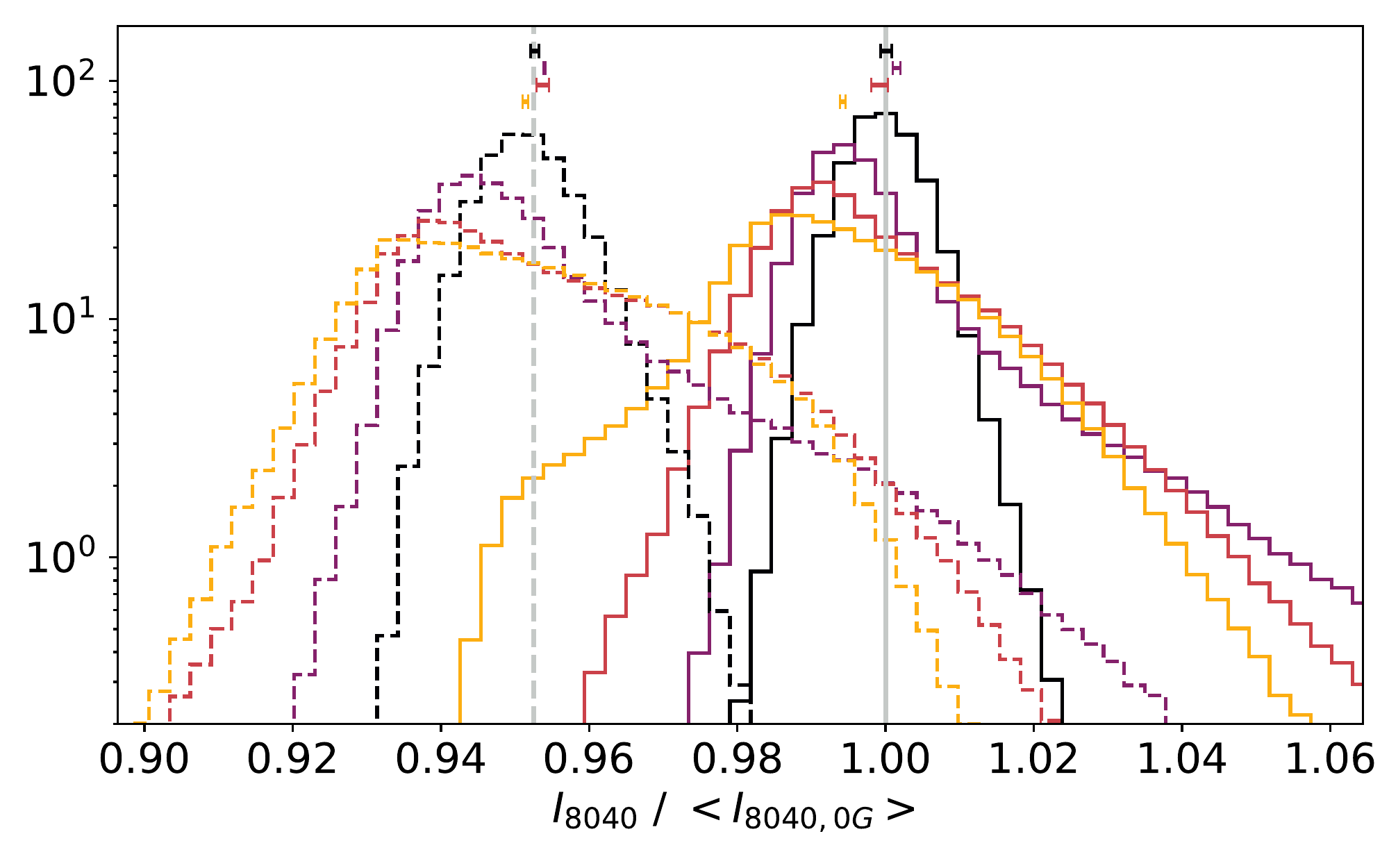}}
\caption[Histograms of intensity for K0, M0 and M2 snapshots at disc-centre and $\mu$ = 0.5.]{Semi-log histograms of intensity for K0, M0 and M2 snapshots (left, middle and right-hand columns, respectively). Black, purple, red and yellow lines indicate field free, $\langle B_z\rangle=$\,100\,G, 300\,G and 500\,G simulations, respectively. Intensities have been normalised to the mean intensities of the field-free disc-centre snapshots. Minimum and maximum values of the mean intensities of the individual snapshots that are used to generate these combined histograms are indicated by the small coloured horizontal bars. Histograms are shown at two different limb positions, $\mu=1.0$ (solid lines) and $\mu=0.5$ (dashed lines). From top to bottom, graphs show wavelengths of 388\,nm, 602\,nm, 1\,610\,nm, and 8\,040\,nm. The histograms have been normalised such that the sum over all bins equals unity.
}
\label{fig:hist1D}
\end{figure*}

For a more detailed view of intensity distributions for different spectral types, and to highlight their limb dependence, Fig.~\ref{fig:hist1D} shows histograms for K0 (left column), M0 (middle column) and M2 (right column) main-sequence stars at four different wavelengths (388~nm., 602~nm, 1610~nm and 8040~nm from top to bottom). Intensity bin sizes of the histograms vary across wavelength and spectral type, and are given in the caption of the figure.  Disc-centre values are shown with solid lines, while $\mu = 0.5$ histograms are shown by dashed lines. All intensities have been normalised by the field-free disc-centre intensity at the given wavelength. 

The histograms are obtained by combining data from all snapshots. The coloured horizontal bars at the top of each graph in Fig.~\ref{fig:hist1D} show the range of means of all snapshots at a given $\mu$ and $\langle B_z \rangle$. These represent the temporal variations of the simulations. Very little temporal variation in mean snapshot intensity is seen compared to the width of the distribution of pixel values. The limited time sampling of our snapshots means that these temporal variations are possibly underestimated due to granules generally re-appearing in the same positions over these time-scales. We note that the shapes of the histograms are insensitive to these small variations in the mean.
 
The mean intensities of the magnetic field-free snapshots (indicated by vertical grey lines in Fig.~\ref{fig:hist1D}) decrease from disc centre (solid lines) to $\mu = 0.5$ (dashed lines) in all plots. This demonstrates that limb darkening is occurring for these wavelengths for all chosen spectral types. The histogram shapes change dramatically as a function of wavelength as illustrated by the four example wavelengths shown here. The 602\,nm histograms at disc centre closely resemble the disc-centre bolometric intensity histograms presented in \cite{Beeck2015b} and \cite{Salhab+2018}, in particular the narrowing of the distributions for later spectral types and the change in the peak asymmetry between early and late-type K / early M stars (the histograms in \cite{Salhab+2018} are for mean magnetic fluxes of 50\,G and show slightly less developed high-intensity shoulders compared to the histograms with \meanBz $=$ 100~G here). \par 
At disc centre, the central distributions of the histograms tend to show only small differences between magnetic field-free and $\langle B_z \rangle$~=~100\,G histograms, though all magnetic snapshots show an increase in the number of bright pixels. The largest differences appear at 388~nm and, for the M2 star in particular, also at 8040~nm. The magnetic features emerging in the {\meanBz\equals}100\,G simulations are generally small and therefore only a few pixels are moved from the peak to the high-intensity tail when radiation emerges from lower atmospheric layers. When radiation emerges from higher in the atmosphere (for example at 8040~nm) where the pressure is lower, magnetic features take up a larger area, leading to a larger high-intensity tail, and fewer pixels in the peak. Differences become more marked for larger magnetisations: for {\meanBz\equals}300\,G and 500\,G, dark features form that lead to extended low-intensity tails. At many wavelengths there is also an increase in the number of bright pixels, leading to most histograms developing more pronounced shoulders at larger intensity values. The following paragraphs discuss peculiarities for different spectral types.\par
\textbf{\underline{K0V}:}
Fig.~\ref{fig:hist1D} shows that the histograms for K0V stars mainly have single asymmetric peaks at both disc centre and $\mu = 0.5$. This is somewhat different from what was seen for G2V stars \citep[see figure 7 in][for ease of comparison, our histograms are replotted on a linear scale in Fig.~\protect{\ref{fig:hist1D3lin}} in the Appendix]{Norris2017}. This is likely due to the less discrete transition between granules and inter-granular lanes for K0 stars compared to G2 stars that leads to less separation between the intensities seen in each feature type. Disc-centre histograms in the NIR (see, e.g., 1610~nm histograms in Fig.~\ref{fig:hist1D}) are the only exception as a double peak is seen for all but the $\langle B_z \rangle$~=~500\,G simulations; this is due to the granules being more defined at this wavelength.\par 
When the mean magnetic field is increased, the number of small-scale (predominantly bright) and larger-scale (predominantly dark) magnetic features increases (see Fig.~\ref{fig:images_K0}) which  leads to a broadening of the histograms. At disc centre, increasingly prominent low-intensity tails are seen that can be associated with larger magnetic features. Due to their larger diameter, the surrounding atmosphere is unable to heat the larger features sufficiently, so they appear dark by comparison. At the same time, high-intensity tails develop due to the increased number of high-contrast small features. At disc centre in the NIR (where radiation emerges from deeper atmospheric layers) the area taken up by bright magnetic features is negligible and cannot be discerned on the histograms shown here. The combination of bright and dark features results in a flattening of the peak (or peaks in the case of NIR). The relative importance of bright and dark features changes with wavelength, for example, the distribution is close to symmetric at 388~nm, while there is a large low-intensity tail in the NIR.\par 
Looking towards the limb, at $\mu = 0.5$, the histogram peak height decreases with increasing magnetic field as more magnetic concentrations form in the intergranular lanes. These magnetic features allow radiation from deeper layers of the hot granular walls to emerge when the viewing angle is increased, lengthening the high-intensity tails. In the visible and NIR the main histogram peaks become more symmetric for larger \meanBz; in addition, there is a strengthening of the low-intensity tails. This increase in low intensities is due to larger, cool, magnetic features, which are dark at these wavelengths. When viewed at an angle, these features allow more of the hot walls to be visible (hence the corresponding larger high intensity tails), but, due to their size, some dark areas of the features will still remain visible, increasing the number of dark pixels as $\langle B_z \rangle$ increases.\par

\textbf{\underline{M stars}:} As with K0V stars, Fig.~\ref{fig:hist1D} shows that the M0V and M2V snapshots tend to have singly peaked distributions of intensity (an exception are the non-magnetic snapshots at 388\,nm, in particular for the M2V star).  Contrary to the K0 simulations, the low-intensity tails increase more strongly than the high-intensity tails as $\langle B_z \rangle$ increases from 100\,G to 500\,G. This is due to the emergence of large features that are dark at the chosen wavelengths, as illustrated in the images in Fig.~\ref{fig:images_M2} (right-hand column). Small-scale magnetic features are less conspicuous, except around 8~$\mu$m where radiation emerges from high atmospheric layers; at these heights, the granulation signature is very weak with some intergranular lanes showing weak positive contrasts, a sign of beginning reversed granulation. At 1.6$\mu$m where radiation emerges from deeper atmospheric layers, the granules appear at very uniform brightness, while the intergranular lanes extend over a range of brightness. This leads to very asymmetric histograms with a peak at the granular brightness and a sharp drop at the high-intensity side. This sharp drop remains present even in the magnetic simulations as small-scale features have very low contrast and are almost invisible. For $\langle B_z \rangle$~=~500~G, the larger dark features stand out very clearly and result in extended low-intensity tails. 

At 388~nm, 602~nm and 1\,610~nm, these dark features remain very pronounced even at inclined viewing angles, resulting in prominent low-intensity tails. At the same time, and similar to K0V stars, the higher intensity tails increase for $\mu=0.5$ as the hot and bright walls of the magnetic features are revealed. This effect is less pronounced for the cooler M-type stars where the Wilson depression is smaller \citep[see][]{Beeck2015,Salhab+2018}. Comparatively little brightening is seen at 1.6$\mu$m where there is little change in the asymmetry of the histogram shape, though the dramatic decrease at the high-intensity side of the peak is softened for larger \meanBz.\par
In summary, the histograms in Fig.~\ref{fig:hist1D} show that the distribution of intensities seen in a stellar atmosphere changes significantly across spectral type, wavelength and magnetic field strength. The strong spectral type and wavelength responses show that for high resolution, the appearance of spectral features cannot be simply scaled from solar values.

\section[Contrast spectra of magnetic regions for G, K and early M stars]{Contrast spectra of magnetic regions for G, K and early M stars}
\label{sect:contrasts}
The following section presents investigations into the effect of spectral type, magnetic field and wavelength on the intensities emerging from these simulated regions of a star. As the spatial resolution is low in observations of other stars, the high resolution of intensities obtained by simulations is often not required when using these outputs. Therefore, in this section we present the mean emergent intensities across a simulation box. To account for the varying total intensities emitted by different spectral types, we compare the contrasts of the magnetised regions with respect to the hydrodynamic snapshots of the same spectral type.\par
Contrasts are calculated using the total mean of all snapshots of a given spectral type with a given initial magnetic field. These are taken to represent the mean that would be observed for an area of the size of the simulation boxes. The means thus include pixels of very low magnetic field and of dark magnetic features, that are not strictly facular (but which we count to faculae, as these magnetic structures are much smaller than starspots and typically smaller than granules). The contrast of a magnetic region is calculated as
\begin{equation} \label{eq:1}
C(\langle B_z\rangle, \lambda,\mu)\,=\, \frac{I(\langle B_z\rangle, \lambda, \mu) - I(0,\lambda,\mu)}{I(0,\lambda,\mu)}.
\end{equation} 
Here, $C(\langle B_z\rangle,\lambda,\mu)$ is the spectral contrast for a given average vertical magnetic field, {\meanBz}, and limb angle, $\mu$; $I(\langle B_z\rangle, \lambda,\mu)$ is the mean spectral intensity over all pixels in all snapshots for a given {\meanBz} and $\mu$; and $I(0,\lambda, \mu)$ is the mean spectral intensity over all pixels in all hydrodynamic boxes at a given $\mu$.

\begin{figure*}
  \centering
   \begin{overpic}[width=.95\textwidth,trim={0 40 0 0},clip=]{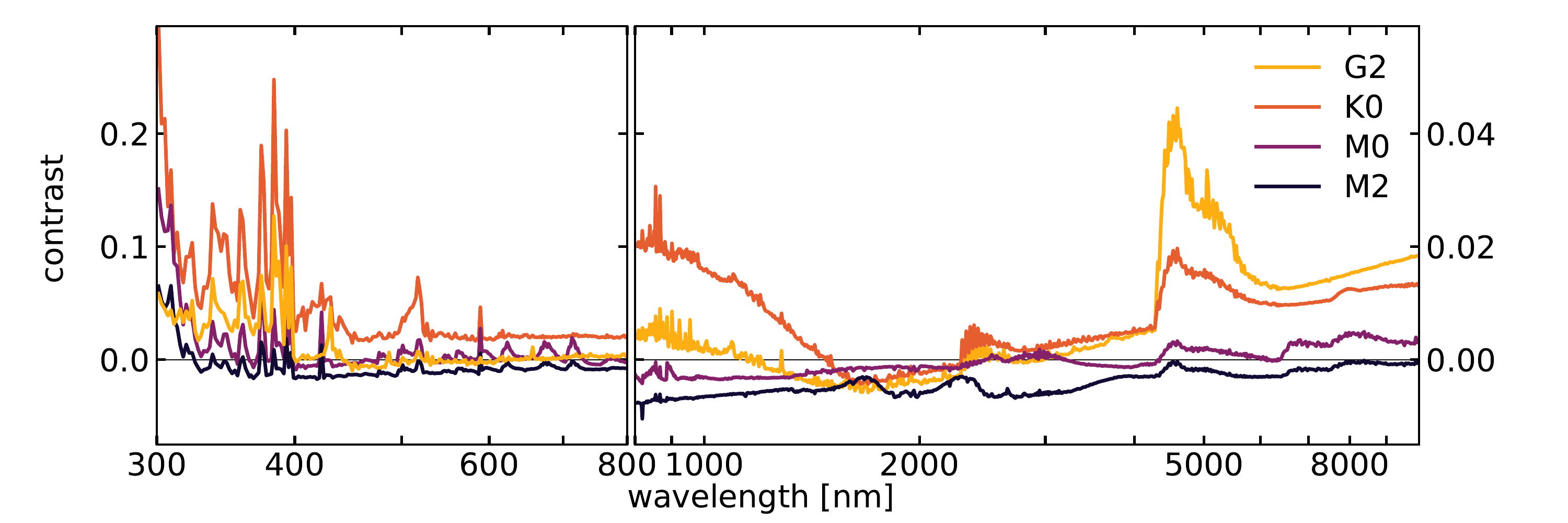}
      \put(30,24){\textsf{\Large $\mu$ = 1.0}}
   \end{overpic}
   \begin{overpic}[width=.95\textwidth,trim={0 0 0 10},clip=]{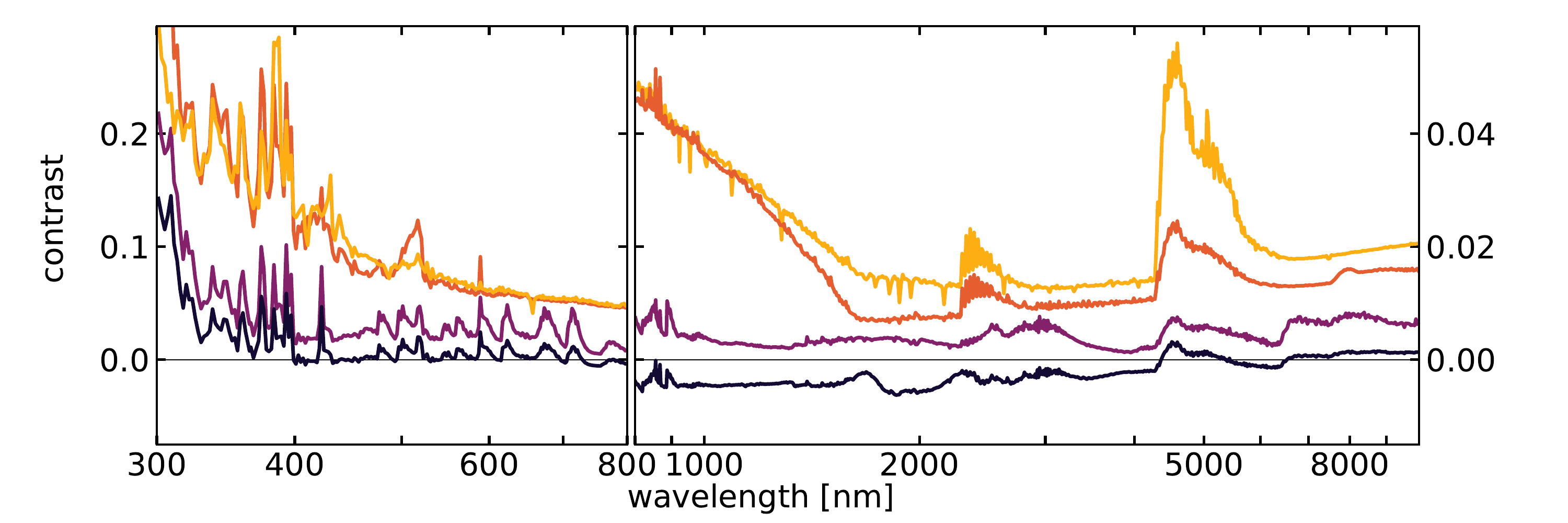}
      \put(30,28.5){\textsf{\Large $\mu$ = 0.5}}
   \end{overpic}
\caption[Mean contrast spectra for all MURaM spectral types with {\meanBz}{\equals}300\,G.]{Average contrast spectra relative to field-free simulations for all MURaM spectral types for snapshots with {\meanBz}{\equals}300\,G. Contrasts are shown at disc centre (top), and $\mu = 0.5$ (bottom) for wavelengths between 300\,nm and 10\,000\,nm using logarithmic $x$ axes. The  $y$-axis scale changes (by a factor of 5) at 800\,nm to bring out the spectral shape of the contrast in the visible and IR.}
\label{fig:contspec_300G}
\end{figure*}

\begin{figure*}
  \centering
   \begin{overpic}[width=.95\textwidth,trim={0 40 0 0},clip=]{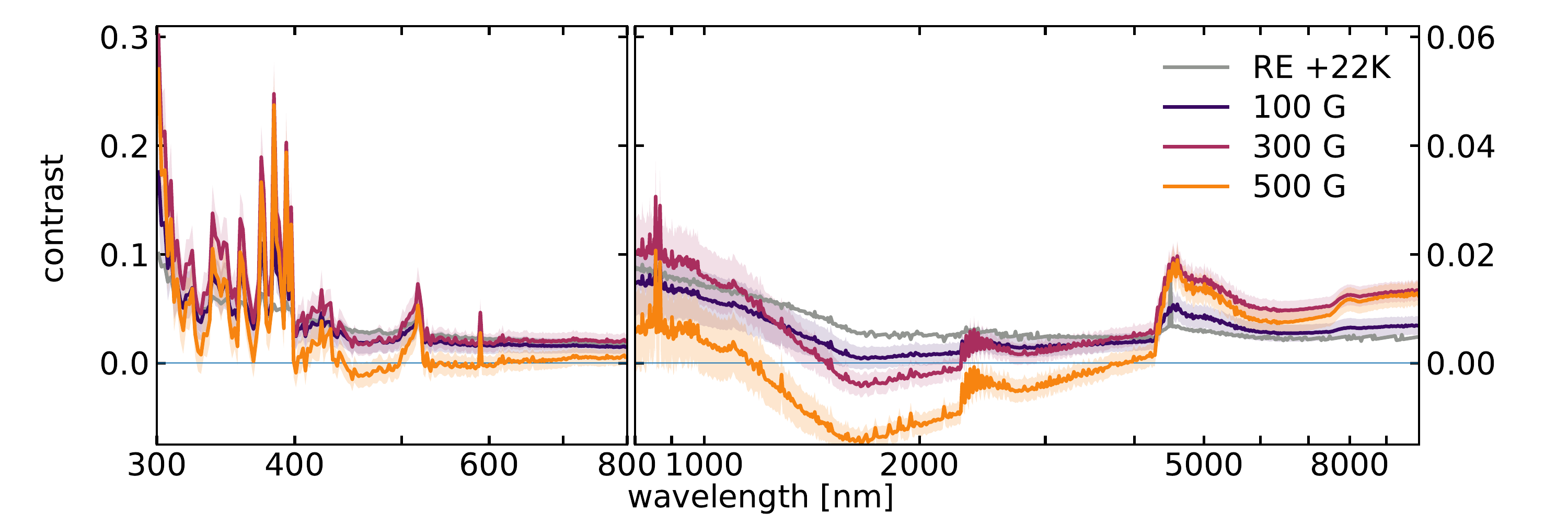}
      \put(30,24){\textsf{\Large $\mu$ = 1.0}}
   \end{overpic}
   \begin{overpic}[width=.95\textwidth,trim={0 0 0 10},clip=]{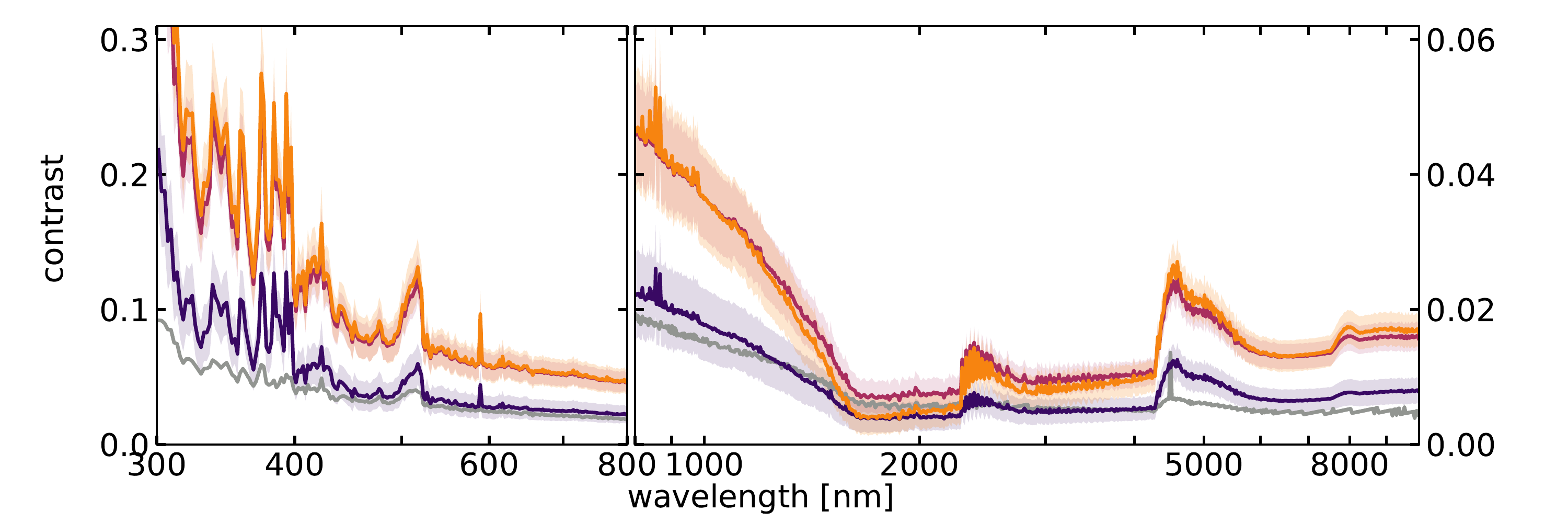}
      \put(30,28.5){\textsf{\Large $\mu$ = 0.5}}
   \end{overpic}
\caption[Mean contrast spectra for K0 spectral types for different magnetic flux densities]{Average contrast spectra relative to field-free simulations for K0 spectral types with {\meanBz}{\equals} $\{100, 300, 500\}$\,G (see also Fig.~\protect{\ref{fig:contspec_300G}}). Grey lines show contrasts for 1D radiative equilibrium atmospheres with approximately the same bolometric contrasts as the {\meanBz}{\equals}100\,G magnetic simulations (see Table~\protect{\ref{tab:contrast}}). The shading indicates the standard deviation of the contrast derived from the differences between the mean spectra calculated from individual snapshots of the same magnetisation; they thus reflect the temporal variations in the MURaM simulations.}
\label{fig:contspec_K0}
\end{figure*}

\begin{figure*}
   \centering
   \begin{overpic}[width=.95\textwidth,trim={0 40 0 0},clip=]{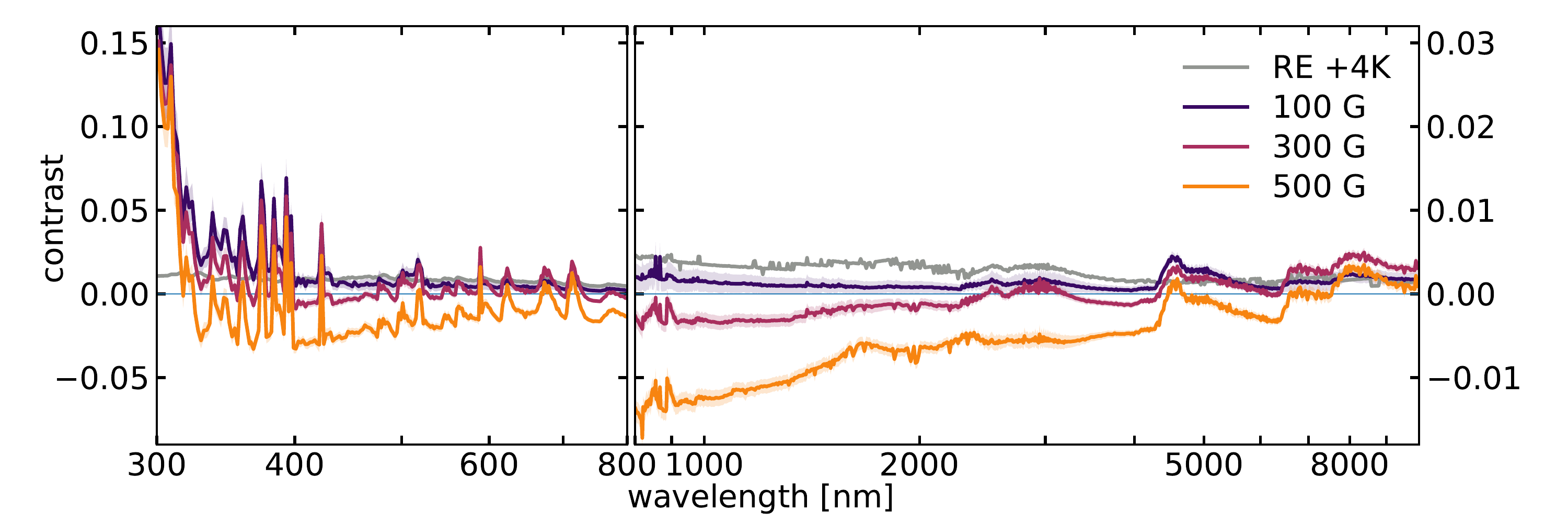}
      \put(30,24){\textsf{\Large $\mu$ = 1.0}}
   \end{overpic}
   \begin{overpic}[width=.95\textwidth,trim={0 0 0 10},clip=]{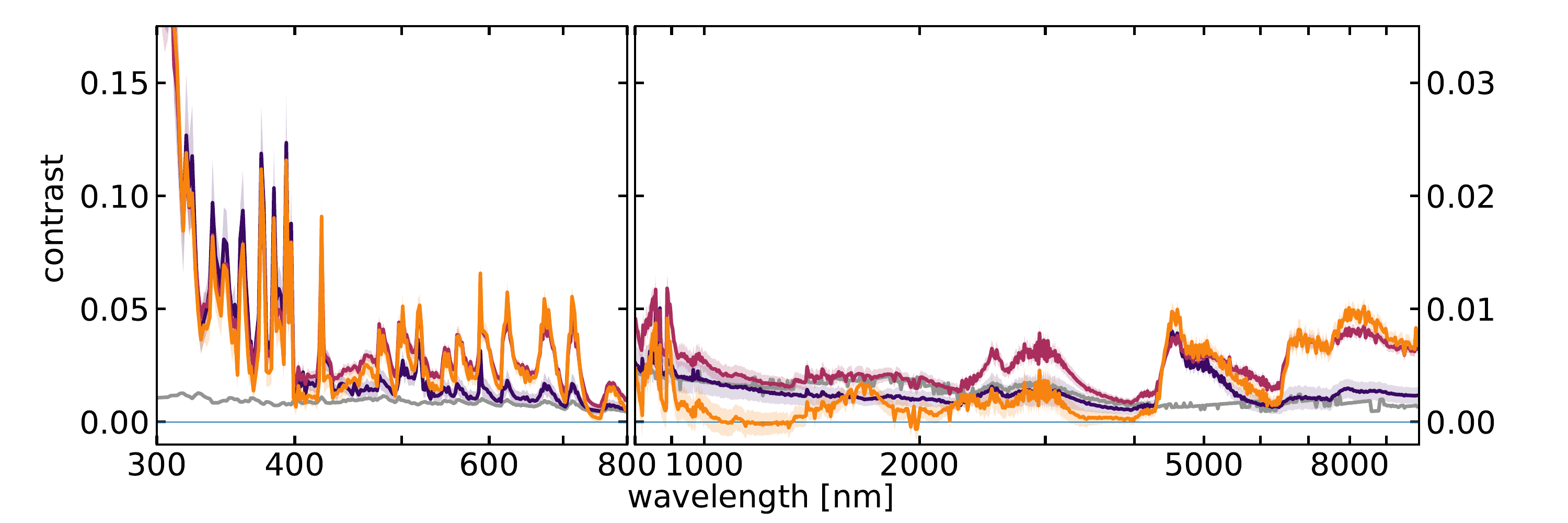}
      \put(30,28.5){\textsf{\Large $\mu$ = 0.5}}
   \end{overpic}
\caption[Mean contrast spectra for M0 spectral types for different magnetic flux densities]{Average contrast spectra as in Fig.\,\ref{fig:contspec_K0} but for spectral type M0. The $y$-axis scales differ at disc centre and $\mu=0.5$.}
\label{fig:contspec_M0}
\end{figure*}

\begin{figure*}
   \centering
   \begin{overpic}[width=.95\textwidth,trim={0 40 0 0},clip=]{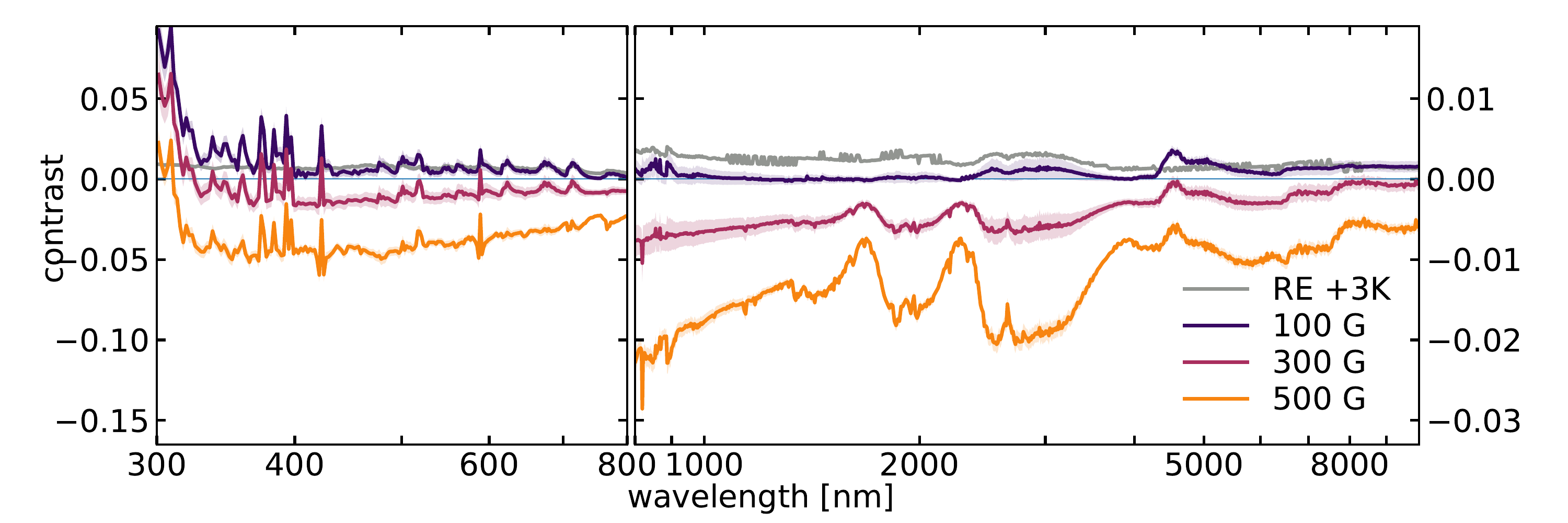} 
      \put(30,24){\textsf{\Large $\mu$ = 1.0}}
   \end{overpic}
   \begin{overpic}[width=.95\textwidth,trim={0 0 0 10},clip=]{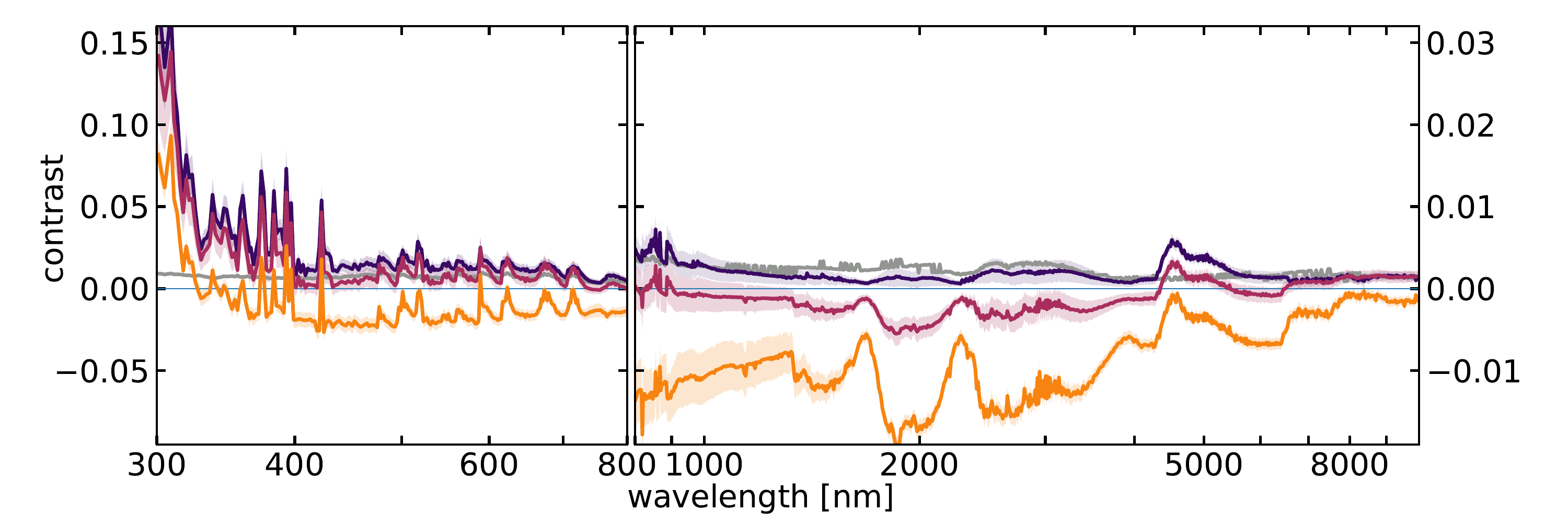}
      \put(30,28.5){\textsf{\Large $\mu$ = 0.5}}
   \end{overpic}
\caption[Mean contrast spectra for M0 spectral types for different magnetic flux densities]{Average contrast spectra as in Fig.\,\ref{fig:contspec_K0} but for spectral type M2.}
\label{fig:contspec_M2}
\end{figure*}

Tab.~\ref{tab:contrast} lists the bolometric (wavelength-integrated) contrasts of the magnetic snapshots for all four spectral types. For a given spectral type, we find that contrasts initially increase as the mean magnetic field {\meanBz} increases, before decreasing again as more dark pores form. The magnetic flux density for which the bolometric contrast peaks decreases with the effective temperature of the star.
The largest disc-integrated bolometric contrasts are seen for the G2 simulations, where they can reach 5\% for \meanBz{\equals}500~G, followed by contrasts for K0. Here the bolometric contrasts are highest for \meanBz{\equals}300~G where they reach 3.3\%. Bolometric contrasts for M-type stars are very low, and can be either positive (e.g., roughly 0.5\% for M0, \meanBz\ at 100~G and 300~G), neutral (M0, \meanBz{\equals}500~G; M2, \meanBz{\equals}100~G and 300~G), or negative (M2, \meanBz{\equals}500~G). These lower M-star contrasts mirror low contrasts seen for granulation \citep[see][]{Beeck2013b} and for spots \citep[see][]{panja+2020} on late-type stars. 

Disc-integrated bolometric contrasts only tell part of the story, as contrasts change significantly with wavelength and limb distance. Figure~\ref{fig:contspec_300G} shows mean contrast spectra for {\meanBz}{\equals}300~G with respect to field-free simulations for all four spectral types. The top panel shows the disc-centre contrasts, while the bottom panel shows contrasts at $\mu = 0.5$. To bring out lower-contrast features, we use an expanded scale for wavelengths above 800\,nm. Generally, contrasts increase towards the limb, though the rate of the increase varies with spectral type, magnetic activity and wavelength region. For example, for M0, we observe a switch in the sign for the visible and NIR contrasts from predominantly dark at disc centre to bright at a limb distance of 0.5.

Even though the disc-integrated bolometric contrasts are largest for the G2-star calculations, faculae at disc centre are brighter for the K0 star compared to the G2 star in the NUV, visible and NIR. This may partly be due to the maximum of the Planck function shifting to redder wavelengths for K stars. Indeed, in the IR the trend is reversed, as expected if this is the cause. Closer to the limb the G2-star contrast catches up with that of the K0 star, even in NUV and visible. This may have to do with the deeper Wilson depression seen for spots on earlier-type stars \citep[see][]{panja+2020}, meaning that a larger area of the hot walls becomes visible \citep[see also][]{Beeck2015}{}{}. 

By and large, contrasts decrease from the UV to the visible and NIR, taking on a minimum near 1.6\,$\mu$m and increasing slightly towards longer wavelengths, with a distinct peak around 4.5 $\mu$m (dominantly caused by rovibrational CO lines). Complexity is introduced by different atmospheric species becoming more prominent in the spectra for different spectral types. For example, contrasts for M0 and M2 stars show strong variations between 400 and 900~nm due to molecular bands, such as the TiO band heads at 517~nm, 619~nm and 709~nm, while other stars vary more smoothly at these wavelengths. Fig.~\ref{fig:contspec_300G} reveals that there is no obvious trend with stellar temperature that could easily be used to scale spectral contrasts. 

Figs~\ref{fig:contspec_K0}, \ref{fig:contspec_M0} and \ref{fig:contspec_M2} show the changes in average contrast for different magnetic flux densities for K0, M0 and M2 spectral types, respectively \citep[results from G2 main-sequence MURaM simulations were presented in][]{Norris2017}. For all spectral types considered here, the contrasts obtained from simulations with higher magnetic flux densities show stronger spectral features and a steeper wavelength dependence. The contrast spectra for weak mean fields, {\meanBz\equals}100\,G, are positive (i.e., bright) at most wavelengths and disc positions and show less variation with wavelength. However, even for such relatively low magnetic flux densities the contrast spectra are very structured and show numerous spectral features. \par
The contrasts derived from the K0 main-sequence simulations show similar spectral features to those derived for the Sun from G2 main-sequence simulations  \citep[see][]{Norris2017}. Two noticeable differences are the strengthening of the feature around 520~nm (most likely due to (0,0) MgH with band head at 521\,nm, combined with the Mg b triplet) and a  weaker response of the CO feature near 4.5\,$\mu$m in the K0-star contrasts. As seen in Fig.\,\ref{fig:images_K0}, dark pore-like features begin to appear in the presence of a strong mean magnetic field. When looking at disc centre (top row in Fig.\,\ref{fig:contspec_K0}) bright features dominate for {\meanBz\equals}300\,G. For {\meanBz\equals}500\,G, bright and dark features largely balance, resulting in very low contrasts in most of the visible and negative contrasts around the opacity minimum at 1.6\,$\mu$m. Closer to the limb (e.g., at $\mu=0.5$ as shown in the lower row of Fig.\,\ref{fig:contspec_K0}) contrasts increase and become positive throughout. The contrast increases more strongly for larger flux densities such that the contrasts at {\meanBz\equals}500\,G almost matches that at 300\,G, even though disc-centre intensities are typically more than a factor of two lower. \par
As shown in Fig.\,\ref{fig:contspec_M0}, contrasts for M0 stars with {\meanBz\equals}100\,G are positive at all wavelengths and limb positions. For \meanBz$=300$\,G and \meanBz$=500$\,G, disc-centre contrasts above $\sim$350~nm are mostly negative, except in some of the molecular features. The negative contrast is due to the magnetic features becoming large relative to the Wilson depression which is smaller for later-type stars, \citep[see][]{Beeck2015}. The radiation from the hot walls is then inefficient at heating the cooled magnetic region, making them appear dark compared to the surrounding atmosphere. Heating of the upper layers of the magnetic atmospheres is strong, however, resulting in a steep increase of UV contrasts below $\sim 320$\,nm. 

Away from disc centre (see lower panel in  Fig.~\ref{fig:contspec_M0}), contrasts are positive for all magnetic flux densities considered, with the \meanBz\equals 500\,G simulations showing the steepest increase of contrasts towards the limb. When integrated over the stellar disc, the flux contrasts for {\meanBz\equals}100\,G and 300\,G are positive at all wavelengths. For \meanBz\equals 500\,G, disc-integrated contrasts are positive in the UV and most of the visible and mid IR, but they are negative for most wavelengths between approximately 800\,nm and 4300\,nm. Indeed, as shown in Tab.\,\ref{tab:contrast}, the bolometric contrast vanishes, i.e., the equivalent \teff\ for the \meanBz\equals 500\,G M0 simulations is equal to that of the field-free simulations. 

At disc centre and for \meanBz\equals100\,G, the M0 and M2 contrasts show many similarities, though the M0 contrasts remain positive throughout, while M2 contrasts are largely negative in the NIR. Many of the prominent spectral features seen in the M0 contrasts also stand out at M2 (see Fig.\,\ref{fig:contspec_M2}), though the strength of the features varies. For example at \meanBz\equals300\,G and 500\,G, the M0 contrasts show stronger features in the visible and relatively flat contrasts in the IR, while M2 contrasts show broad molecular lines at 1.8 and 2.4 $\mu$m. Contrary to what is seen for the hotter spectral types, only little brightening is seen at $\mu=0.5$ for the M2 contrasts at \meanBz\equals300\,G and 500\,G.  

To highlight the effects of the emission properties of the facular features as well as those of the corrugated aspect of the granulation and the geometric shape of the magnetic features, we have also included contrasts derived from 1D radiative equilibrium atmospheres (grey lines in Figs~\ref{fig:contspec_K0} to \ref{fig:contspec_M2}). The 1D radiative equilibrium atmospheres were calculated using ATLAS9 \citep{Kurucz1992} for effective temperatures that closely match (within a degree) the disc-integrated effective temperatures of the magnetic and non-magnetic MURaM simulations as provided in Tabs~\ref{table:SpectralTypes} and \ref{tab:contrast}. In all cases, the contrasts derived from the 1D atmospheres show flatter spectral response with fewer and weaker spectral features compared to the contrasts derived from the 3D simulations. They also show very different centre-to-limb variability. This is particularly noticeable for strong {\meanBz} where the disc-centre contrasts are typically negative, while contrasts closer to the limb become positive, irrespective of the overall bolometric disc-averaged contrast of the simulation box. Contrasts derived from 1D atmospheres do not display such a switch and remain either positive or negative at all limb angles, depending on whether the feature has a higher or lower effective temperature. 

\section{Discussion and Summary}
\label{sec:discussion}
Using 3-dimensional box-in-a-star simulations and spectral synthesis, we have calculated spectral intensities for wavelengths between approximately 250~nm and 160\,000~nm for main-sequence spectral types of G2V (see paper \textsc{i}), K0, M0 and M2V. 
We have observed that, for a given injected magnetic field and spectral type (i.e., \teff\ and {\logg}), features of different sizes, apparent structure and spectral brightness will emerge. The resulting intensity distributions are typically double peaked -- reflecting the stellar granulation -- with extended tails due to small-scale bright and dark magnetic features. 

As these small-scale features are unresolved on stars other than the Sun, contrasts were calculated for mean intensities over a simulation box. These contrasts have been found to be complicated functions of effective temperature, magnetic field, and wavelength, leading to the conclusion that accurate contrast values cannot be obtained using simple effective temperature scaling relations and solar values. Contrasts derived from the magneto-convection simulations are very different from contrasts obtained from 1D radiative-equilibrium atmospheres for effective temperatures that correspond to the overall bolometric contrast of the simulation boxes. This is not surprising given that they are obtained from a collection of atmospheres that represent a large variety of different magnetic and non-magnetic features. In particular, the ``magnetic contrasts'' show steeper increases towards shorter (NUV) wavelengths and much more pronounced spectral features than the equivalent radiative-equilibrium contrasts. 
In addition, many of the NIR contrasts are negative (dark) at disc centre, but become positive (bright) closer to the limb. Such behaviour is only observed in models that account for the geometric effects of the granulation \citep[see also][]{witzke+2022faculae}. 

The calculations presented here will help improve stellar reconstructions for variability studies and stellar noise parameterizations in planetary transits. The spectral data can be used to represent small-scale magnetic features, as was done for the Sun by \cite{Yeo2017}. Different spatial scales, from one pixel to the full simulation box, can be selected and used to interpret observations, or as parameters in a larger-scale stellar simulation to obtain more accurate spectral outputs (including for revised centre-to-limb variations) for typical active-region surface distributions \citep[see, e.g.,][]{Nemec+2022}.

Detailed centre-to-limb calculations from field-free stellar granulation simulations are now available for a large range of spectral types \citep[e.g.,][]{Magic2015} and can be used to model exoplanetary transits \citep{Morello2017}. Calculations such as the ones presented in this paper will allow accounting for the presence of active regions and their effects on the centre-to-limb variation of the stellar intensity. 

In addition, they can also be used to estimate the stellar contamination of exoplanet transits. A small number of transits show evidence of exoplanets passing in front of bright features \citep[e.g.,][]{Kirk2016,Espinoza+2019}. These allow tentative estimates of the contrasts and filling factors of the occulted features, though measurements so far mainly cover the visible wavelength range where contrasts show a relatively flat spectral response. As a result there is a strong degeneracy between feature contrast and filling factor. 

Assuming the transit-profile deformations to be due to the presence of bright stellar surface features, \cite{Kirk2016} found contrasts of the order of 12\% and 9\% in the $u'$ and $g'$ bands for WASP-52, a K2 host star. In their model, the bright feature's centre is at a limb distance of $\mu=0.78$. For our closest comparison spectral type, K0, 300G contrasts at $\mu=0.8$ are 11\% and 5\% for the $u'$ and $g'$ filters, respectively\footnote{Contrasts have been estimated from the filter responses only; i.e., camera and atmospheric effects have been neglected.}. While the $g'$ values in particular are lower, these contrasts are in reasonably good agreement with the contrasts presented in \cite{Kirk2016}. 

From transit observations of WASP-19b, where the host star is a mid G-type star with \teff$\sim$5460~K, \cite{Espinoza+2019} determined noticeably larger contrasts of the order of 10\% (in the red and NIR) up to 20\% near 450\,nm for a large bright feature close to $\mu=0.8$. Such large contrasts are not seen in our K0 or G2 simulations except at very large viewing angles where $\mu \le 0.5$ ($\theta \ge 60^\circ$), see, e.g., the lower panel in Fig.~\ref{fig:contspec_300G}). We note, however, that the contrast spectra shown here are derived from means over a full simulation snapshot. Individual features show much stronger brightness enhancements as demonstrated by the histograms and images such as, e.g.,  Figs\,\ref{fig:images_K0} and \ref{fig:images_muK0} for K0. 

In addition to direct occultations and thus deformations of the transit lightcurve, the presence of unocculted active regions can bias measurements of the lightcurve depth (and hence the inferred extent of exoplanet atmospheres) through the transit light-source effect, TLSE \citep[][]{rackham+2018TLSE1,rackham2019TLSE2}. Most current exoplanet retrieval algorithms account for the effect of active regions, though they tend to model faculae using ``hot-star'' spectra. 
\cite{witzke+2022faculae} showed that 1D radiative-equilibrium stellar atmosphere models fail to capture the spectral slope and limb dependence of the facular contrasts for G2 main-sequence stars. The calculations here confirm these findings for K0, M0 and M2 main-sequence stars: 1D radiative equilibrium models (shown as grey lines on Figs~\ref{fig:contspec_K0} to \ref{fig:contspec_M2}) show a much weaker UV excess slope, fewer spectral features and different centre-to-limb behaviour than facular contrasts derived from magneto-convection models. 

Recent examples of exoplanet atmosphere retrievals  where radiation from unocculted bright features affects the transit depth  include GJ 1214b \citep{rackham2017gj1214b}, WASP-79b \citep{rathcke+2021}, and WASP-103b \citep{kirk+2021wasp103b}. Assuming hot-star spectra, the derived excess temperatures were of the order of 350~K for GJ 1214 (M4.5{\sc v}, \teff\,$\simeq$\,3300~K), 450~K to 500~K for WASP-79 (F5{\sc v}, \teff \,$\simeq$\,6600~K) and 350~K for WASP 103 (F8{\sc v}, \teff\,$\simeq$\,6100~K). All of the host stars lie outside the effective temperature range considered here so that direct comparisons are not possible. 

Given the trends observed for early M stars, we would expect faculae to manifest quite differently on the M4.5 star GJ~1214 compared to what is seen on the Sun. For M2 stars, we derive very low or even negative  contrasts at 300\,G (i.e. features darker than the quiet stellar surface), except close to the limb and in the UV. It is thus difficult to see how unocculted active regions akin to the ones modelled here can mimic a spectrally flat brightening in excess of 300 K. However, as the TLSE is due to a difference in emission between the occulted and unocculted region of the star, a lower transit depth would also ensue if the planet were to pass in front of an activity belt with an even coverage of dark features that are small enough so as not to show up as distinct ``bumps'' in the transit lightcurve.   

The G2 simulations can act as a rough indicator of expected contrasts for the late-F stars WASP 79 and WASP 103, though simulations for hotter stars are expected to yield larger temperature differences than the cooler G2 simulations \citep[see][]{Beeck2015}. We derive an effective temperature difference of 55\,K and 71\,K between the mean disc-integrated bolometric fluxes for 300\,G and 500\,G simulations relative to non-magnetic simulations. While these bolometric temperatures are much lower than inferred from transit spectroscopy, brightness temperatures away from disc centre can be much higher, especially in the UV or in specific bands, such as in the G-band ($\sim $430\,nm) where the mean brightness temperature at a limb distance of $\mu=0.5$ is approximately 180\,K higher for the 500-G simulations compared to the non-magnetic simulations. For the four wavelength regions shown in Figs~\ref{fig:images_K0} to \ref{fig:hist1D}, mean brightness temperature differences at $\mu=0.5$ are approximately 260\,K (388\,nm), 120\,K (602\,nm), 80\,K (1.61\,$\mu$m) and 112\,K (8.04\,$\mu$m). As part of the active-region temperature fit in exoplanet retrievals is driven by the spectral slope of the contrast, it is unlikely that the mean temperature offsets derived from hot-star spectra can be compared directly to brightness temperatures derived from magnetoconvection simulations. 

Indeed, our calculations confirm that it is difficult to unambiguously detect bright features in the visible or NIR with current instrumentation; observations in the UV offer more spectral leverage due to the larger facular contrasts. At high signal-to-noise ratios and spectral resolution it should become possible to trace facular signatures in some of the strong spectral features (e.g., TiO bands for M-type and the CN band for K-type stars). As these spectral domains are also strongly spot affected, detailed modelling of spot contrasts will become necessary, in particular for the later spectral types where observations \citep[see, e.g.,][]{Berdyugina2005} and models \citep{panja+2020} infer smaller temperature differences between starspots and their surrounding photospheres. 

The spectral synthesis calculations presented here rely on the assumption of LTE. While LTE calculations are sufficient for visible and IR wavelengths, they provide a poor approximation of the emergent intensity in the UV. For instance, \cite{Shapiro+2010} showed that quiet-Sun LTE continuum fluxes are typically 40\% too low between 210\,nm and 250\,nm. Non-LTE effects on the emergent intensities will be explored in a future paper \citep{tagirov2023}.

All results in this paper assume solar metallicity. Metallicity plays an important role in the opacities in a stellar atmosphere, as it affects both line and continuum opacity sources (particularly H$^-$). Spectral lines have been found to be the main source of variability on time-scales of a day and longer on the Sun \citep{Shapiro+2015}. As such, understanding the role of metallicity is important for accurate modelling of limb darkening, facular contrasts and stellar variability. Higher metallicity has been observed to be associated with higher variability  in \citet{Karoff2018} and attributed to the larger facular contrasts resulting from the metallicity increase. 1D atmospheric models have also been used to demonstrate this increase of facular contrast and variability with increased metallicity \citep{Witzke2018}. To properly account for these changes with metallicity in stellar and planetary studies, these 3D calculations should be reproduced for a variety of metallicities.\par
Finally, this paper does not fully explore the changes in centre-to-limb variations with spectral type, as these data will allow. Nor does it select individual magnetic features to explore how the typical structures and emergent intensities vary. These issues will be explored in future papers.

\section*{Acknowledgments}
The authors would like to thank the referee for their helpful comments and swift response. 
CMN acknowledges support through studentship funding of the UK Science and Technology Facilities Council (STFC). This work was also supported by the German Federal Ministry of 
Education and Research under project 01LG1209A. YCU would like to thank James Kirk for useful discussions, and acknowledges financial support through STFC Grants ST/S000372/1 and ST/W000989/1. We would also like to thank the International Space Science Institute, Bern, for their support of science teams 335 and 373 and the resulting helpful discussions.

For the purpose of open access, the author has applied a Creative Commons Attribution (CC BY) license to any Author Accepted Manuscript version arising.

\section*{Data Availability}
The data underlying this article will be shared on reasonable request to the corresponding author.

\bibliographystyle{mnras} 
\bibliography{PaperII.bib}

\begin{appendix}

\section{Tables of brightness temperatures} 
As the snapshots shown in Figs~\ref{fig:images_K0} to \ref{fig:images_muM2} have been normalised, we list their original brightness temperatures, $T_b$, in Tabs~\ref{tab:SpecInt_SN} and \ref{tab:SpecInt_muSN} at disc centre ($\mu=$\,1.0) and at $\mu=$\,0.5, respectively. The values given in the first line of each set (for the non-magnetic simulations) correspond to $T_b = \frac{hc}{\lambda k} \left(\ln \left(1 + \frac{2hc^3}{\lambda^5 I_\lambda}\right)\right)^{-1}$, where $I_\lambda$ is the monochromatic intensity of the given snapshot at wavelength $\lambda$. 
The values in subsequent lines list the brightness temperature increase or decrease relative to the non-magnetic runs. The tables illustrate clearly that the spectral response of active regions is very far from that expected from radiative equilibrium atmospheres. 
\begin{table}
    \caption{Disc-centre brightness temperatures and temperature offsets for the snapshots shown in Figs~\ref{fig:images_K0} to \ref{fig:images_M2}. The values listed for the non-magnetic runs are brightness temperatures; values for the magnetic runs correspond to the brightness temperature increase (positive contrasts) or decrease (negative contrasts) relative to the non-magnetic runs. See text for further description.}
    \label{tab:SpecInt_SN}
    \centering
    
       \begin{tabular}{ccrrrr}
       \hline\hline
        spectral & \meanBz & 388nm & 602nm & 1.61$\mu$m & 8.04$\mu$m \\
         type & [G] & [K] \ & [K] \ & [K] \ & [K]\hspace*{.8ex} \\ \hline
        K0 & -- & 4472 & 5219  &  5971 &  4530 \\
        K0 & 100 & $+42$ & $+18$ &  $+4$ &  $+25$ \\
        K0 & 300 & $+67$ & $+22$ &  $-11$ &  $+47$ \\
        K0 & 500 & $+53$ & $-1$ &  $-44$ &  $+44$ \\ \hline
        M0 & --  & 3603 & 3943 &  4654 &  3586 \\
        M0 & 100 & $+10$ & $+3$ & $+2$ & $+6$ \\
        M0 & 300 & $+5$ & $+3$ & $-3$ &  $+12$ \\
        M0 & 500 & $-2$ & $-6$ &  $-3$ & $+8$ \\ \hline
        M2 & -- &  3440 & 3628 &  4260 & 3317 \\
        M2 & 100 & $+5$ & $+4$ & $\pm 0$ & $+4$ \\
        M2 & 300 & $-2$ & $-4$ &  $-7$ & $-2$ \\
        M2 & 500 & $-15$ & $-21$ & $-18$ & $-15$ \\
    \hline
    \end{tabular}
\end{table}

\begin{table}
    \caption{Brightness temperatures and brightness temperature differences as in Tab.~\protect{\ref{tab:SpecInt_SN}}, but for images at $\mu=0.5$ shown in Figs~\ref{fig:images_muK0} to \ref{fig:images_muM2}; see text for description.}
    \label{tab:SpecInt_muSN}
    \centering    
       \begin{tabular}{ccrrrr}
              \hline\hline
         spectral & \meanBz & 388nm & 602nm & 1.61$\mu$m & 8.04$\mu$m \\
         type & [G] & [K] \ & [K] \ & [K] \ & [K]\hspace*{.8ex} \\ \hline
         K0 &  --  & 4181 & 4787 & 5541 & 4333 \\
         K0 & 100 & $+43$ & $+26$ & $+12$ & $+27$ \\
         K0 & 300 & $+83$ & $+56$ & $+23$ & $+57$ \\
         K0 & 500 & $+86$ & $+56$ & $+14$ & $+62$ \\
         \hline
         M0 & -- & 3480 & 3701 & 4390 & 3428 \\
         M0 & 100 & $+19$ & $+7$ & $+4$ & $+8$ \\
         M0 & 300 & $+16$ & $+18$ & $+8$ & $+20$ \\
         M0 & 500 & $+14$ & $+17$ & $+5$ & $+24$ \\
         \hline
         M2 & --  & 3335 & 3407 & 4062 & 3195 \\
         M2 & 100 & $+11$ & $+7$ & $+2$ & $+4$ \\
         M2 & 300 & $+6$ & $+5$ & $-4$ & $+3$ \\
         M2 & 500 & $-3$ & $-7$ & $-15$ & $-3$ \\
         \hline
     \end{tabular}
\end{table}

\section{Additional Histograms}
\label{sec:histo_additional}
For ease of comparison with previous work, histograms of intensity for K0, M0 and M2 snapshots (left, middle and right-hand columns, respectively) as in Fig.~\protect{\ref{fig:hist1D}} are reproduced here using a linear $y$ scale.
\begin{figure*}
\vbox{\begin{overpic}[height=.15\textheight,trim={5 10 5 5},clip]{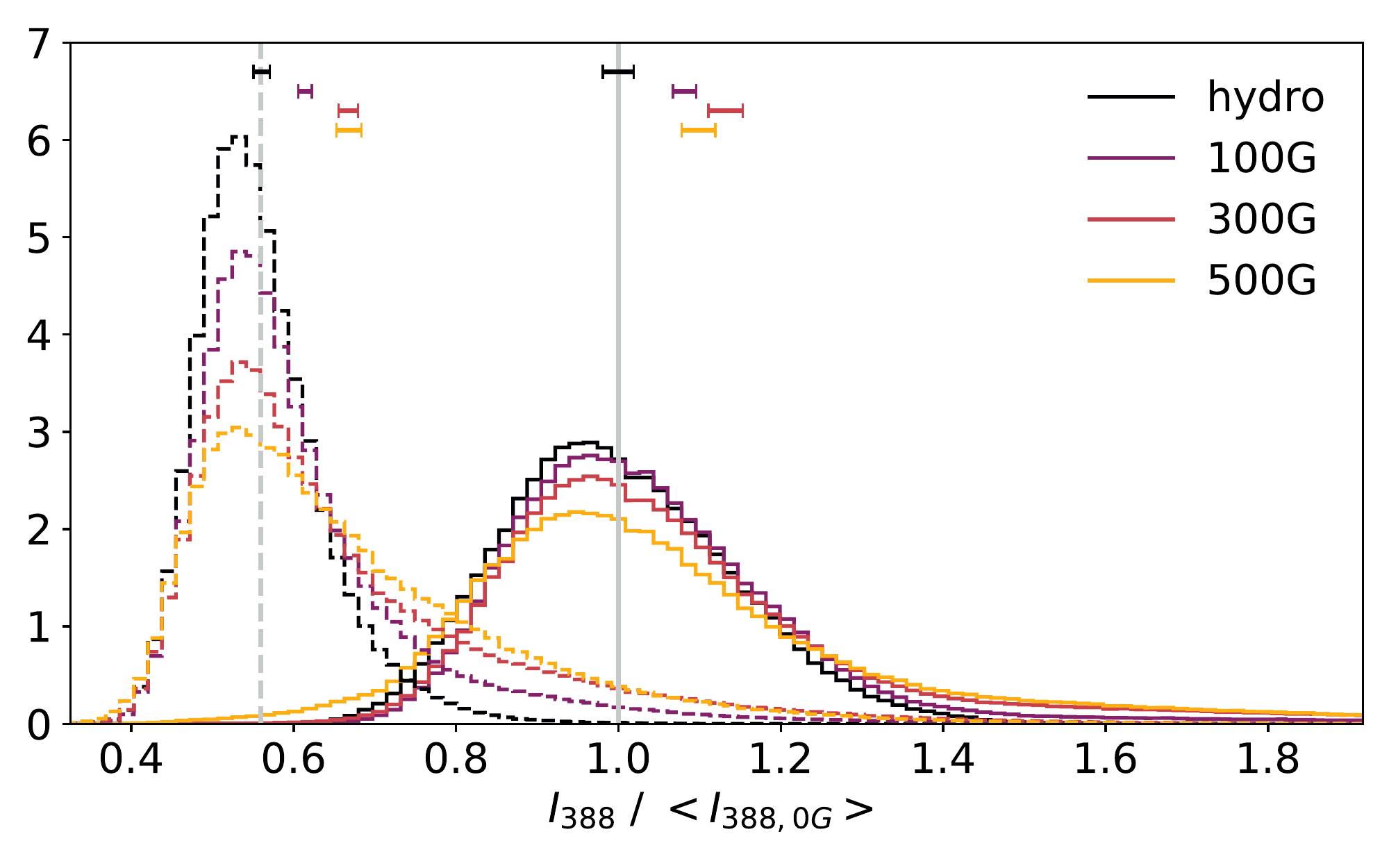}
    \put(52,61){\textsf{\Large K0}}
\end{overpic}
\begin{overpic}[height=.15\textheight,trim={5 10 5 0},clip]{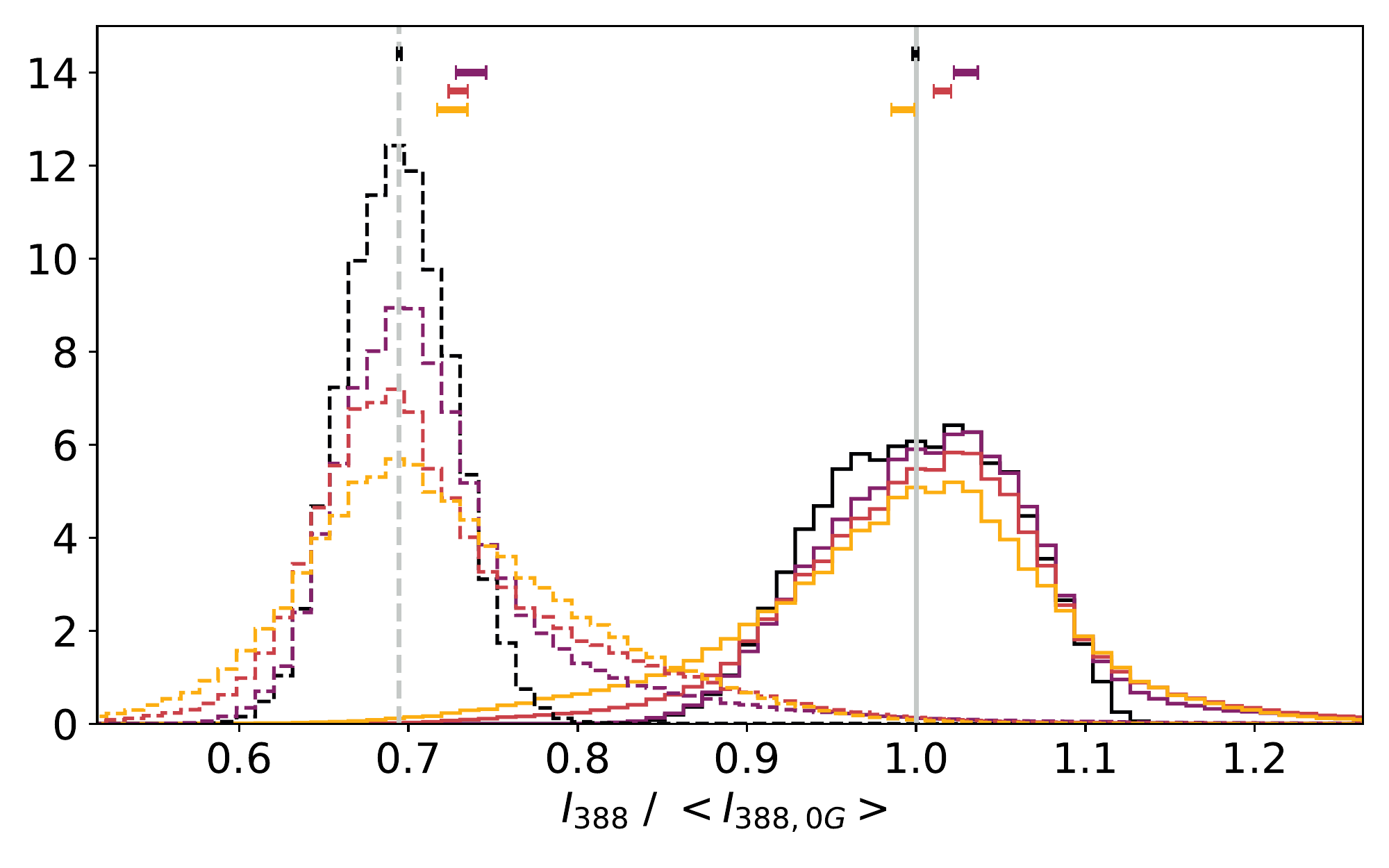}
    \put(52,61){\textsf{\Large M0}}
\end{overpic}
\begin{overpic}[height=.15\textheight,trim={5 10 5 5},clip]{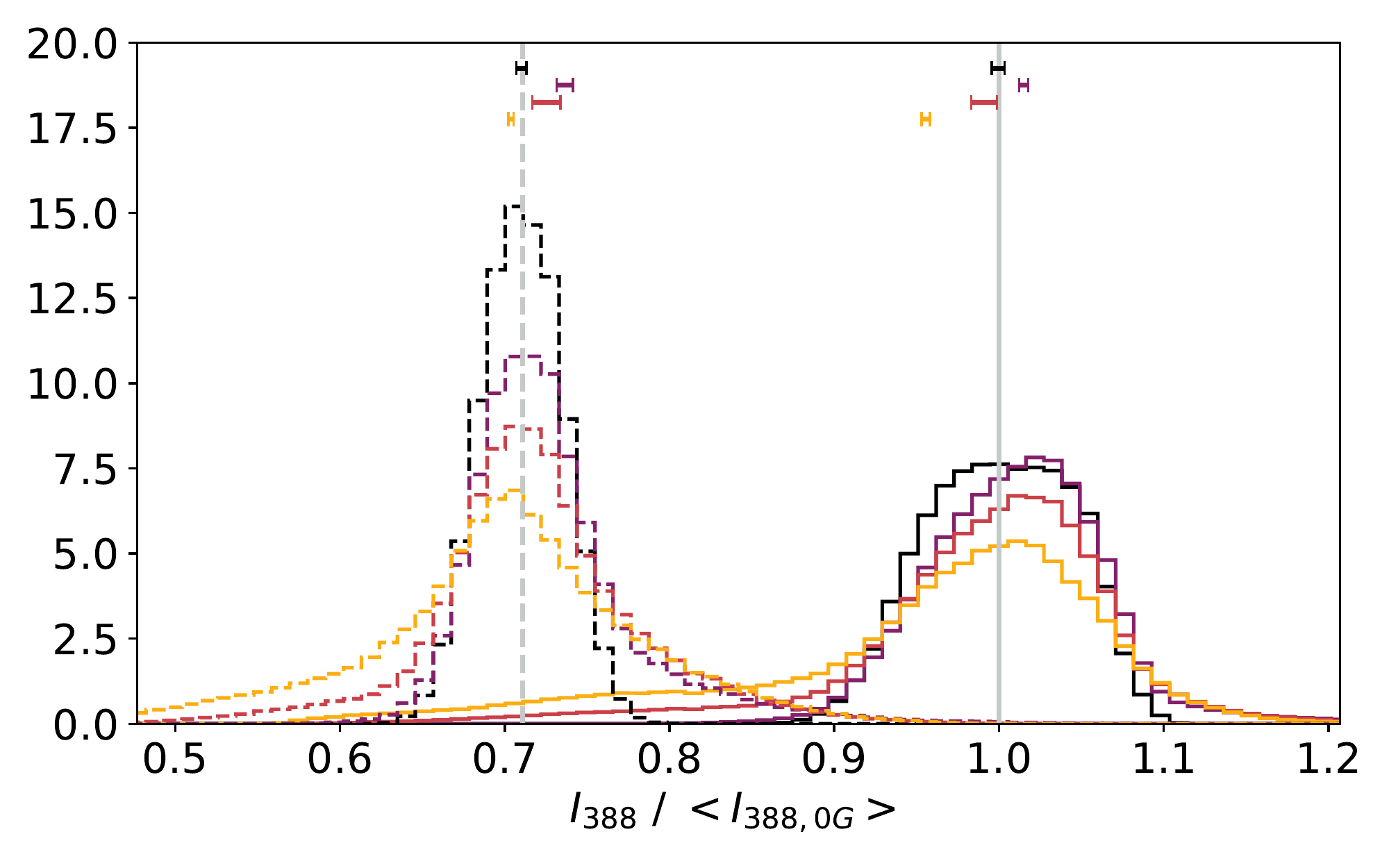}
    \put(52,61){\textsf{\Large M2}}
\end{overpic}
}
\vbox{\includegraphics[height=.15\textheight,trim={5 10 5 5},clip]{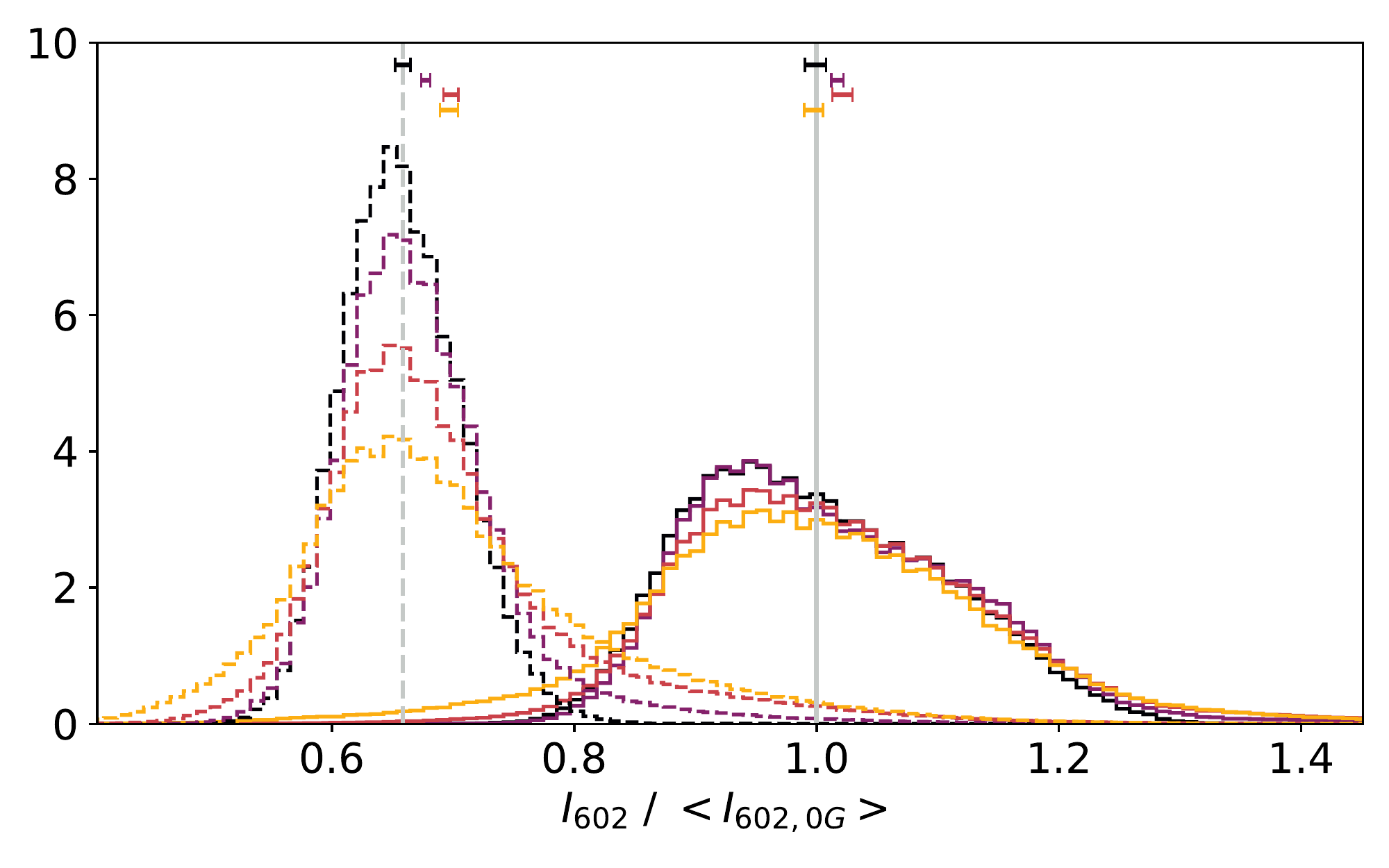}
\includegraphics[height=.15\textheight,trim={5 10 5 5},clip]{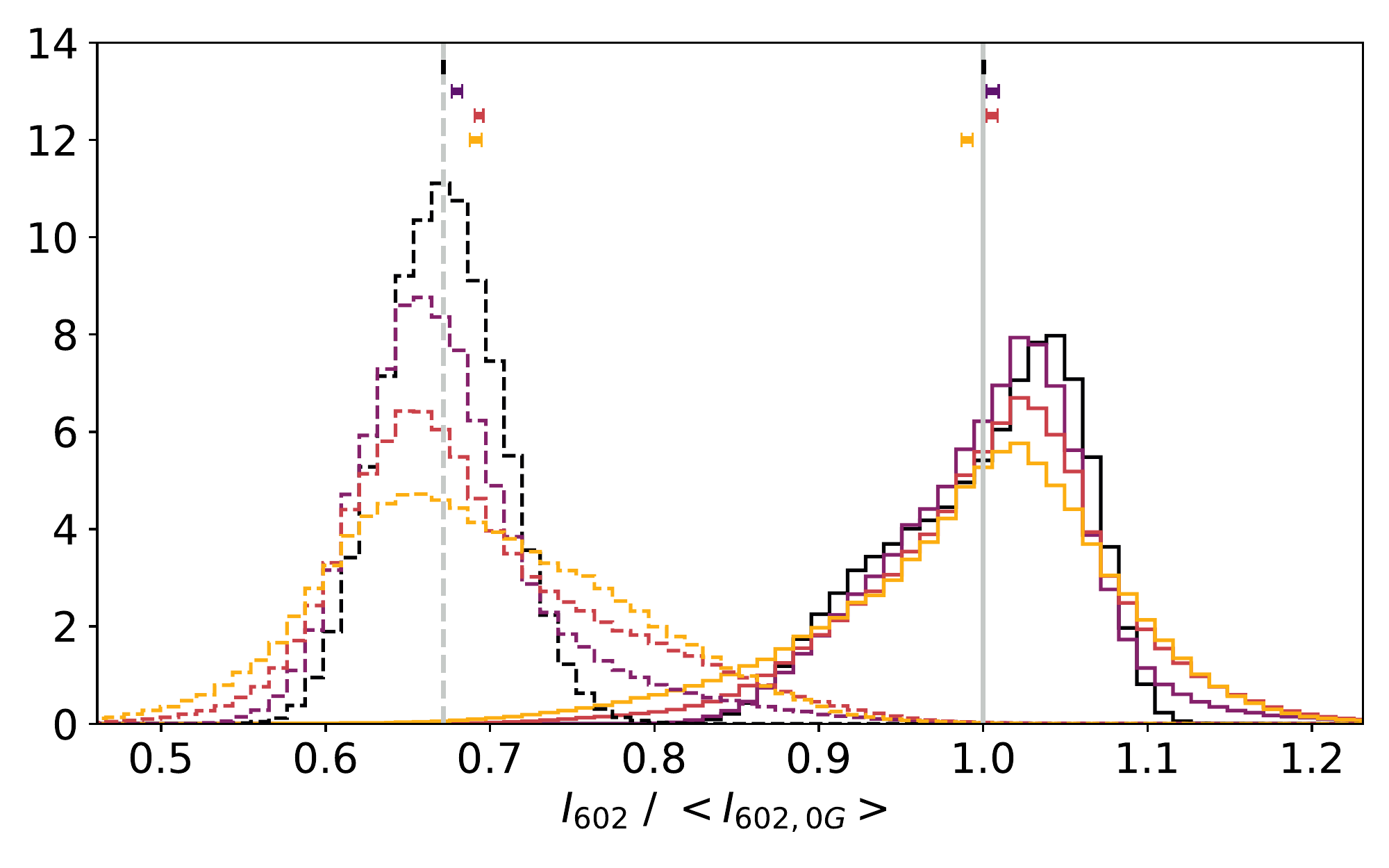}
\includegraphics[height=.15\textheight,trim={5 10 5 5},clip]{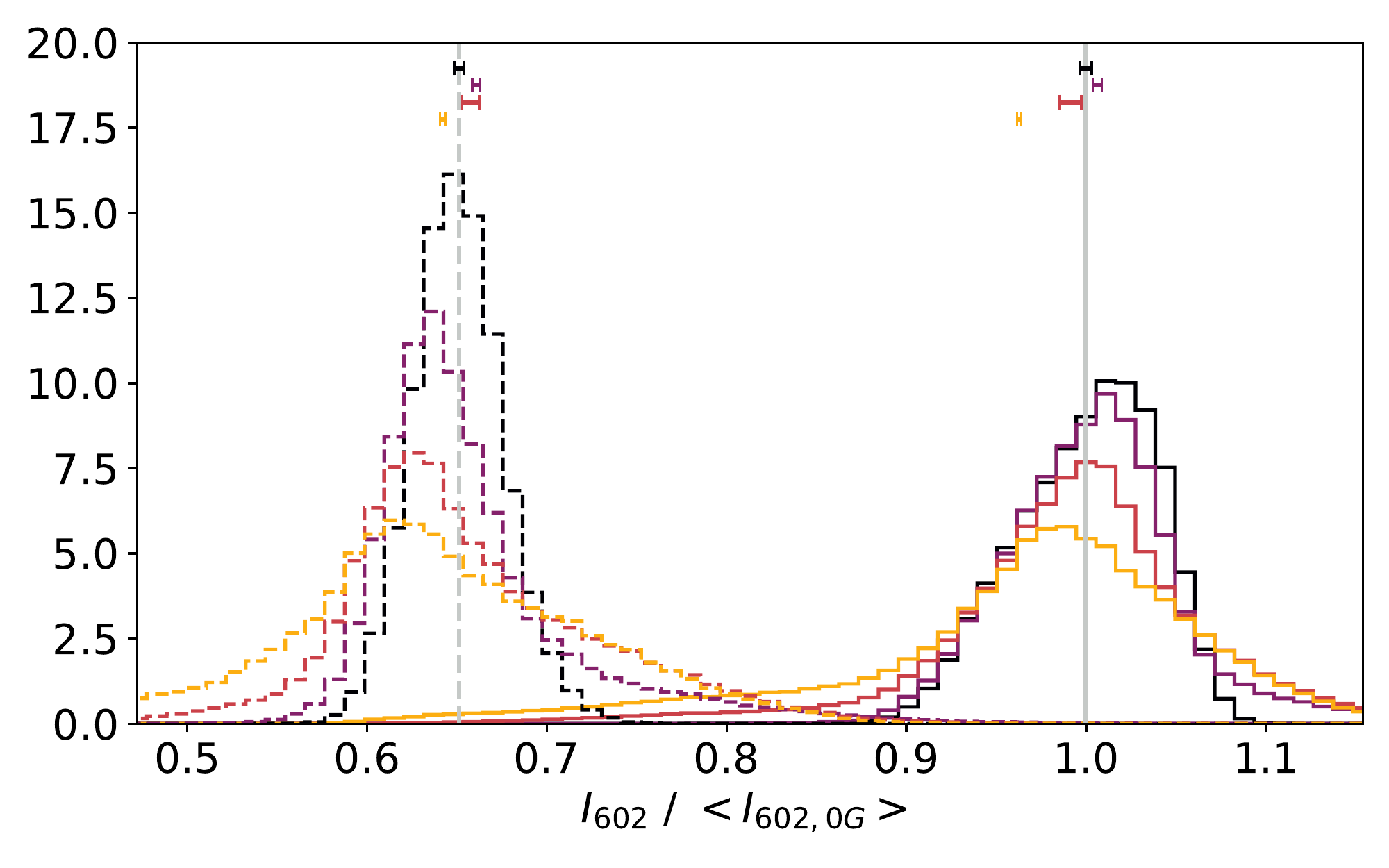}}
\vbox{\includegraphics[height=.15\textheight,trim={5 10 5  5},clip]{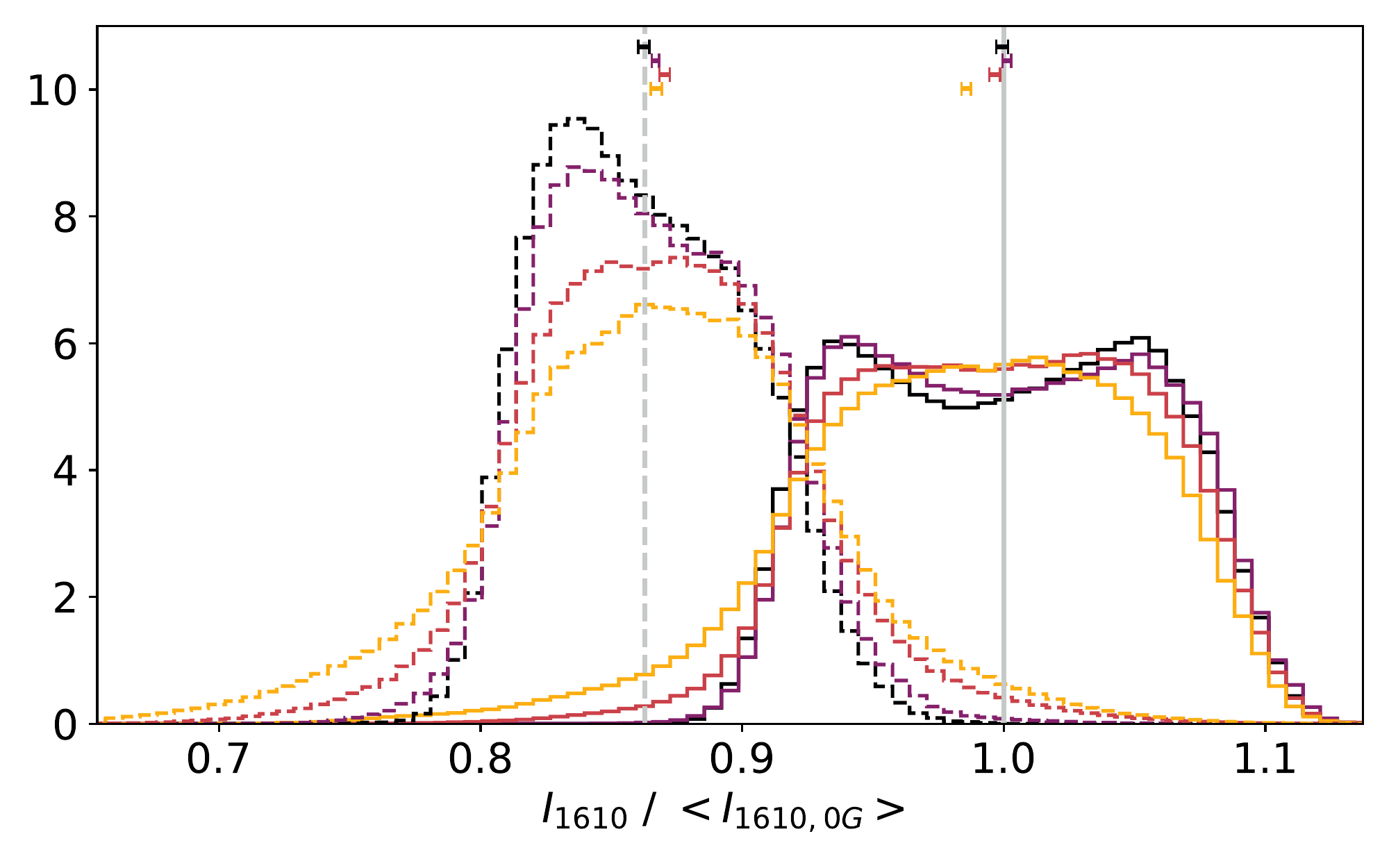} \includegraphics[height=.15\textheight,trim={5 10 5 5},clip]{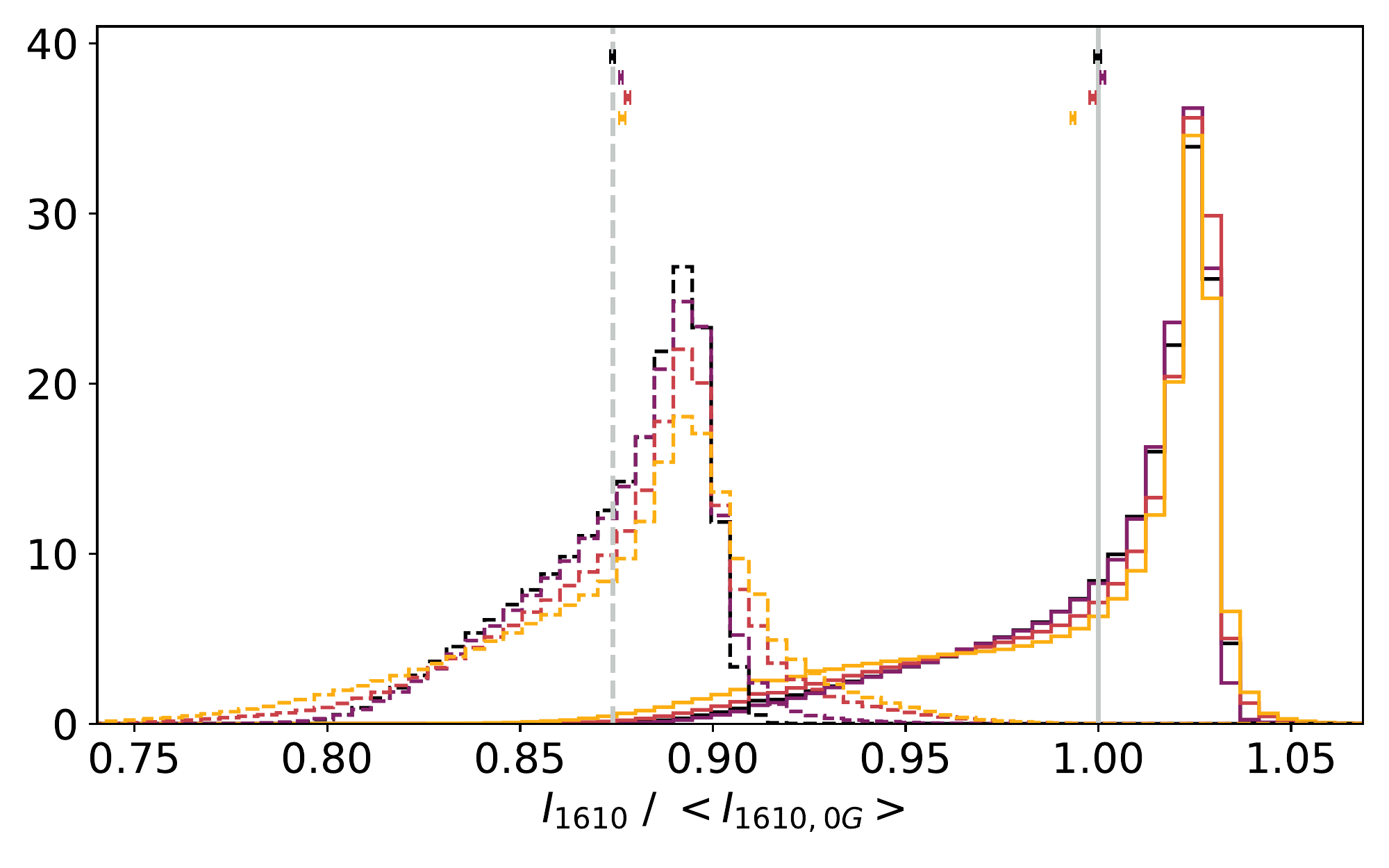}  \includegraphics[height=.15\textheight,trim={5 10 5 5},clip]{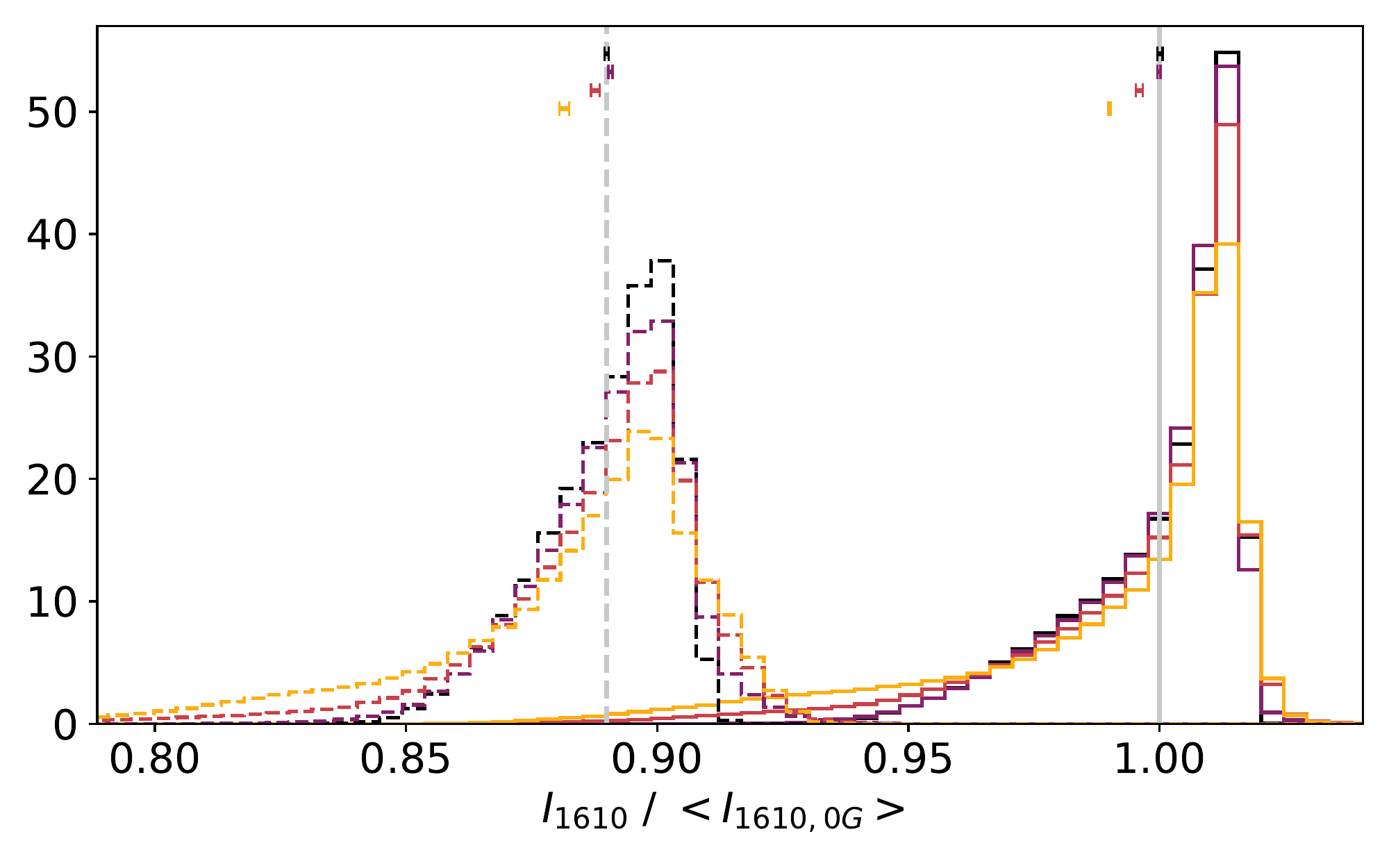}
\vbox{
\includegraphics[height=.15\textheight,trim={5 10 5 5},clip]{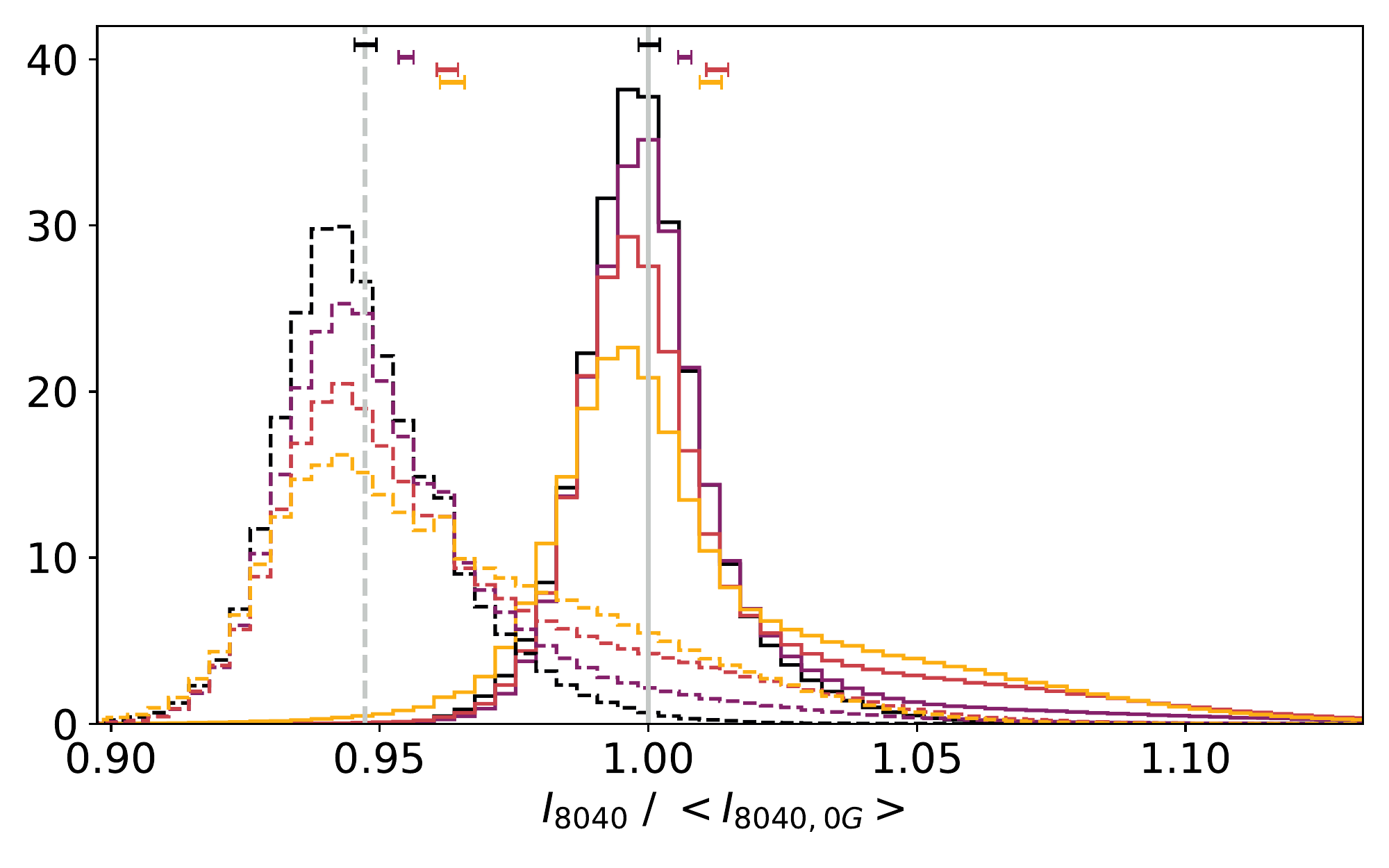}  \includegraphics[height=.15\textheight,trim={5 10 5 5},clip]{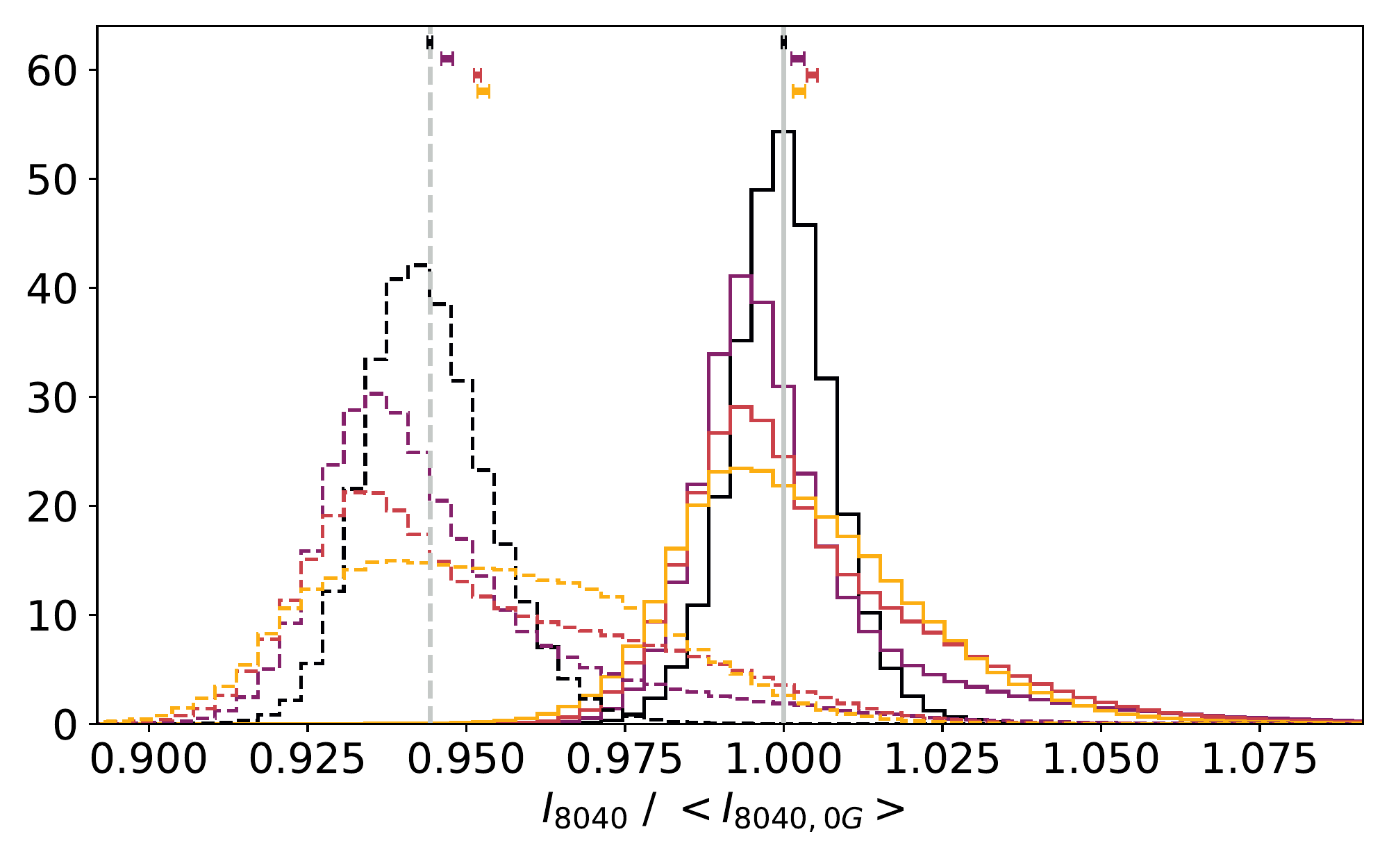}  \includegraphics[height=.15\textheight,trim={5 10 5 5},clip]{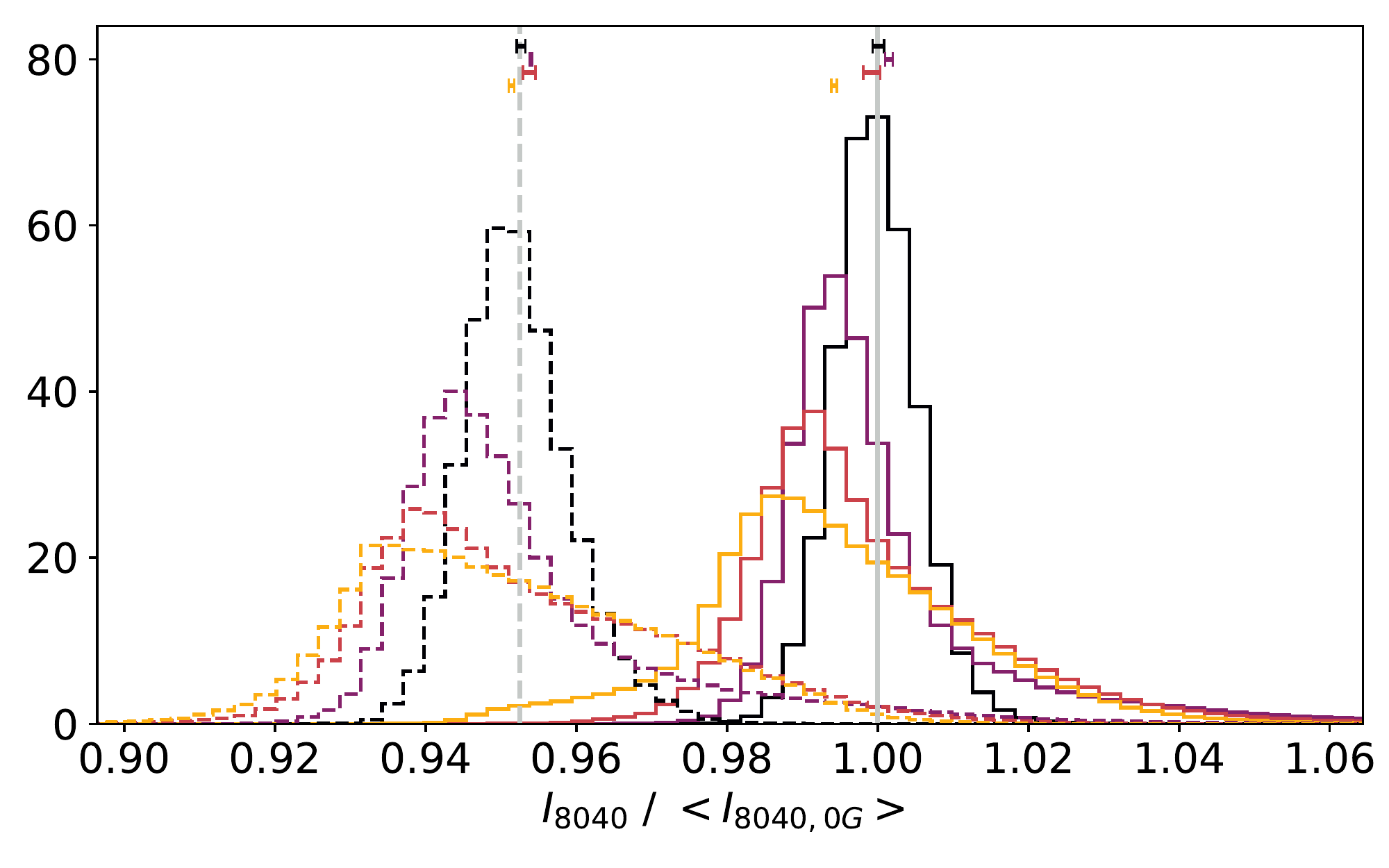}}
\caption[Histograms of intensity for K0, M0 and M2 snapshots at disc-centre and $\mu$ = 0.5.]{Histograms of intensity for K0, M0 and M2 snapshots (left, middle and right-hand columns, respectively). These are equivalent to the ones shown in Fig.~\protect{\ref{fig:hist1D}}, but displayed using a linear $y$ scale.}
}
\label{fig:hist1D3lin}
\end{figure*}
\end{appendix}
\end{document}